\documentclass[10pt]{article}
\usepackage[a4paper,nohead]{geometry}
\usepackage[utf8x]{inputenc}
\usepackage[T1]{fontenc}
\usepackage{graphicx}
\usepackage{amsmath}
\usepackage[sort&compress,numbers]{natbib}
\usepackage{amsfonts}
\usepackage{amssymb}
\usepackage{amsmath}
\usepackage{latexsym}
\usepackage{color}
\usepackage{ntheorem}
\usepackage{braket}
\usepackage[colorlinks=true]{hyperref}
\usepackage{authblk}
\usepackage{tensor}

\def\volform{\mbox{$\eta$}}

\newcounter{mnotecount}[section]

\renewcommand{\themnotecount}{\thesection.\arabic{mnotecount}}
\newcommand{\mnote}[1]
{\protect{\stepcounter{mnotecount}}$^{\mbox{\footnotesize
$
\bullet$\themnotecount}}$ \marginpar{
\raggedright\tiny\em
$\!\!\!\!\!\!\,\bullet$\themnotecount: #1} }

\newtheorem{theorem}{\sc  Theorem\rm}[section]

\newtheorem{corollary}[theorem]{\sc  Corollary\rm}

\newtheorem{lemma}[theorem]{\sc Lemma\rm}

\newtheorem{proposition}[theorem]{\sc Proposition\rm}

\newtheorem{remark}[theorem]{\sc Remark\rm}

\newcommand{\jlcax}[1]{}
\newcommand{\eean}{\nonumber\end{eqnarray}}

\newcommand{\kk}[1]{}

\newcommand{\beq}{\begin{equation}}

\newcommand{\FS}       
                  {F}

\newcommand{\HS} 
       {H_{\mbox{\scriptsize volume}}}

{\ptc{this should be removed in the oberwolfach version}}%

\newcommand{\eeal}[1]{\label{#1}\end{eqnarray}}
\newcommand{\bed}{\begin{deqarr}}
\newcommand{\eed}{\end{deqarr}}
\newcommand{\bedl}[1]{\begin{deqarr}\label{#1}}
\newcommand{\eedl}[2]{\arrlabel{#1}\label{#2}\end{deqarr}}

\newcommand{\bel}[1]{\begin{equation}\label{#1}}
\newcommand{\bea}{\begin{eqnarray}}
\newcommand{\bean}{\begin{eqnarray}\nonumber}
\newcommand{\beal}[1]{\begin{eqnarray}\label{#1}}
\newcommand{\eea}{\end{eqnarray}}

\newcommand{\be}{\begin{equation}}
\newcommand{\eeq}{\end{equation}}
\newcommand{\ee}{\end{equation}}
\newcommand{\beqa}{\begin{eqnarray}}
\newcommand{\eeqa}{\end{eqnarray}}
\newcommand{\beqan}{\begin{eqnarray*}}
\newcommand{\eeqan}{\end{eqnarray*}}
\newcommand{\ba}{\begin{array}}
\newcommand{\ea}{\end{array}}

\newcommand{\mcD}{{\mycal D}}

\newcommand{\scri}{{\mycal I}}%

\newcommand{\warn}[1]
{\protect{\stepcounter{mnotecount}}$^{\mbox{\footnotesize
$
\bullet$\themnotecount}}$ \marginpar{
\raggedright\tiny\em
$\!\!\!\!\!\!\,\bullet$\themnotecount: {\bf Warning:} #1} }

\newcommand{\R}{\mathbb R}

\newcommand{\eq}[1]{(\ref{#1})}

\newcommand{\ptc}[1]{\mnote{{\bf ptc:}#1}}

\newcommand{\mcL}{{\mycal L}}

\newcommand{\beqar}{\begin{deqarr}}
\newcommand{\eeqar}{\end{deqarr}}

\newcommand{\beaa}{\begin{eqnarray*}}
\newcommand{\eeaa}{\end{eqnarray*}}

\newcommand{\hmetr}{\mathfrak{h}}
\newcommand{\hvolf}{\epsilon}
\newcommand{\rnabla}{\not\hspace{-.2em}\nabla}
\newcommand{\reta}{\not\hspace{-.2em}\eta}
\newcommand{\Eqref}[1]{Eq.~\eqref{#1}}
\newcommand{\Eqsref}[1]{Eqs.~\eqref{#1}}
\newcommand{\Sectionref}[1]{Section~\ref{#1}}
\newcommand{\Sectionsref}[1]{Sections~\ref{#1}}

\newcommand{\Propref}[1]{Proposition~\ref{#1}}
\newcommand{\Lemref}[1]{Lemma~\ref{#1}}
\newcommand{\Theoremref}[1]{Theorem~\ref{#1}}

\newcommand{\keyword}[1]{\textit{#1}}
\newcommand{\imagi}{\mathrm{i}}

\DeclareFontFamily{OT1}{rsfs}{}
\DeclareFontShape{OT1}{rsfs}{m}{n}{ <-7> rsfs5 <7-10> rsfs7 <10-> rsfs10}{}
\DeclareMathAlphabet{\mycal}{OT1}{rsfs}{m}{n}

{\catcode `\@=11 \global\let\AddToReset=\@addtoreset}
\AddToReset{equation}{section}

{\catcode `\@=11 \global\let\AddToReset=\@addtoreset}
\AddToReset{figure}{section}

{\catcode `\@=11 \global\let\AddToReset=\@addtoreset}
\AddToReset{table}{section}

\begin{document}

\title{Analysis of a Bianchi-like equation satisfied by the Mars-Simon tensor
\thanks{Preprint UWThPh-2017-14.}
}
\author[1]{Florian Beyer}
\author[2]{Tim-Torben Paetz}
\affil[1]{Department of Mathematics and Statistics, University of Otago, P.O. Box 56, Dunedin 9054, New Zealand}
\affil[2]{Gravitational Physics, University of Vienna, Boltzmanngasse 5, 1090 Vienna, Austria}

\maketitle

\vspace{-0.2em}

\begin{abstract}
The Mars-Simon tensor (MST), which e.g.\ plays a crucial role to provide gauge invariant characterizations of the Kerr-NUT-(A)(dS) family, satisfies a Bianchi-like equation.
In this paper we analyze this equation in close analogy to the Bianchi equation, in particular
it will be shown that the constraints are preserved supposing that a generalized Buchdahl condition holds.
This permits the systematic construction of solutions to this equation in terms of a well-posed Cauchy problem.
A particular emphasis  lies on the asymptotic Cauchy problem,  where data are prescribed on a spacelike $\scri$ (i.e.\ for $\Lambda >0$).
In contrast to the Bianchi equation,  the MST equation  is of Fuchsian type at $\scri$, for which existence and uniqueness results are derived.
\end{abstract}



\section{Introduction}

The Kerr-NUT-(A)(dS) family plays a distinguished role among  solutions to Einstein's field equations: Not merely is it an explicitly known family
of vacuum solutions but it is also expected that e.g.\ the Kerr subfamily  satisfies certain  black hole uniqueness and stability results.
In fact for the Kerr-de Sitter family stability has recently been established \cite{hintz_vasy}, while uniqueness is still open.

While the explicit form of the solutions employs coordinates which are adapted to the symmetries, it is equally important to have gauge invariant
characterizations at hand.
One approach to accomplish this  is based on the so-called \emph{Mars-Simon tensor (MST)} \cite{IK, mars, mars1, mars-senovilla, simon}, which may be regarded as a generalization of the conformal Weyl tensor
in spacetimes which admit a Killing vector:
Given a vacuum solution of the Einstein equations with cosmological constant $\Lambda$, the vanishing of the Weyl tensor shows that the spacetime
is locally Minkowski or (Anti-)de Sitter, respectively.
Similarly, the vanishing of the MST implies that the spacetime is locally given by a certain explicitly known family of solutions, classified in \cite{mars-senovilla},
which contains in particular the Kerr-NUT-(A)dS family.

It turns out that the MST satisfies an equation very similar to the Bianchi equation which is satisfied by the Weyl tensor.
This ``MST equation'' , which has been derived in \cite{IK} for $\Lambda=0$ and in \cite{mpss} for arbitrary signs of $\Lambda$,
played an important role in the derivation of   Kerr uniqueness results \cite{aik1, IK}, and in a characterization
of $\Lambda>0$-vacuum spacetimes in terms of their asymptotic Cauchy data on $\scri$ \cite{mpss}.
In fact, the motivation for this paper comes from the latter one: In contrast to the Bianchi equation the MST equation
(or rather its analog in an appropriately conformally rescaled spacetime)  is not regular near a smooth spacelike
$\scri$. Instead, the equation is shown there to be  of \emph{Fuchsian type}.

The  MST equation can be split into a symmetric hyperbolic system of evolution equations, and a system of constraint equations \cite{mpss}.
As for the Bianchi equation, we will  show that the constraints are preserved under evolution, supposing that a certain \emph{generalized Buchdahl condition} holds.
Given  a solution to the constraints on some Cauchy surface, the existence of a solution to the full system follows from standard results.
When the initial hypersurface is a spacelike $\scri$, though, the Cauchy  problem becomes more involved due to the Fuchsian character of the system at $\scri$.
In \cite{mpss} uniqueness of solutions which extend smoothly through $\scri$ with regular data at $\scri$ has been proved.
It  is a main goal of this paper to analyze in detail the behavior of solutions near $\scri$ and establish a well-posedness result for this 
\emph{singular initial value problem}
(in contrast, it is easy to solve the constraints on $\scri$, whereas they  are more involved on an ordinary  Cauchy surface of the physical spacetime).

We tackle the singular initial value problem by means of the \emph{Fuchsian technique}. Earlier Fuchsian
techniques by Rendall and
co-authors, which were later
applied to problems in general relativity
\cite{Isenberg:1999ba,Isenberg:2002ku,ChoquetBruhat:2006fc,Andersson:2001fa,Damour:2002exa,lrr-2008-1}, were restricted to the real-analytic setting.
The first attempts to overcome the analyticity restriction were made
in \cite{Claudel:1998tt,Rendall:2000ki}. A series of papers
\cite{Beyer:2010foa,Beyer:2011ce,Ames:2012vz,Ames:2013uh, beyer11, Beyer:2015fhs,Ames:2016uy}
led to a version of the Fuchsian technique which applies to a general class of
quasilinear hyperbolic equations without the analyticity
restriction.

We study the MST  equation on a given background, and a  solution thereof  will  generally not coincide with the MST of the background spacetime.
The reasons why this is interesting are basically the same as for the Bianchi equation: A solution to the Bianchi equation
 is in one-to-one correspondence  with  a solution to the linearized  Einstein equations (at least for $\Lambda=0$),
and a similar result
should be expected for a solution to  the MST
equation on a background with vanishing MST (we will only partially address this issue).
Moreover, the importance of the Bianchi equation arises from its appearance as part of Friedrich's conformal field equations \cite{F3};
one might think of a similar system of equations where the MST equation replaces the Bianchi equation in order to
study certain classes of perturbations of
spacetimes which admit a Killing vector field (e.g.\ since the (conformally rescaled) MST is much better behaved near the poles of a Kerr-de Sitter $\scri$ than
the (conformally rescaled) Weyl tensor). Analyzing  this equation on a given background then  provides a toy model for this.

While the Bianchi  equation is meaningful on any background spacetime,  the MST equation can only be posed on a background which admits a Killing vector field.
However, due to the Buchdahl condition  well-posedness results for the Bianchi equation are only available on locally  conformally flat backgrounds.
Corresponding results for the MST equation permit a larger class of background spacetimes, namely those  with vanishing MST such as e.g.\ the Kerr-NUT-(A)dS family.

The structure of the paper is as follows: In Section~\ref{sec_MST} we recall the definition of the MST.
In Section~\ref{sec_Bianchi} we review a couple of results on the Bianchi equation as a guideline for our analysis of the MST equation.
Section~\ref{sec_MST_eqn} is devoted to the study of the MST equation. More precisely, in Section~\ref{sec_lin_MST}
we will prove  that the linearized MST satisfies the MST equation on a background with vanishing MST.
In the  remainder of that section  we will establish a generalized Buchdahl condition which ensures that the constraints are preserved under evolution.

In Section~\ref{sec_conf_MST_eqn}
 we will consider the analog of the MST equation in Penrose's conformally rescaled spacetimes.
We then restrict attention to the case where the cosmological constant is positive, so that $\scri$ is spacelike.
The behavior of the constraint and evolution equations near $\scri$ is unveiled in Section~\ref{sec_asymp_behavior}.

In  \Sectionref{sec:fuchsian}, we solve the singular initial value problem of the MST equations with data on ${\scri^-}$. In \Sectionref{sec:prelim} we establish some basic notation, list our assumptions and formulate the main theorem, \Theoremref{thm:maintheorem}. The remainder of \Sectionref{sec:fuchsian} is devoted to the proof of that theorem. \Sectionref{sec:spectraldecomp} provides a detailed study of a particular matrix in  the evolution equations which largely determines the leading singular behavior of solutions at ${\scri^-}$. In \Sectionref{sec:SIVPevol} we investigate the evolution equation of the remainder of the singular initial value problem and establish certain decay conditions. The leading-order term for the singular initial value problem is derived in \Sectionref{sec:deriveLOT}. By that stage we will have identified the singular data and established a local existence theory for the singular initial value problem of the evolution equations. In \Sectionref{sec_constraints}, we then find conditions on the singular data which guarantee that the asserted family of solutions of the evolution equations also satisfies the constraints. Finally, we show in \Sectionref{sec:smoothextendibility} that the same conditions also imply that the solutions extend smoothly through ${\scri^-}$ once an overall singular term  expected for generic MSTs has been subtracted. Notice here that according to the general structure of the singular data, generic solutions of the evolution equations do not have this property.

\subsection{Notation}

For the convenience of the reader let us finish the introduction with some remarks concerning the notation.
The ``physical'' spacetime, solution to Einstein's vacuum field equations with cosmological constant $\Lambda$, will be denoted by $(M,g)$. 
Throughout the paper we assume $( M , g)$ to be a smooth $3+1$-dimensional connected, oriented and time-oriented Lorentzian manifold with signature $(-,+,+,+)$.
Its associated Levi-Civita covariant derivative,
connection coefficients and volume form are denoted by $\nabla_{\mu}$, $\Gamma^{\alpha}_{\mu\nu}$ and $\eta_{\mu\nu\sigma\rho}$, respectively.
Our conventions concerning Riemann tensor, Ricci tensor etc.\ follow those in \cite{wald}.

The conformally rescaled counterpart of  $(M,g)$  will be denoted by $(\widetilde M, \widetilde g)$. Correspondingly, objects associated to $\widetilde g$
are decorated with a tilde,  $\widetilde\nabla_{\mu}$, $\widetilde\Gamma^{\alpha}_{\mu\nu}$ and $\widetilde\eta_{\mu\nu\sigma\rho}$.

 Spacetime indices are Greek. Coordinates in $3+1$ splits are denoted by $\{x^{\mu}\} =\{t\equiv x^0, x^i\}$
 with corresponding tensorial indices.
Objects associated to the family $t\mapsto g_{ij}(t, x^k)$ of Riemannian metrics are marked with a slash, $\rnabla_i$, $\not\hspace{-.2em}\Gamma^k_{ij}$ 
and $\reta_{ijk}$. The action of $\rnabla_i$ on spacetime tensors is defined as follows:
\begin{eqnarray}
\rnabla_i v_0 &:=& \partial_i v_0
\,,
\\
\rnabla_i v_j &:=& \partial_iv_j -\not\hspace{-.2em}\Gamma^k_{ij}v_k
\,.
\end{eqnarray}
and similarly for tensors of higher rank.
Again, the corresponding objects associated to the family  $t\mapsto \widetilde g_{ij}(t, x^k)$ in the conformally rescaled spacetime
are decorated with a tilde.

For $\Lambda>0$ and in spacetimes with an appropriate fall-off behavior of the gravitational field, (past) null infinity is represented in $(\widetilde M, \widetilde g)$
by a spacelike hypersurface $\scri^-$.
The Riemannian metric induced by $\widetilde g_{\mu\nu}$ on $\scri^-$ will be denoted by $\hmetr_{ij}$, its covariant derivative by $\mcD_i$
 and the volume form by $\hvolf_{ijk}$ (the only exception will be Section~\ref{sec_Cauchy_Buchdahl}, where $\mcD_i$ and  $\hvolf_{ijk}$ denote  covariant derivative and volume form of the induced metric  on an arbitrary Cauchy surface).

\section{Mars-Simon tensor (MST)}
\label{sec_MST}

Let $( M , g)$
 be a $\Lambda$-vacuum spacetime which admits a Killing vector field $X$.
The  \emph{Mars-Simon tensor (MST)}, cf.\ \cite{IK, mars, mars1, mars-senovilla, simon}, is defined as follows:
\begin{equation}
\mathcal{S}_{\mu\nu\sigma\rho} := \mathcal{C}_{\mu\nu\sigma\rho} + Q\, \mathcal{U}_{\mu\nu\sigma\rho}
\,,
\label{dfn_MST}
\end{equation}
where
\begin{eqnarray}
 \mathcal{C}_{\mu\nu\sigma\rho} &:=&  C_{\mu\nu\sigma\rho} + \imagi C^{\star}_{\mu\nu\sigma\rho}
\,,
\\
 \mathcal{U}_{\mu\nu\sigma\rho} &:=&  \mathcal{F}_{\mu\nu}\mathcal{F}_{\sigma\rho} +\frac{1}{3}\mathcal{F}^2 \mathcal{I}_{\mu\nu\sigma\rho}
\,,
\\
 \mathcal{I}_{\mu\nu\sigma\rho} &:=& \frac{1}{4}(2g_{\mu[\sigma}g_{\rho]\nu} + \imagi \volform_{\mu\nu\sigma\rho})
\,,
\\
\mathcal{F}_{\mu\nu} &:=& F_{\mu\nu} + \imagi F^{\star}_{\mu\nu}
\,,
\\
\mathcal{F}^2 &:=& \mathcal{F}_{\mu\nu} \mathcal{F}^{\mu\nu}
\,,
\\
F_{\mu\nu} &:=& \nabla_{\mu}X_{\nu}
\,.
\end{eqnarray}
In these expressions $\volform_{\mu\nu\sigma\rho}$ is the volume form of $g$ and $\star$ the corresponding Hodge dual.
$\mathcal{C}_{\mu\nu\sigma\rho}$ and $\mathcal{F}_{\mu\nu}$ are the self-dual Weyl tensor and the self-dual Killing form. They satisfy
$\mathcal{C}^{\star}_{\mu\nu\sigma\rho}=-\imagi \mathcal{C}_{\mu\nu\sigma\rho}$ and $\mathcal{F}_{\mu\nu}^{\star} =-\imagi\mathcal{F}_{\mu\nu}$.
The symmetric double two-form $ \mathcal{I}_{\mu\nu\sigma\rho} $ provides a metric in the space of self-dual two-forms in the sense that
$ \mathcal{I}_{\mu\nu\sigma\rho} \mathcal{W}^{\sigma\rho}=\mathcal{W}_{\mu\nu}$ for any self-dual two-form $\mathcal{W}_{\mu\nu}$.
Some useful identities satisfied by self-dual tensors can be found e.g.\ in \cite{IK, israel}.

The MST is a Weyl field, i.e.\ it has all the algebraic symmetries of the Weyl tensor,
\begin{equation}
\mathcal{S}_{(\mu\nu)\sigma\rho}= 0\,, \quad
\mathcal{S}_{\mu\nu\sigma\rho}= \mathcal{S}_{\sigma\rho\mu\nu}\,, \quad
g^{\nu\rho}\mathcal{S}_{\mu\nu\sigma\rho}=0\,, \quad
\mathcal{S}_{[\mu\nu\sigma]\rho}=0
\,.
\end{equation}
Moreover, its Lie derivative along the associated Killing  vector $X$ vanishes
\begin{equation}
\mcL_X\mathcal{S}_{\mu\nu\sigma\rho} =0
\,,
\end{equation}
supposing that $Q$ satisfies $\mcL_X Q=0$ (as it is the case for the $Q$ defined below).

In the literature different definitions of the function $Q:  M \mapsto \mathbb{C}$ have proven to be advantageous in different contexts \cite{IK, mars-senovilla, mpss, kerr_char}.
All the different choices for $Q$ are obtained by requiring a certain component of the MST (or a derivative thereof) to vanish, and are therefore equivalent
in spacetimes where the MST vanishes (supposing that certain quantities are non-zero).
Here we are interested in the Bianchi-like equations satisfied by the MST, and this requires a particular choice for the function $Q$ \cite{mars-senovilla}:
\begin{eqnarray}
Q &:=& \frac{3J}{R}-\frac{\Lambda}{R^2}
\,,
\label{dfn_Q}
\\
R &:=& -\frac{\imagi}{2}\sqrt{\mathcal{F}^2}
\,,
\\
J &:=& \frac{R + \sqrt{R^2-\Lambda \sigma}}{\sigma}
\,,
\label{dfn_J}
\end{eqnarray}
and where $\sigma$ denotes the Ernst potential of the closed 1-form
\begin{equation}
\sigma_{\beta} :=2X^{\alpha}\mathcal{F}_{\alpha\beta}
\,.
\end{equation}
In general the Ernst potential exists only locally and is defined only up to an additive complex constant.
In this paper we will eventually restrict attention to background spacetimes with vanishing MST and in that case
the Ernst 1-form is exact and the additive constant is determined by the requirement on the MST to vanish \cite{mars-senovilla}.

The definition of the complex square roots depends on the sign of the cosmological constant.
For $\Lambda>0$ (the case which we are primarily  interested in) the square root is preferably chosen  in such a way that the  real part of $R$
approaches minus infinity at $\scri$, in
agreement with the usual behavior for e.g.\ the Kerr-de Sitter family in Boyer-Lindquist type coordinates  (cf.\ \cite{mpss} for more details).
With the signs chosen in \eq{dfn_Q}-\eq{dfn_J} this is achieved by taking the positive branch, i.e.\ the one that takes positive real numbers
and gives positive real values.

In \cite{mpss} this choice for $Q$ was denoted by $Q_{\mathrm{ev}}$. Since we  assume $Q=Q_{\mathrm{ev}}$ throughout the paper, no confusion
 arises and we will simply write $Q$ henceforth.

\begin{proposition}[\cite{mpss}, cf.\ \cite{IK} for the $\Lambda=0$-case]

Let $( M , g, X)$ be a $\Lambda$-vacuum spacetime with Killing vector field $X$ such that
\begin{equation}
\mathcal{F}^2\ne 0\,, \quad Q\mathcal{F}^2 + 8\Lambda\ne0\,, \quad \sigma \ne 0
\,.
\label{ineqs}
\end{equation}
 Then the MST \eq{dfn_MST} with $Q$ given by \eq{dfn_Q}
 satisfies the Bianchi-like equation
\begin{equation}
\nabla_{\rho}\mathcal{S}_{\alpha\beta\mu}{}^{\rho} =\mathcal{J}( {\mathcal{S}})_{\alpha\beta\mu}
\,,
\label{MST_evolution_eqn}
\end{equation}
where
\begin{eqnarray}
\mathcal{J}( {\mathcal{S}})_{\alpha\beta\mu}
&=&
4 \Lambda  \frac{  5  Q\mathcal{F}^2  +4\Lambda }{Q\mathcal{F}^2 + 8\Lambda}
 \mathcal{F}_{\alpha\beta}\mathcal{F}_{\mu\rho}
   \mathcal{F}^{-4}  X^{\sigma}  \mathcal{F}^{\gamma\delta}  {\mathcal{S}}_{\gamma\delta\sigma}{}^{\rho}
-QX^{\sigma} \mathcal{F}_{\mu\rho} {\mathcal{S}}_{\alpha\beta\sigma}{}^{\rho}
\nonumber
\\
&&
+ \frac{2}{3}  \mathcal{F}^{-2}  \Big(  Q \mathcal{F}^{2}
-2\Lambda  \frac{  5  Q\mathcal{F}^2  +4\Lambda }{Q\mathcal{F}^2 + 8\Lambda}
\Big) \mathcal{I}_{\alpha\beta\mu\rho}  X^{\sigma} \mathcal{F}^{\gamma\delta}  {\mathcal{S}}_{\gamma\delta\sigma}{}^{\rho}
\,.
\label{MST_rhs}
\end{eqnarray}
\end{proposition}

\begin{remark}
\label{properties_I}
{\rm
$\mathcal{J}( {\mathcal{S}})_{\alpha\beta\mu}$ has the following properties \cite{mpss}:
$$
\mathcal{J}( {\mathcal{S}})_{\alpha\beta\mu}=\mathcal{J}( {\mathcal{S}})_{[\alpha\beta]\mu}
\,, \quad
\mathcal{J}( {\mathcal{S}})_{[\alpha\beta\mu]}=0\,,
\quad
\mathcal{J}( {\mathcal{S}})^{\mu}{}_{\beta\mu}=0
\,.
$$
It is further self-dual in the first pair of anti-symmetric indices.
}
\end{remark}


\section{Bianchi equation}
\label{sec_Bianchi}

The MST has the same algebraic symmetries as the Weyl tensor, and  fulfills an equation similar, though more complicated, to the Bianchi equation for the  Weyl tensor.
In fact, the MST was introduced as a generalization of the Weyl tensor in $\Lambda$-vacuum spacetimes which admit a Killing vector field:
Its vanishing (supplemented by certain additional conditions) provides a local gauge-independent characterization of Kerr-(A)(dS) family
in much the same way as the vanishing of the Weyl tensor  provides a local characterization of Minkowski and (Anti-)de Sitter spacetime, respectively \cite{mars, mars1, mars-senovilla}.

While the MST equation comes along with some additional features such as the Fuchsian behavior near a spacelike $\scri$, a couple of results
which  hold for the Bianchi equation can be derived for the MST equation, as well. Even more, the analysis of the Bianchi equation  provides
a guideline how to analyze the MST equation and what results should be expected.
It is the aim of this section to review some crucial results on the  Bianchi equation.

Consider a spacetime $( M , g)$ which satisfies the vacuum equations $R_{\mu\nu}=\Lambda g_{\mu\nu}$. Then the Weyl tensor of $g$ satisfies the \emph{Bianchi equation}
\begin{equation}
\nabla_{\rho} C_{\mu\nu\sigma}{}^{\rho} =0
\,.
\label{Bianchi_eqn}
\end{equation}
Now, given a $\Lambda$-vacuum  spacetime $( M , g)$, it is of interest to analyze this equation on that given background  and construct solutions from appropriate initial surfaces:
\begin{enumerate}
\item[(i)]
First of all the Bianchi equation is part of Friedrich's conformal field equations  (cf.\ e.g.\ \cite{F3})
which replace  Einstein's vacuum field equations in a setting where
the metric is conformally rescaled. Analyzing \eq{Bianchi_eqn} on a background therefore provides  a kind of toy model for these equations.
\item[(ii)] 
On a conformally  flat background, solutions to the  linearized field  equations provide solutions  to the Bianchi equation,
cf.\ Lemma~\ref{lemma_Bianchi1} below.
Even more, solutions to \eq{Bianchi_eqn} turn out to be in one-to-one correspondence with solutions to the linearized field  equations (to our knowledge, this result has only been established on a flat background so far, cf. Lemma~\ref{lemma_Bianchi2} below, though an analog result should be expected on an (Anti-)de Sitter background, as well).
\end{enumerate}

\subsection{Linearized gravity}
\label{sec_lin_grav}

Let us elaborate somewhat more detailed on (ii).
We consider a Lorentzian metric $\eta$ and denote by
\begin{equation}
g= \eta +  h
\end{equation}
a perturbation thereof.
In the following
we will decorate all fields related to the background metric $\eta$ with superscript $(0)$.
The curvature tensor associated to the metric $g$ takes the  form
(cf.\ e.g.\ \cite{c_BHs})
\begin{eqnarray}
R_{\mu\nu\sigma\rho} = R^{(0)} _{\mu\nu\sigma\rho} +\nabla^{(0)}_{\rho}\nabla^{(0)}_{[\mu}h_{\nu]\sigma}-\nabla^{(0)}_{\sigma}\nabla^{(0)}_{[\mu}h_{\nu]\rho}
-h_{[\mu}{}^{\kappa}{R}^{(0)}_{\nu]\kappa\sigma\rho} + O(h^2)
\label{curv1}
\,.
\end{eqnarray}
%

Two perturbations $h$ and $h'$ describe the same physical perturbation if and only if there exists a one-form $\xi$ such that $h'_{\mu\nu}-h_{\mu\nu} = \nabla_{(\mu}\xi_{\nu)}$.
It is convenient to employ this gauge freedom
\begin{equation}
h_{\mu\nu} \mapsto h_{\mu\nu} + \nabla_{(\mu}\xi_{\nu)}
\label{gauge_trafos}
\end{equation}
to  impose the gauge condition
\begin{equation}
\nabla^{(0)}_{\beta} h_{\alpha}{}^{\beta}=\frac{1}{2}\nabla^{(0)}_{\alpha} \mathrm{tr}_{\eta} h
\,.
\label{gauge_condition}
\end{equation}
(It is preserved under evolution of the linearized Einstein equations \eq{linearized_Einstein} below.)
Assuming further that the background metric satisfies the $\Lambda$-vacuum equations, we obtain from \eq{curv1}
($\Box_{\eta}:= \eta^{\mu\nu}\nabla^{(0)}_{\mu}\nabla^{(0)}_{\nu}$)
\begin{eqnarray}
R_{\mu\nu\sigma\rho} &=& \frac{2}{3}\Lambda \eta_{\sigma[\mu}\eta_{\nu]\rho}
+  C^{(0)}_{\mu\nu\sigma\rho} + \nabla^{(0)}_{\rho}\nabla^{(0)}_{[\mu}h_{\nu]\sigma}-\nabla^{(0)}_{\sigma}\nabla^{(0)}_{[\mu}h_{\nu]\rho}
\nonumber
\\
&&
-\frac{\Lambda}{3} h_{\rho[\mu}\eta_{\nu]\sigma}
+ \frac{\Lambda}{3} h_{\sigma[\mu}\eta_{\nu]\rho}
-h_{[\mu}{}^{\kappa} C^{(0)}_{\nu]\kappa\sigma\rho}
+O(h^2)
\,,
\end{eqnarray}
whence
\begin{eqnarray}
R_{\mu\nu}
&=&\Lambda \eta_{\mu\nu}- \frac{1}{2}\Box_{\eta} h_{\mu\nu} +\frac{4}{3}\Lambda h_{\mu\nu}
-\frac{\Lambda}{3}\mathrm{tr}_{\eta}h\, \eta_{\mu\nu}
- C^{(0)}_{\mu\alpha\nu\beta}h^{\alpha\beta}
+ O(h^2)
\,,
\phantom{xxx}
\label{Ricci_expansion}
\\
R &=&4\Lambda  -\frac{1}{2}\big(\Box_{\eta}+2\Lambda \big)  \mathrm{tr}_{\eta} h+ O(h^2)
\,,
\end{eqnarray}
and one easily computes the Weyl tensor associated to $g$,
\begin{eqnarray}
C_{\mu\nu\sigma\rho}
 &\equiv&
R_{\mu\nu\sigma\rho} -g_{\sigma[\mu}R_{\nu]\rho} + g_{\rho[\mu}R_{\nu]\sigma}
 +\frac{1}{3}Rg_{\sigma[\mu}g_{\nu]\rho}
\\
 &=&
  C^{(0)}_{\mu\nu\sigma\rho} + \nabla^{(0)}_{\rho}\nabla^{(0)}_{[\mu} h_{\nu]\sigma}-\nabla^{(0)}_{\sigma}\nabla^{(0)}_{[\mu} h_{\nu]\rho}
\nonumber
\\
&&
+ \frac{1}{2}\eta_{\sigma[\mu}\Box_{\eta}  h_{\nu]\rho}
-  \frac{1}{2}\eta_{\rho[\mu}\Box_{\eta}  h_{\nu]\sigma}
 -\frac{1}{6}\Box_{\eta} \mathrm{tr}_{\eta}  h\,\eta_{\sigma[\mu}\eta_{\nu]\rho}
\nonumber
\\
&& -2\frac{\Lambda}{3}(\eta_{\sigma[\mu}  h_{\nu]\rho}-\eta_{\rho[\mu}  h_{\nu]\sigma})
+\frac{\Lambda}{3}\mathrm{tr}_{\eta} h\,\eta_{\sigma[\mu}\eta_{\nu]\rho}
\nonumber
\\
&&
-h_{[\mu}{}^{\kappa} C^{(0)}_{\nu]\kappa\sigma\rho}
+\eta_{\sigma[\mu} C^{(0)}_{\nu]\alpha\rho\beta}h^{\alpha\beta}
-\eta_{\rho[\mu} C^{(0)}_{\nu]\alpha\sigma\beta}h^{\alpha\beta}
+O(h^2)
\,.
\phantom{xx}
\label{linearized_Weyl}
\end{eqnarray}
It follows from \eq{Ricci_expansion}
 that the \emph{linearized Einstein equations on a vacuum background $( M , \eta)$}  read
\begin{equation}
\Box_{\eta} h_{\mu\nu} =\frac{2}{3}\Lambda \big( h_{\mu\nu}
-\mathrm{tr}_{\eta}h\, \eta_{\mu\nu}\big)
-2  C^{(0)}_{\mu\alpha\nu\beta}h^{\alpha\beta}
\,.
\label{linearized_Einstein}
\end{equation}
Let us now assume that the linearized field equations \eq{linearized_Einstein} are satisfied. Then the  Weyl tensor \eq{linearized_Weyl} has the expansion
\begin{equation}
C_{\mu\nu\sigma\rho} =
 C^{(0)}_{\mu\nu\sigma\rho}+
 C^{(\mathrm{lin})}_{\mu\nu\sigma\rho}
+O(h^2)
\,,
\end{equation}
where
\begin{equation}
C^{(\mathrm{lin})}_{\mu\nu\sigma\rho} =  \nabla^{(0)}_{\rho}\nabla^{(0)}_{[\mu} h_{\nu]\sigma}-\nabla^{(0)}_{\sigma}\nabla^{(0)}_{[\mu} h_{\nu]\rho}
-\frac{\Lambda}{3}(\eta_{\sigma[\mu}  h_{\nu]\rho}-\eta_{\rho[\mu}  h_{\nu]\sigma})
-h_{[\mu}{}^{\kappa} C^{(0)}_{\nu]\kappa\sigma\rho}
\,.
\label{linearized_Einstein_vacuum}
\end{equation}
We compute the divergence. Using the   linearized equations \eq{linearized_Einstein}, the gauge condition \eq{gauge_condition},
and the Bianchi equation for the background metric,
we obtain
\begin{eqnarray}
\nabla^{(0)}_{\rho}C_{\mu\nu\sigma}{}^{\rho} &=&
-2h^{\alpha\beta}\nabla^{(0)}_{[\mu} C^{(0)}_{\nu]\alpha\sigma\beta}
+ C^{(0)}_{\alpha\sigma\beta [\mu}\nabla^{(0)}_{\nu]} h^{\alpha\beta}
- C^{(0)}_{ \sigma[\mu \beta}{}^{\alpha}\nabla^{(0)}_{\alpha}  h_{\nu]}{}^{\beta}
\nonumber
\\
&&
+ C^{(0)}_{\mu\nu\beta}{}^{\alpha}\nabla^{(0)}_{\alpha} h_{\sigma}{}^{\beta}
+O(h^2)
\,.
\end{eqnarray}
That yields the well-known (cf.\ e.g.\ \cite{wald2})

\begin{lemma}
\label{lemma_Bianchi1}
Let $( M ,\eta)$ be a $\Lambda$-vacuum spacetime and let $h$ be a solution to the linearized vacuum equations. Then the linearized Weyl tensor $C^{(\mathrm{lin})}_{\mu\nu\sigma\rho}$  associated to the perturbation
$g=\eta + h$ satisfies the Bianchi equation, $\nabla^{(0)}_{\rho}C^{(\mathrm{lin})}_{\mu\nu\sigma}{}^{\rho} =0$, supposing that
the background metric $\eta$ is locally conformally flat.
\end{lemma}

The converse is also true (at least for vanishing cosmological constant).
The proof can be found e.g.\ in \cite{bc}.
\begin{lemma}
\label{lemma_Bianchi2}
Let $( M ,\eta)$ be a Minkowski background, and let   $C_{\mu\nu\sigma\rho} $ be a solution to the Bianchi equation $\nabla^{(0)}_{\rho}C_{\mu\nu\sigma}{}^{\rho} =0$
with all the algebraic symmetries of the Weyl tensor.
Then there exists a unique (up to gauge transformations \eq{gauge_trafos})  perturbation $h$ which satisfies  the linearized vacuum equations \eq{linearized_Einstein},
such that $C_{\mu\nu\sigma\rho} $ is the linearized Weyl tensor of $g=\eta+h$.
\end{lemma}

\subsection{Cauchy problem and Buchdahl condition}
\label{sec_Cauchy_Buchdahl}

Having a motivation at hand that the Bianchi equation on a given background is of interest by itself, one would to  like establish well-posedness of
the associated Cauchy problem.

Let $\Sigma\subset M $ be a Cauchy surface in $( M , \eta)$, and introduce Gaussian coordinates $(t\equiv x^0, x^i)$ near $\Sigma$. Taking the algebraic symmetries of the Weyl tensor into account, it is easy to check that the Bianchi equation is equivalent to the system ($\star$ denotes the Hodge dual)
\begin{eqnarray}
\nabla^{(0)}_{\rho} C_{0(ij)}{}^{\rho} =0
\,, &&
\nabla^{(0)}_{\rho} C^{\star}_{0(ij)}{}^{\rho} =0
\,,
\label{Bianchi_evolution}
\\
\nabla^{(0)}_{\rho} C_{0i 0}{}^{\rho} =0
\,,&&
\nabla^{(0)}_{\rho} C^{\star}_{0i 0}{}^{\rho} =0
\,.
\label{Bianchi_constraint}
\end{eqnarray}
One checks that  \eq{Bianchi_evolution} forms a symmetric hyperbolic system of evolution equations, for which standard well-posedness results are available.
The system  \eq{Bianchi_constraint} does not contain transverse derivatives and therefore provides a system of constraint equations.
It requires the initial data $C_{\mu\nu\sigma\rho}|_{\Sigma}$ to satisfy the following constraint equations
\begin{equation}
\nabla^{(0)}_{k} C_{0i 0}{}^{k}|_{\Sigma} =0
\,, \quad \nabla^{(0)}_{k} C^{\star}_{0i0}{}^{k}|_{\Sigma} =0
\,.
\label{Bianchi_constraint2}
\end{equation}
This is equivalent to the existence  of a complex symmetric trace-free tensor $\mathcal{E}_{ij}:= C_{0i0j} + \imagi C^{\star}_{0i0j}|_{\Sigma}$ which solves
\begin{equation}
\label{eq:blubbbb}
\mcD_{k}\mathcal{E}_i{}^k
=
 \imagi\epsilon_{i}{}^{lm}K_{kl}\mathcal{E}_{m }{}^{k}
\,,
\end{equation}
where $\mcD_i$ and $\epsilon_{ijk}$  denote the covariant derivative and  the volume form of the induced metric  on $\Sigma$, while   $K_{ij}=\frac{1}{2}\mcL_{\partial_t}g_{ij}$
  denotes   the second fundamental form.
To solve this  constraint one may employ York's splitting method: Here  $\mathcal{E}_{ij}$ is written in the form
\begin{equation}
\mathcal{E}_{ij} = \mathcal{B}_{ij} + 2\mcD_{(i} \mathcal {Y}_{j)}-\frac{2}{3}\eta_{ij}\mcD_k\mathcal{Y}^k
\,,
\end{equation}
where the symmetric, trace-free tensor $\mathcal{B}_{ij} $ provides certain prescribed ``seed'' data.
The constraint equation \eq{eq:blubbbb} then becomes a linear elliptic PDE-system for the complex vector field  $\mathcal{Y}$.
In the time-symmetric case, for instance,
solving the constraint equations simply  amounts to the construction of symmetric trace-free and divergence-free tensors on the initial surface.

One then needs to make sure that \eq{Bianchi_constraint2} is preserved under evolution.
It turns out that this is the case if and only if the \emph{Buchdahl condition} \cite{buchdahl} holds,
\begin{equation}
C_{[0}{}^{\alpha\beta\gamma} C^{(0)}_{i]\alpha\beta\gamma} =0
\,,
\quad
C_{[0}{}^{\alpha\beta\gamma} C^{(0)\star}_{i]\alpha\beta\gamma} =0
\label{Buchdahl_cond}
\end{equation}
which is certainly the case whenever the background metric $\eta$ is locally conformally flat, or when $C_{\mu\nu\sigma\rho}$ is the Weyl tensor associated to $\eta$. In the latter case this is obvious for the first condition in \eq{Buchdahl_cond}, for the
second we have
\begin{equation*}
C^{(0)}_{[0}{}^{\alpha\beta\gamma} C^{(0)\star}_{i]\alpha\beta\gamma} =
\frac{\imagi}{2}\eta^{\beta\gamma\delta\rho}C^{(0)}_{[0}{}^{\alpha}{}_{\beta\gamma} C^{(0)}_{i]\alpha\delta\rho} 
=\frac{\imagi}{2}\eta^{\beta\gamma\delta\rho}C^{(0)}_{[0}{}^{\alpha}{}_{\delta\rho} C^{(0)}_{i]\alpha\beta\gamma} 
=-C^{(0)}_{[0}{}^{\alpha\beta\gamma} C^{(0)\star}_{i]\alpha\beta\gamma}
\,.
\end{equation*}

To sum it up, the Bianchi equation, regarded as an equation on a given background $( M , \eta)$ is of particular relevance  whenever
$( M , \eta)$  is locally conformally flat.
In that case the Buchdahl condition is satisfied and  solutions are obtained from data on some
Cauchy surface $\Sigma\subset  M $ if and only if these data satisfy the constraint equations  \eq{Bianchi_constraint2}.

\section{Bianchi-like equation for the MST}
\label{sec_MST_eqn}

In this section we will carry out a similar analysis for the  Bianchi-like equation \eq{MST_rhs}
satisfied by  the MST as we did for the Bianchi equation in the previous section.

\subsection{Linearized MST}
\label{sec_lin_MST}

First of all we would like to derive an analog of Lemma~\ref{lemma_Bianchi1}.
This is to explore the analogies between the Bianchi and the MST equation, but, even more important, we would like to provide
a motivation to analyze  the   MST equation on a given background.
It seems likely that there is an analog of Lemma~\ref{lemma_Bianchi2} which we have not explored here.

Since the computations turn out to be fairly lengthy we will restrict attention to the case of a vanishing cosmological constant
\begin{equation}
\Lambda=0
\,.
\end{equation}
Nonetheless, we expect a corresponding result to hold for $\Lambda \ne 0$ as well. This is to be analyzed elsewhere.

Let us assume we have been given a $\Lambda=0$-vacuum spacetime $( M , \eta)$ which admits a Killing vector field $X^{(0)}$.
In this setting \eq{MST_evolution_eqn}-\eq{MST_rhs} simplifies to
\begin{equation}
\nabla^{(0)}_{\rho}\mathcal{S}_{\alpha\beta\mu}{}^{\rho}
+{Q}^{(0)}  X^{(0)\sigma}  {\mathcal{F}}^{(0)}_{\mu\rho}{\mathcal{S}}_{\alpha\beta\sigma}{}^{\rho}
- \frac{2}{3}{Q}^{(0)}  X^{(0)\sigma} {\mathcal I}^{(0)}_{\alpha\beta\mu}{}^{\rho} {\mathcal F}^{(0)\gamma\delta} {\mathcal{S}}_{\gamma\delta\sigma\rho}  =0
\,.
\label{MST_eqn_background}
\end{equation}
We want to analyze under which conditions   the linearized MST $\mathcal{S}^{(\mathrm{lin})}_{\mu\nu\sigma\rho}$ of the perturbed metric $g=\eta + h$ satisfies this equation, when  $h$ is some perturbation which satisfies the linearized vacuum equations
\begin{equation}
\Box_{\eta} h_{\mu\nu} =-2  C^{(0)}_{\mu\alpha\nu\beta}h^{\alpha\beta}
\,.
\label{linearized_Einstein2}
\end{equation}
As before we will impose the gauge condition \eq{gauge_condition}
\begin{equation}
\nabla^{(0)}_{\beta} h_{\alpha}{}^{\beta}=\frac{1}{2}\nabla^{(0)}_{\alpha} \mathrm{tr}_{\eta} h
\,.
\label{gauge_condition2}
\end{equation}

To define the MST of the perturbed metric $g$, it needs to admit a Killing vector field $ X$. In fact, to determine a linearization it is sufficient to have
a vector field which satisfies the Killing equation up to and including the linear order in the perturbation $h$.
As before, we denote by $  X^{(0)}$ the Killing vector of  $( M , \eta)$.
We write $X$ as  $ X=  X^{(0)} + X^{(\mathrm{lin})} + O(h^2)$. Since
\begin{equation}
\nabla_{(\mu}X_{\nu)}= \frac{1}{2}\mcL_{ X^{(0)}} h_{\mu\nu}   +  \nabla^{(0)}_{(\mu} X^{(\mathrm{lin})} _{\nu)}
+ O(h^2)
\end{equation}
we  need to assume that  the perturbation $\eta+ h$ admits a vector field $X^{(\mathrm{lin})}$  such that
\begin{equation}
\mcL_{   X^{(0)}} h_{\mu\nu}   + 2 \nabla^{(0)}_{(\mu} X^{(\mathrm{lin})} _{\nu)}=0
\,,
\label{cond_perturbation}
\end{equation}
 which we assume henceforth.

%

A solution to \eq{MST_eqn_background} will be required to be self-dual w.r.t.\ the background $\eta$, while the
MST associated to $g$ will be self-dual w.r.t.\  $g$. In general these two notions only  coincide   in the leading order.
In other words, the linearized MST
of $g$ will certainly be self-dual w.r.t.\ $\eta$ if   the MST associated to the background metric $\eta$ vanishes
(we have indicated in the formula w.r.t.\ which metric the Hodge dual is taken),
\begin{equation*}
\mathcal{S}^{(0)} +\mathcal{S}^{(\mathrm{lin})}+O(h^2) =\mathcal{S} = \imagi\mathcal{S}^{*_g}
= \Big(1+\frac{1}{2}\mathrm{tr}_{\eta}h\Big)\underbrace{\imagi \mathcal{S}^{(0)*_\eta}}_{=\mathcal{S}^{(0)}}+\imagi\mathcal{S}^{(\mathrm{lin})*_{\eta}}+O(h^2)
\,.
\end{equation*}
For the remainder of this section we will thus assume
\begin{equation}
\mathcal{S}^{(0)}_{\mu\nu\sigma\rho}=0
\,.
\label{cond_vanishing_MST}
\end{equation}
This is in accordance with the Bianchi equation where we needed to assume the vanishing of the Weyl tensor of the background metric in order to derive
an analog result (note that the vanishing of the Weyl tensor is equivalent to the vanishing of the self-dual Weyl tensor).

\begin{remark}
{\rm
$\Lambda$-vacuum spacetimes with vanishing MST \eq{cond_vanishing_MST}  have been classified in \cite{mars-senovilla, mars-senovilla-null}.
This class includes in particular the Kerr-NUT-(A)(dS) family.
So while, compared to the Bianchi equation, the class of admissible perturbations is restricted to those which preserve a symmetry, the class of
admissible background spacetimes is much larger.
}
\end{remark}

Let us now determine the linearization of the MST
\begin{equation}
\mathcal{S}_{\mu\nu\sigma\rho} =\mathcal{C}_{\mu\nu\sigma\rho} - Q\Big(\mathcal{F}_{\mu\nu}\mathcal{F}_{\sigma\rho} - \frac{1}{3}\mathcal{F}^2\mathcal{I}_{\mu\nu\sigma\rho} \Big)
\,.
\end{equation}
The linearized Weyl tensor has already been given  in \eq{linearized_Einstein_vacuum}. We deduce that its self-dual counterpart satisfies (for $\Lambda=0$)
\begin{eqnarray}
\mathcal{C}_{\mu\nu\sigma\rho} &=& {\mathcal{C}}^{(0)}_{\mu\nu\sigma\rho} -h_{[\mu}{}^{\kappa} {\mathcal{C}}^{(0)}_{\nu]\kappa\sigma\rho}
- \imagi ( h_{\kappa}{}^{\gamma})_{\mathrm{tf}} \volform^{(0)}_{\sigma\rho}{}^{\kappa\delta}  C^{(0)}_{\mu\nu\gamma\delta}
\nonumber
\\
&&
 -4{\mathcal{I}}^{(0)}_{\sigma\rho}{}^{\gamma\delta}\nabla^{(0)}_{\gamma}\nabla^{(0)}_{[\mu} h_{\nu]\delta}
+O(h^2)
\,,
\label{expansion_Weyl}
\end{eqnarray}
where $(v_{\alpha\beta})_{\mathrm{tf}}$ denotes the $\eta$-trace-free part of the corresponding two-tensor.
Moreover,
\begin{eqnarray}
\volform_{\mu\nu\sigma\rho} &=& \Big(1 + \frac{1}{2}\mathrm{tr}_{\eta} h\Big) \volform^{(0)}_{\mu\nu\sigma\rho}+ O(h^2)
\,,
\\
\mathcal{I}_{\mu\nu\sigma\rho} &=& {\mathcal{I}}^{(0)}_{\mu\nu\sigma\rho}+
\frac{1}{2}\eta_{\sigma[\mu }h_{\nu]\rho}
-\frac{1}{2} \eta_{\rho[\mu} h_{\nu]\sigma}+\frac{\imagi}{8}\mathrm{tr}_{\eta} h\,\volform^{(0)}_{\mu\nu\sigma\rho}
+ O(h^2)
\,,
\end{eqnarray}
and, employing \eq{cond_perturbation},
\begin{eqnarray}
F_{\mu\nu}
&=&  F^{(0)}_{\mu\nu}-  h_{\sigma[\mu} F^{(0)}_{\nu]}{}^{\sigma}
+ X^{(0)\alpha}\nabla^{(0)}_{[\mu}h_{\nu]\alpha}
+ \nabla^{(0)}_{[\mu} X^{\mathrm{lin}}_{\nu]}
+ O(h^2)
\,,
\\
\mathcal{F}_{\mu\nu}
&=&   {\mathcal{F}}^{(0)}_{\mu\nu} -  h_{\alpha[\mu}{\mathcal  F}^{(0)}_{\nu]}{}^{\alpha}
+2{\mathcal{I}}^{(0)}_{\mu\nu}{}^{\sigma\rho}  X^{(0)\alpha}\nabla^{(0)}_{\sigma}h_{\rho\alpha}
+{\mathcal{G}}^{(\mathrm{lin})}_{\mu\nu}
+ O(h^2)
\,,
\\
\mathcal{F}^2 &=&({\mathcal{F}^{(0)}})^2
+4 X^{(0)\alpha}{ \mathcal{F}}^{(0)\mu\nu}\nabla^{(0)}_{\mu}h_{\nu\alpha}
+2  {\mathcal{F}}^{(0)}_{\mu\nu} {\mathcal{G}}^{(\mathrm{lin})\mu\nu}
+ O(h^2)
\,,
\end{eqnarray}
where
\begin{equation}
{\mathcal{G}}^{(\mathrm{lin})}_{\mu\nu}:= \nabla^{(0)}_{[\mu} X^{(\mathrm{lin})}_{\nu]}+ \imagi ( \nabla^{(0)}_{[\mu} X^{(\mathrm{lin})}_{\nu]})^{\star}=O(h)
\end{equation}
 denotes the self-dual two-form associated to  $X^{(\mathrm{lin})}$.

It remains to determine the linearization of $Q= Q^{(0)} + Q^{(\mathrm{lin})} + O(h^2)$. Unfortunately, this term cannot be computed explicitly.
It follows from the definition of $Q$ \eq{dfn_Q} that
\begin{eqnarray}
\nabla_{\mu} Q &=& -\frac{1}{6} Q^2\sigma_{\mu}= -\frac{1}{3} Q^2 X^{\alpha}\mathcal{F}_{\alpha\mu}
\\
 &=& -\frac{1}{6}  (Q^{(0)})^2\sigma^{(0)}_{\mu}
-\frac{1}{3} Q^{(0)}  Q^{(\mathrm{lin})}\sigma^{(0)}_{\mu}
  +\frac{1}{3} ( Q^{(0)})^2 X^{(0)\alpha}  h_{\gamma[\alpha}{\mathcal  F}^{(0)}_{\mu]}{}^{\gamma}
\nonumber
\\
&&
-\frac{2}{3} ( Q^{(0)})^2{\mathcal{I}}^{(0)}_{\alpha\mu}{}^{\sigma\rho}  X^{(0)\alpha} X^{(0)\gamma}\nabla^{(0)}_{\sigma}h_{\rho\gamma}
 -\frac{1}{3}  (Q^{(0)})^2 X^{(\mathrm{lin})\alpha}{\mathcal{F}}^{(0)}_{\alpha\mu}
\nonumber
\\
&&
 -\frac{1}{3} (Q^{(0)})^2  X^{(0)\alpha}{\mathcal{G}}^{(\mathrm{lin})}_{\alpha\mu}
+ O(h^2)
\,,
\end{eqnarray}
whence $ Q^{(\mathrm{lin})}$ satisfies the differential equation
\begin{eqnarray}
 \nabla^{(0)}_{\mu}  Q^{(\mathrm{lin})}
&=& -\frac{1}{3}  Q^{(0)}  Q^{(\mathrm{lin})}\sigma^{(0)}_{\mu}
  +\frac{1}{3} ( Q^{(0)})^2 X^{(0)\alpha}  h_{\gamma[\alpha}{\mathcal  F}^{(0)}_{\mu]}{}^{\gamma}
\nonumber
\\
&&
-\frac{2}{3} ( Q^{(0)})^2{\mathcal{I}}^{(0)}_{\alpha\mu}{}^{\sigma\rho}  X^{(0)\alpha} X^{(0)\gamma}\nabla^{(0)}_{\sigma}h_{\rho\gamma}
 -\frac{1}{3} (Q^{(0)})^2 X^{(\mathrm{lin})\alpha}{\mathcal{F}}^{(0)}_{\alpha\mu}
\nonumber
\\
&&
 -\frac{1}{3}  (Q^{(0)})^2  X^{(0)\alpha}{\mathcal{G}}^{(\mathrm{lin})}_{\alpha\mu}
+ O(h^2)
\,.
\label{PDE_Qlin}
\end{eqnarray}

From these expansions we determine the linearization of the  MST.
After some simplifications we find that
\begin{eqnarray}
\mathcal{S}_{\mu\nu\sigma\rho} &=&
-4 {\mathcal{I}}^{(0)}_{\sigma\rho}{}^{\alpha\beta}\nabla^{(0)}_{[\mu} \nabla^{(0)}_{|\alpha|}h_{\nu]\beta}
+ Q^{(0)}{\mathcal{U}}^{(0)}_{\sigma\rho[\mu }{}^{\alpha} h_{\nu]\alpha}
\nonumber
\\
&&-2  Q^{(0)} {\mathcal{F}}^{(0)}_{\mu\nu} {\mathcal{I}}^{(0)}_{\sigma\rho}{}^{\alpha\beta} X^{(0)\gamma}\nabla^{(0)}_{\alpha}h_{\beta\gamma}
-2  Q^{(0)}  {\mathcal{F}}^{(0)}_{\sigma\rho} {\mathcal{I}}^{(0)}_{\mu\nu}{}^{\alpha\beta}  X^{(0)\gamma}\nabla^{(0)}_{\alpha}h_{\beta\gamma}
\nonumber
\\
&&
+ \frac{4}{3} Q^{(0)} X^{(0)\alpha}{ \mathcal{F}}^{(0)\gamma\delta}\nabla^{(0)}_{\gamma}h_{\delta\alpha}{\mathcal{I}}^{(0)}_{\mu\nu\sigma\rho}
+ Q^{(\mathrm{lin})} {\mathcal{U}}^{(0)}_{\mu\nu\sigma\rho}
\nonumber
\\
&&
-  Q^{(0)} {\mathcal{F}}^{(0)}_{\mu\nu}{\mathcal{G}}^{(\mathrm{lin})}_{\sigma\rho}
-  Q^{(0)} {\mathcal{G}}^{(\mathrm{lin})}_{\mu\nu}{\mathcal{F}}^{(0)}_{\sigma\rho}
+\frac{2}{3}  Q^{(0)} {\mathcal{F}}^{(0)}_{\alpha\beta} {\mathcal{G}}^{(\mathrm{lin})\alpha\beta}{ \mathcal{I}}^{(0)}_{\mu\nu\sigma\rho}
+ O(h^2)
\,,
\label{linearized_MST}
\end{eqnarray}
and  $ Q^{(\mathrm{lin})}$ is given by \eq{PDE_Qlin}.
With regard to \eq{MST_eqn_background}
we also compute the following contraction,
\begin{eqnarray}
{\mathcal{F}}^{(0)\sigma\rho}\mathcal{S}_{\mu\nu\sigma\rho}
&=&
-4 {\mathcal{F}}^{(0)\sigma\rho}\nabla^{(0)}_{[\mu}  \nabla^{(0)}_{|\sigma|}h_{\nu]\rho}
-\frac{2}{3} Q^{(0)} {\mathcal{F}}^{(0)}_{[\mu }{}^{\lambda} h_{\nu]\lambda}
\nonumber
\\
&&
- \frac{2}{3} Q^{(0)} X^{(0)\gamma}{\mathcal{F}}^{(0)}_{\mu\nu}{ \mathcal{F}}^{(0)\alpha\beta}\nabla^{(0)}_{\alpha}h_{\beta\gamma}
\nonumber
\\
&&
- 2 Q^{(0)} ( {\mathcal{F}}^{(0)})^2 {\mathcal{I}}^{(0)}_{\mu\nu}{}^{\alpha\beta}  X^{(0)\gamma}\nabla^{(0)}_{\alpha}h_{\beta\gamma}
-\frac{2}{3}Q^{(\mathrm{lin})}({\mathcal{F}}^{(0)})^2 {\mathcal{F}}^{(0)}_{\mu\nu}
\nonumber
\\
&&
-  Q^{(0)}({\mathcal{F}}^{(0)})^2 {\mathcal{G}}^{(\mathrm{lin})}_{\mu\nu}
-\frac{1}{3}  Q^{(0)} {\mathcal{F}}^{(0)}_{\alpha\beta} {\mathcal{G}}^{(\mathrm{lin})\alpha\beta}{ \mathcal{F}}^{(0)}_{\mu\nu}
+ O(h^2)
\,,
\label{linearized_MST_contr}
\end{eqnarray}
where we employed one more time the vanishing of the MST of the background metric.

Before we proceed, let us collect a couple of useful relations satisfied by ${\mathcal{F}}^{(0)}_{\mu\nu}$ in any $\Lambda=0$-vacuum spacetime $( M , \eta,   X^{(0)})$ with vanishing MST \cite{mars-senovilla}
\begin{eqnarray}
\nabla^{(0)}_{\kappa}{\mathcal{F}}^{(0)}_{\alpha\beta} &=& \frac{1}{2} Q^{(0)} \sigma^{(0)}_{\kappa}{ \mathcal{F}}^{(0)}_{\alpha\beta}
+\frac{1}{3}  Q^{(0)}({\mathcal{F}}^{(0)})^2  X^{(0)\lambda}{\mathcal{I}}^{(0)}_{\alpha\beta\kappa\lambda}
\,,
\label{relation_F1}
\\
\nabla^{(0)}_{\beta}{\mathcal{F}}^{(0)}_{\alpha}{}^{\beta} &=& 0
\,,
\\
\nabla^{(0)}_{\mu}({\mathcal{F}}^{(0)})^2 &=&  \frac{2}{3} Q^{(0)} ({\mathcal{F}}^{(0)})^2   \sigma^{(0)}_{\mu}
\,,
\\
\sigma^{(0)\alpha}{\mathcal{F}}^{(0)}_{\alpha\mu} &=& -\frac{1}{2}({\mathcal{F}}^{(0)})^2  X^{(0)}_{\mu}
\\
  \nabla^{(0)}_{\kappa} Q^{(0)}
&=& -  \frac{1}{6} ( Q^{(0)})^2  \sigma^{(0)}_{\kappa}
\,.
\label{relation_F5}
\end{eqnarray}
Let $V_{\alpha\beta}$ be an  antisymmetric tensor, $\mathcal{V}_{\alpha\beta}=V_{\alpha\beta} + \imagi V^{\star}_{\alpha\beta}$ its self-dual counterpart,  and $\mathcal{W}_{\alpha\beta}$ self-dual, then (using the properties of the Levi-Civita symbol)
\begin{equation}
\mathcal{W}_{\gamma[\alpha}\mathcal{V}_{\beta]}{}^{\gamma} = \mathcal{W}_{\gamma[\alpha}V_{\beta]}{}^{\gamma}  
+ \frac{\imagi}{2} \mathcal{W}_{\gamma[\alpha}\eta_{\beta]}{}^{\gamma\delta\rho}V_{\delta\rho} 
= \mathcal{W}_{\gamma[\alpha}V_{\beta]}{}^{\gamma}  
- \frac{1}{4}\eta_{\gamma[\alpha}{}^{\sigma\kappa}\eta_{\beta]}{}^{\gamma\delta\rho} \mathcal{W}_{\sigma\kappa}V_{\delta\rho} 
=2 \mathcal{W}_{\gamma[\alpha}V_{\beta]}{}^{\gamma} 
\,.
\end{equation}

From \eq{cond_perturbation} we deduce
\begin{eqnarray}
0 &=& \nabla^{(0)}_{[\alpha}(\mcL_{   X^{(0)}} h_{\beta]\rho}   + \nabla^{(0)}_{\beta]} X^{(\mathrm{lin})} _{\rho}+  \nabla^{(0)}_{|\rho|} X^{(\mathrm{lin})} _{\beta]})
\nonumber
\\
&=&
\nabla^{(0)}_{\rho}  X^{(0)\gamma} \nabla^{(0)}_{[\alpha} h_{\beta]\gamma}
+\nabla^{(0)}_{\alpha}  X^{(0)\gamma}\nabla^{(0)}_{[\gamma}h_{\beta]\rho}
-\nabla^{(0)}_{\beta}  X^{(0)\gamma}\nabla^{(0)}_{[\gamma}h_{\alpha]\rho}
\nonumber
\\
&&
+ X^{(0)\kappa} \nabla^{(0)}_{\kappa}\nabla^{(0)}_{[\alpha}h_{\beta]\rho}
+  C^{(0)}_{\alpha\beta\rho}{}^{\kappa}X^{(\mathrm{lin})} _{\kappa}
+  \nabla^{(0)}_{\rho} \nabla^{(0)}_{[\alpha}  X^{(\mathrm{lin})} _{\beta]}
\,,
\end{eqnarray}
whence
\begin{eqnarray}
{\mathcal{I}}^{(0)}_{\mu\nu}{}^{\alpha\beta}{\mathcal{F}}^{(0)}_{\sigma}{}^{\rho}\nabla^{(0)}_{\rho} X^{(0)\gamma}\nabla^{(0)}_{\alpha}h_{\beta\gamma}
&=&
-{\mathcal{I}}^{(0)}_{\mu\nu}{}^{\alpha\beta}{\mathcal{F}}^{(0)}_{\alpha}{}^{\gamma}{\mathcal{F}}^{(0)}_{\sigma}{}^{\rho} \nabla^{(0)}_{\gamma}h_{\beta\rho}
+\frac{1}{8}{\mathcal{F}}^{(0)}_{\sigma}{}^{\rho}{\mathcal{F}}^{(0)}_{\mu\nu} \nabla^{(0)}_{\rho}\mathrm{tr}_{\eta} h
\nonumber
\\
&&
-{\mathcal{I}}^{(0)}_{\mu\nu}{}^{\alpha\beta}{\mathcal{F}}^{(0)}_{\sigma}{}^{\rho}  X^{(0)\kappa}  \nabla^{(0)}_{\kappa} \nabla^{(0)}_{\alpha}h_{\beta\rho}
\nonumber
\\
&&
-\frac{1}{2}{\mathcal{F}}^{(0)}_{\sigma}{}^{\rho}   {\mathcal{C}}^{(0)}_{\mu\nu\rho}{}^{\kappa}X^{(\mathrm{lin})} _{\kappa}
-\frac{1}{2}{\mathcal{F}}^{(0)}_{\sigma}{}^{\rho}   \nabla^{(0)}_{\rho}{\mathcal{G}}^{(\mathrm{lin})}_{\mu\nu}
\label{nabla_X_relation1}
\\
{\mathcal{I}}^{(0)}_{\sigma}{}^{\rho\alpha\beta}\nabla^{(0)}_{\rho} X^{(0)\gamma}\nabla^{(0)}_{\alpha}h_{\beta\gamma}
&=&
-{\mathcal{I}}^{(0)}_{\sigma\rho}{}^{\alpha\beta}{\mathcal{F}}^{(0)}_{\alpha}{}^{\gamma} \nabla^{(0)}_{\gamma}h_{\beta}{}^{\rho}
+\frac{1}{8}{\mathcal{F}}^{(0)}_{\sigma}{}^{\rho} \nabla^{(0)}_{\rho}\mathrm{tr}_{\eta} h
\nonumber
\\
&&
-{\mathcal{I}}^{(0)}_{\sigma\rho}{}^{\alpha\beta}  X^{(0)\gamma}  \nabla^{(0)}_{\gamma} \nabla^{(0)}_{\alpha}h_{\beta\rho}
-\frac{1}{2}   \nabla^{(0)}_{\rho} {\mathcal{G}}^{(\mathrm{lin})} _{\sigma}{}^{\rho}
\\
{\mathcal{I}}^{(0)}_{\mu\nu\sigma}{}^{\rho}\nabla^{(0)}_{\rho} X^{(0)\alpha}{ \mathcal{F}}^{(0)\gamma\delta}\nabla^{(0)}_{\gamma}h_{\delta\alpha}
&=&
- {\mathcal{I}}^{(0)}_{\mu\nu\sigma\rho}{ \mathcal{F}}^{(0)\gamma\delta} X^{(0)\kappa} \nabla^{(0)}_{\kappa} \nabla^{(0)}_{\gamma}h_{\delta}{}^{\rho}
\nonumber
\\
&&
-\frac{1}{3} Q^{(0)} {\mathcal{I}}^{(0)}_{\mu\nu\sigma}{}^{\rho}{ \mathcal{F}}^{(0)}_{\rho}{}^{\kappa}X^{(\mathrm{lin})} _{\kappa}
-\frac{1}{2} {\mathcal{I}}^{(0)}_{\mu\nu\sigma}{}^{\rho}{ \mathcal{F}}^{(0)\gamma\delta}  \nabla^{(0)}_{\rho}{\mathcal{G}}^{(\mathrm{lin})}_{\gamma\delta}
\label{nabla_X_relation3}
\end{eqnarray}

Finally, we need to  determine the linearization of the divergence of the MST.
First of all we deduce from the  linearized Einstein equations \eq{linearized_Einstein} and  the gauge condition \eq{gauge_condition}
\begin{eqnarray}
\nabla^{(0)}_{\rho}( {\mathcal{I}}^{(0)}_{\sigma}{}^{\rho\alpha\beta}\nabla^{(0)}_{[\mu} \nabla^{(0)}_{|\alpha|}h_{\nu]\beta})
&=&
-\frac{1}{8}h^{\alpha\beta} \nabla^{(0)}_{\alpha}  {\mathcal{C}}^{(0)}_{\mu\nu\sigma\beta}
-\frac{1}{4}{\mathcal{C}}^{(0)}_{\alpha\sigma\beta[\mu}\nabla^{(0)}_{\nu]} h^{\alpha\beta}
-\frac{1}{4} {\mathcal{C}}^{(0)}_{\alpha\beta \sigma [\mu}\nabla^{(0)\alpha}h_{\nu]}{}^{\beta}
\nonumber
\\
&&
+ \frac{1}{8}{\mathcal{C}}^{(0)}_{\mu\nu \alpha\beta}\nabla^{(0)\alpha} h_{\sigma}{}^{ \beta}
+ \frac{1}{16}{\mathcal{C}}^{(0)}_{\mu\nu \sigma \rho}\nabla^{(0)\rho} \mathrm{tr}_{\eta} h
\,.
\end{eqnarray}
Then a rather lengthy calculation which employs \eq{linearized_Einstein}, \eq{gauge_condition},
the vanishing of the background MST (to express the self-dual-Weyl tensor  in terms of $\mathcal{F}_{\alpha\beta}$) as well as the relations
\eq{relation_F1}-\eq{relation_F5} and \eq{nabla_X_relation1}-\eq{nabla_X_relation3}
reveals that
\begin{eqnarray}
\nabla^{(0)}_{\rho}\mathcal{S}_{\mu\nu\sigma}{}^{\rho}
&=&
\frac{\imagi}{2} Q^{(0)} X^{(0)\gamma}{\mathcal{F}}^{(0)}_{\sigma}{}^{\rho}\Big(\volform^{(0)}_{\mu\nu}{}^{\alpha\beta}{C}^{(0)}_{\alpha\beta[\gamma}{}^{\kappa}h_{\rho]\kappa}
-{\volform}^{(0)}_{\gamma\rho}{}^{\alpha\beta}  C^{(0)}_{\alpha\beta[\mu }{}^{\lambda} h_{\nu]\lambda}
\nonumber
\\
&&\qquad\qquad
+2{\volform}^{(0)}_{\mu\nu}{}^{\alpha\beta} \nabla^{(0)}_{[\gamma}\nabla^{(0)}_{|\alpha}h_{\beta|\rho]}
- 2{\volform}^{(0)}_{\gamma\rho}{}^{\alpha\beta}\nabla^{(0)}_{[\mu} \nabla^{(0)}_{|\alpha}h_{\beta|\nu]}\Big)
\nonumber
\\
&& \hspace{-4em} +  Q^{(0)} Q^{(\mathrm{lin})} ({\mathcal{F}}^{(0)})^2  \Big(\frac{1}{12}  X^{(0)}_{\sigma} {\mathcal{F}}^{(0)}_{\mu\nu}
-\frac{1}{18}{\mathcal{I}}^{(0)}_{\mu\nu\sigma}{}^{\kappa}\sigma^{(0)}_{\kappa}\Big)
\nonumber
\\
&&
 \hspace{-4em}
+ 4 {Q}^{(0)}   X^{(0)\kappa}  {\mathcal{F}}^{(0)}_{\sigma}{}^{\rho}{\mathcal{I}}^{(0)}_{\kappa\rho}{}^{\alpha\beta}\nabla^{(0)}_{[\mu} \nabla^{(0)}_{|\alpha|}h_{\nu]\beta}
-\frac{8}{3} {Q}^{(0)}   X^{(0)\kappa}{\mathcal{F}}^{(0)\alpha\beta} {\mathcal I}^{(0)}_{\mu\nu\sigma}{}^{\rho} \nabla^{(0)}_{\alpha}\nabla^{(0)}_{[\kappa} h_{\rho]\beta}
\nonumber
\\
&&
 \hspace{-4em}
-\frac{4}{3} {Q}^{(0)}  X^{(0)\kappa}{\mathcal{F}}^{(0)\alpha\beta} {\mathcal I}^{(0)}_{\mu\nu\sigma}{}^{\rho}C_{\kappa \rho\alpha}{}^{\lambda}  h_{\lambda\beta}
\nonumber
\\
&&
 \hspace{-4em}
+\frac{4}{3}(  Q^{(0)})^2   X^{(0)\alpha}  X^{(0)\kappa}\Big(  \frac{2}{3}{\mathcal{F}}^{(0)}_{\kappa\rho}{ \mathcal{F}}^{(0)\gamma\delta}
- ({\mathcal{F}}^{(0)})^2{\mathcal{I}}^{(0)}_{\kappa\rho}{}^{\gamma\delta}
\Big){\mathcal{I}}^{(0)}_{\mu\nu\sigma}{}^{\rho}\nabla^{(0)}_{\gamma}h_{\delta\alpha}
\nonumber
\\
&&
 \hspace{-4em}
- (  Q^{(0)})^2  X^{(0)\gamma}\Big( \frac{2}{3}  X^{(0)}_{\sigma}  {\mathcal{F}}^{(0)}_{\mu\nu}{\mathcal{F}}^{(0)\alpha\beta}
-\frac{1}{2} ({\mathcal{F}}^{(0)})^2   X^{(0)}_{\sigma}{\mathcal{I}}^{(0)}_{\mu\nu}{}^{\alpha\beta}
-2   X^{(0)\lambda}{\mathcal{F}}^{(0)}_{\mu\nu}{\mathcal{F}}^{(0)}_{\sigma}{}^{\rho}{\mathcal{I}}^{(0)}_{\lambda\rho}{}^{\alpha\beta} \Big) \nabla^{(0)}_{\alpha}h_{\beta\gamma}
\nonumber
\\
&&
 \hspace{-4em}
+\frac{1}{4}( Q^{(0)})^2 ({\mathcal{F}}^{(0)})^2   X^{(0)}_{\sigma} {\mathcal{F}}^{(0)}_{[\mu }{}^{\alpha} h_{\nu]\alpha}
-\frac{1}{3}( Q^{(0)})^2({\mathcal{F}}^{(0)})^2  X^{(0)\kappa}  {\mathcal{F}}^{(0)}_{\sigma}{}^{\rho}{\mathcal{I}}^{(0)}_{\kappa\rho[\mu }{}^{\alpha} h_{\nu]\alpha}
\nonumber
\\
&&
 \hspace{-4em} +  \frac{1}{6} ( Q^{(0)})^2   X^{(0)}_{\sigma}  {\mathcal{F}}^{(0)}_{\gamma\delta} {\mathcal{G}}^{(\mathrm{lin})\gamma\delta}{\mathcal{F}}^{(0)}_{\mu\nu}
+\frac{2}{9} (  Q^{(0)})^2  \sigma^{(0)\rho} {\mathcal{F}}^{(0)}_{\alpha\beta} {\mathcal{G}}^{(\mathrm{lin})\alpha\beta}{ \mathcal{I}}^{(0)}_{\mu\nu\sigma\rho}
\nonumber
\\
&&
 \hspace{-4em}
+\frac{1}{4}(  Q^{(0)})^2({\mathcal{F}}^{(0)})^2   X^{(0)}_{\sigma}{\mathcal{G}}^{(\mathrm{lin})}_{\mu\nu}
-\frac{1}{2}(  Q^{(0)})^2  \sigma^{(0)\rho}{\mathcal{F}}^{(0)}_{\mu\nu}{\mathcal{G}}^{(\mathrm{lin})}_{\sigma\rho}
\nonumber
\\
&&
 \hspace{-4em}
-\frac{2}{3}(  Q^{(0)})^2({\mathcal{F}}^{(0)})^2  X^{(0)\lambda}{\mathcal{G}}^{(\mathrm{lin})}_{\lambda}{}^{\rho}{\mathcal{I}}^{(0)}_{\mu\nu\sigma\rho}
+ O(h^2)
\,.
\end{eqnarray}
Let us consider the  term in brackets in the first two lines somewhat more carefully.
The tensor
\begin{equation}
\Xi_{\alpha\beta\mu\nu}:=\nabla^{(0)}_{[\mu}\nabla^{(0)}_{|[\alpha} h_{\beta]|\nu]}
-\frac{1}{2}h_{[\alpha}{}^{\kappa} C^{(0)}_{\beta]\kappa\mu\nu}
\end{equation}
is manifestly antisymmetric in its first and last pair of indices. One checks that it is also trace-free,
\begin{equation}
\Xi_{\alpha\beta\mu}{}^{\beta}=\frac{1}{4}\nabla^{(0)}_{\alpha}\nabla^{(0)}_{\mu} \mathrm{tr}_{\eta} h
- \frac{1}{2}\nabla^{(0)}_{(\alpha}\nabla^{(0)}_{|\nu|} h_{\mu)}{}^{\nu}
+ \frac{1}{4}\Box_{\eta} h_{\alpha\mu}
+\frac{1}{2}h^{\kappa\beta} C^{(0)}_{\alpha\kappa\mu\beta}=0
\end{equation}
by the   linearized field equations \eq{linearized_Einstein} and   the gauge condition \eq{gauge_condition}.
%
%
This implies that
\begin{equation}
\volform^{(0)}_{\alpha\beta}{}^{\gamma\delta}\Xi_{\mu\nu\gamma\delta} =\volform^{(0)}_{\mu\nu}{}^{\gamma\delta}\Xi_{\alpha\beta\gamma\delta}
\,.
\label{useful_relation}
\end{equation}
To see this, we observe that, applying $\volform^{(0)\mu\nu\kappa\rho}$,  \eq{useful_relation} is equivalent to 
$\volform^{(0)\mu\nu\kappa\rho}\volform^{(0)}_{\alpha\beta\gamma\delta}\Xi_{\mu\nu}{}^{\gamma\delta} =4\Xi_{\alpha\beta}{}^{\kappa\rho}$,
which in turn follows from the tracelessness of $\Xi_{\alpha\beta\mu\nu}$,
\begin{equation*}
\volform^{(0)\mu\nu\kappa\rho}\volform^{(0)}_{\alpha\beta\gamma\delta}\Xi_{\mu\nu}{}^{\gamma\delta} 
= 24\delta_{[\alpha}{}^{\mu}\delta_\beta{}^{\nu}\delta_{\gamma}{}^{\kappa}\delta_{\delta]}{}^{\rho}\Xi_{\mu\nu}{}^{\gamma\delta} 
= 24\delta_{[\gamma}{}^{\kappa}\delta_{\delta}{}^{\rho}\Xi_{\alpha\beta]}{}^{\gamma\delta} 
=4\Xi_{\alpha\beta}{}^{\kappa\rho}
\,.
\end{equation*}

It follows that the term in brackets vanishes
\begin{eqnarray*}
&&\hspace{-3em}\volform^{(0)}_{\mu\nu}{}^{\alpha\beta}{C}^{(0)}_{\alpha\beta[\gamma}{}^{\kappa}h_{\rho]\kappa}
-{\volform}^{(0)}_{\gamma\rho}{}^{\alpha\beta}  C^{(0)}_{\alpha\beta[\mu }{}^{\lambda} h_{\nu]\lambda}
+2{\volform}^{(0)}_{\mu\nu}{}^{\alpha\beta} \nabla^{(0)}_{[\gamma}\nabla^{(0)}_{|\alpha}h_{\beta|\rho]}
- 2{\volform}^{(0)}_{\gamma\rho}{}^{\alpha\beta}\nabla^{(0)}_{[\mu} \nabla^{(0)}_{|\alpha}h_{\beta|\nu]}
\\
&\equiv&
\underbrace{\volform^{(0)}_{\mu\nu}{}^{\alpha\beta}{C}^{(0)}_{\alpha\beta[\gamma}{}^{\kappa}h_{\rho]\kappa}-{\volform}^{(0)}_{\mu\nu}{}^{\alpha\beta} C^{(0)}_{\gamma\rho\alpha}{}^{\kappa} h_{\beta\kappa}
-{\volform}^{(0)}_{\gamma\rho}{}^{\alpha\beta}  C^{(0)}_{\alpha\beta[\mu }{}^{\lambda} h_{\nu]\lambda}
+ {\volform}^{(0)}_{\gamma\rho}{}^{\alpha\beta} C^{(0)}_{\mu\nu\alpha}{}^{\kappa} h_{\beta\kappa}}_{=0}
\\
&&
+2{\volform}^{(0)}_{\mu\nu}{}^{\alpha\beta}\Xi_{\alpha\beta\gamma\rho}
- 2{\volform}^{(0)}_{\gamma\rho}{}^{\alpha\beta}\Xi_{\alpha\beta\mu\nu}
\\
&=&0
\,.
\end{eqnarray*}

Finally, from  \eq{linearized_MST}-\eq{linearized_MST_contr} we compute
\begin{eqnarray}
&&\hspace{-3em}  X^{(0)\kappa} \Big( {\mathcal{F}}^{(0)}_{\sigma}{}^{\rho} {\mathcal I}^{(0)}_{\mu\nu}{}^{\gamma\delta}
- \frac{2}{3} {\mathcal I}^{(0)}_{\mu\nu\sigma}{}^{\rho} {\mathcal F}^{(0)\gamma\delta} \Big) {\mathcal{S}}_{\gamma\delta\kappa\rho}
\nonumber
\\
 &=&
Q^{(\mathrm{lin})}   X^{(0)\kappa} \Big( {\mathcal{F}}^{(0)}_{\sigma}{}^{\rho} {\mathcal{U}}^{(0)}_{\mu\nu\kappa\rho}
+\frac{4}{9}({\mathcal{F}}^{(0)})^2  {\mathcal{F}}^{(0)}_{\kappa}{}^{\rho}
 {\mathcal I}^{(0)}_{\mu\nu\sigma\rho}
\Big)
-4   X^{(0)\kappa}  {\mathcal{F}}^{(0)}_{\sigma}{}^{\rho}{\mathcal{I}}^{(0)}_{\kappa\rho}{}^{\alpha\beta}\nabla^{(0)}_{[\mu} \nabla^{(0)}_{|\alpha|}h_{\nu]\beta}
\nonumber
\\
&&
+\frac{8}{3}    X^{(0)\kappa}{\mathcal{F}}^{(0)\alpha\beta} {\mathcal I}^{(0)}_{\mu\nu\sigma}{}^{\rho} \nabla^{(0)}_{\alpha}\nabla^{(0)}_{[\kappa} h_{\rho]\beta}
+\frac{4}{3}  X^{(0)\kappa}{\mathcal{F}}^{(0)\alpha\beta} {\mathcal I}^{(0)}_{\mu\nu\sigma}{}^{\rho}{C}^{(0)}_{\kappa \rho\alpha }{}^{\gamma} h_{\gamma\beta}
\nonumber
\\
&&
-\frac{1}{4} Q^{(0)} ({\mathcal{F}}^{(0)})^2  X^{(0)}_{\sigma} {\mathcal{F}}^{(0)}_{[\mu }{}^{\alpha} h_{\nu]\alpha}
+\frac{1}{3} Q^{(0)}({\mathcal{F}}^{(0)})^2 X^{(0)\kappa}  {\mathcal{F}}^{(0)}_{\sigma}{}^{\rho}{\mathcal{I}}^{(0)}_{\kappa\rho[\mu }{}^{\alpha} h_{\nu]\alpha}
\nonumber
\\
&&
+ Q^{(0)} X^{(0)\gamma} \Big( \frac{2}{3} X^{(0)}_{\sigma} {\mathcal{F}}^{(0)}_{\mu\nu} { \mathcal{F}}^{(0)\alpha\beta}
-\frac{1}{2}  X^{(0)}_{\sigma}({\mathcal{F}}^{(0)})^2{\mathcal{I}}^{(0)}_{\mu\nu}{}^{\alpha\beta}
-
2  X^{(0)\kappa}  {\mathcal{F}}^{(0)}_{\mu\nu} {\mathcal{F}}^{(0)}_{\sigma}{}^{\rho} {\mathcal{I}}^{(0)}_{\kappa\rho}{}^{\alpha\beta} \Big) \nabla^{(0)}_{\alpha}h_{\beta\gamma}
\nonumber
\\
&&
-  \frac{4}{3}{Q}^{(0)} X^{(0)\kappa}  X^{(0)\gamma}\Big(\frac{2}{3} {\mathcal{F}}^{(0)}_{\kappa\rho}{ \mathcal{F}}^{(0)\alpha\beta}
- ( {\mathcal{F}}^{(0)})^2 {\mathcal{I}}^{(0)}_{\kappa\rho}{}^{\alpha\beta} \Big) {\mathcal I}^{(0)}_{\mu\nu\sigma}{}^{\rho}\nabla^{(0)}_{\alpha}h_{\beta\gamma} 
\nonumber
\\
&&
-  \frac{1}{6}  Q^{(0)}  X^{(0)}_{\sigma}  {\mathcal{F}}^{(0)}_{\gamma\delta} {\mathcal{G}}^{(\mathrm{lin})\gamma\delta}{\mathcal{F}}^{(0)}_{\mu\nu}
+\frac{1}{2}  Q^{(0)}  \sigma^{(0)\rho}{\mathcal{F}}^{(0)}_{\mu\nu}{\mathcal{G}}^{(\mathrm{lin})}_{\sigma\rho}
-\frac{1}{4}  Q^{(0)}({\mathcal{F}}^{(0)})^2   X^{(0)}_{\sigma}{\mathcal{G}}^{(\mathrm{lin})}_{\mu\nu}
\nonumber
\\
&&
+\frac{2}{3}  Q^{(0)}\Big(({\mathcal{F}}^{(0)})^2   X^{(0)\lambda}{\mathcal{G}}^{(\mathrm{lin})}_{\lambda}{}^{\rho}{\mathcal{I}}^{(0)}_{\mu\nu\sigma\rho}
-\frac{1}{3}   \sigma^{(0)\rho} {\mathcal{F}}^{(0)}_{\alpha\beta} {\mathcal{G}}^{(\mathrm{lin})\alpha\beta}\Big){ \mathcal{I}}^{(0)}_{\mu\nu\sigma\rho}
+ O(h^2)
\,,
\end{eqnarray}
 and one readily checks that
\begin{equation}
\nabla^{(0)}_{\rho}\mathcal{S}_{\alpha\beta\mu}{}^{\rho}
+{Q}^{(0)} X^{(0)\sigma} \Big( {\mathcal{F}}^{(0)}_{\mu}{}^{\rho} {\mathcal I}^{(0)}_{\alpha\beta}{}^{\gamma\delta}
- \frac{2}{3} {\mathcal I}^{(0)}_{\alpha\beta\mu}{}^{\rho} {\mathcal F}^{(0)\gamma\delta} \Big) {\mathcal{S}}_{\gamma\delta\sigma\rho}  = O(h^2)
\,.
\end{equation}

We have proved:

\begin{proposition}
\label{prop_lin}
Let $( M , \eta,   X^{(0)})$ be a $\Lambda=0$-vacuum spacetime with a Killing vector $ X^{(0)}$ such that the associated MST vanishes.
Let $h$ be a vacuum  perturbation of $\eta$ such that $\eta+ h$ admits a Killing vector which is a perturbation of $  X^{(0)}$, i.e.\ which satisfies $\mcL_{   X^{(0)}} h_{\mu\nu}   + 2 \nabla^{(0)}_{(\mu} X^{(\mathrm{lin})} _{\nu)}=0 $
for some perturbation $ X^{(\mathrm{lin})}$ of $  X^{(0)}$.
Then  the linearized MST associated to $( M , \eta+h,   X^{(0)}+X^{(\mathrm{lin})})$ satisfies the MST equation \eq{MST_eqn_background} on the background $( M , \eta,  X^{(0)})$.
\end{proposition}

\begin{remark}
{\rm
As indicated above, one should expect this  result to hold for any sign of the cosmological constant.
}
\end{remark}

\begin{remark}
{\rm
\label{expectation}
We have not attempted to derive an analog of Lemma~\ref{lemma_Bianchi2}. Nevertheless, we  expect that  a solution  $\mathcal{S}_{\alpha\beta\mu\nu}$
to the MST equation on a vacuum background  $( M , \eta,   X^{(0)})$ with vanishing MST  defines a unique (up to gauge transformations \eq{gauge_trafos}) perturbation $h$ of $\eta$ and $X^{(\mathrm{lin})}$ of $ X^{(0)}$ such  that the linearized vacuum equations \eq{linearized_Einstein} and  $\mcL_{  X^{(0)}} h_{\mu\nu}   + 2 \nabla^{(0)}_{(\mu} X^{(\mathrm{lin})} _{\nu)}=0 $  are fulfilled, and such
 that $\mathcal{S}_{\alpha\beta\mu\nu}$ is the linearized MST of $( M , \eta+h,  X^{(0)}+X^{(\mathrm{lin})})$.
Note, though,  that  in Lemma~\ref{lemma_Bianchi2}  a vanishing cosmological constant has been assumed, so before considering this issue for the MST with $\Lambda\ne0 $, the corresponding issue for the Bianchi equation should be analyzed.
}
\end{remark}

\subsection{Constraint and evolution equations}
\label{sec_constr_ev}

Let $\Sigma\subset M $ be a Cauchy surface in some $\Lambda$-vacuum spacetime $( M , g)$ which admits a Killing vector $X$.
Later on, when we work in a  conformally rescaled spactime $\widetilde g = \Theta^2 g$ we will introduce Gaussian coordinates,
at this stage, though, we merely assume coordinates $(t\equiv x^0, x^i)$ near $\Sigma$ where
\begin{equation}
g_{0i} = 0\,.
\label{phys_gauge_cond}
\end{equation}

\begin{remark}
{\rm
In Section~\ref{sec_lin_grav} we have  denoted the background spacetime by $( M , \eta,   X^{(0)})$, and the associated fields have been decorated with
superscript $(0)$. Since from now on the background spacetime will be the only one we are working with, we will simply denote it by $( M , g,   X)$ henceforth  without
any superscripts  marking the associated fields.
However, we will still denote   any solution of the MST equation \eq{MST_evolution_eqn} by $\mathcal{S}_{\alpha\beta\mu\nu}$.
The  MST associated to the spacetime  $( M , g, X)$  will be denoted by $\mathfrak{S}_{\alpha\beta\mu\nu}$.
}
\end{remark}

It follows readily from the algebraic symmetries of $\mathcal{S}_{\alpha\beta\mu\nu}$ and its self-duality that
the MST equation is equivalent to the system (cf.\ also Remark~\ref{properties_I}),
\begin{eqnarray}
\nabla_{\rho}\mathcal{S}_{0(ij)}{}^{\rho}& =&\mathcal{J}( {\mathcal{S}})_{0(ij)}
\,,
\label{MST_evolution}
\\
\nabla_{\rho}\mathcal{S}_{0i0}{}^{\rho} &=&\mathcal{J}( {\mathcal{S}})_{0i0}
\,.
\label{MST_constraint}
\end{eqnarray}
As for the Bianchi equation one shows that \eq{MST_evolution} forms a regular
 symmetric hyperbolic system of evolution equations (supposing that there are no blow-ups in the denominator of \eq{MST_rhs}), for which standard well-posedness results are available, cf.\ \cite{mpss}.
Equation \eq{MST_constraint} does not contain transverse derivatives and therefore provides a set of constraint equations.

\subsection{Generalized Buchdahl condition}
\label{sec:generalizedbuchner}

Let us devote attention to the issue whether the constraint equations \eq{MST_constraint} are preserved under evolution.
The main aim of this section is to derive an analog of the Buchdahl condition \eq{Buchdahl_cond}.

Employing the algebraic Weyl symmetries of the fields involved as well as self-duality  we find
(the vacuum equations are not needed at this stage)
\begin{eqnarray*}
\nabla_0\nabla_{\rho}\mathcal{S}_{0i0}{}^{\rho}
&=&
-g_{00}\nabla^{j}\nabla_0\mathcal{S}_{0ij}{}^0 + R_{0j0}{}^{k}\mathcal{S}_{k i0}{}^{j}
+ R_{0ji}{}^{k}\mathcal{S}_{0k 0}{}^{j}
-R_{0j}\mathcal{S}_{0i0}{}^{j}
\\
&=&
-g_{00}\nabla^{j}\nabla_{\rho}\mathcal{S}_{0ij}{}^{\rho}
+g_{00}\nabla_{j}\nabla_k\mathcal{S}_{0i}{}^{jk}
 + R_{0j0}{}^{k}\mathcal{S}_{k i0}{}^{j}
+ R_{0ji}{}^{k}\mathcal{S}_{0k 0}{}^{j}
\\
&&
-R_{0j}\mathcal{S}_{0i0}{}^{j}
\\
&=&
-g_{00}\rnabla^{j}\nabla_{\rho}\mathcal{S}_{0ij}{}^{\rho}
+g_{00}\Gamma^0_{0j}\nabla_{\rho}\mathcal{S}_{0 i}{}^{j\rho}
+g_{00}\Gamma^k_{0j}\nabla_{\rho}\mathcal{S}_{k i}{}^{j\rho}
-\Gamma^j_{j0}\nabla_{\rho}\mathcal{S}_{0i0}{}^{\rho}
\\
&&
+\frac{1}{2}g_{00} R_{jk0}{}^{l}\mathcal{S}_{l i}{}^{jk}
+ \frac{1}{2}g_{00} R_{jki}{}^{l}\mathcal{S}_{0l}{}^{jk}
 + R_{0j0}{}^{k}\mathcal{S}_{k i0}{}^{j}
+ R_{0ji}{}^{k}\mathcal{S}_{0k 0}{}^{j}
\\
&=&
-g_{00} \rnabla^{j}\nabla_{\rho}\mathcal{S}_{0ij}{}^{\rho}
+g_{00}\Gamma^0_{0j}\nabla_{\rho}\mathcal{S}_{0 i}{}^{j\rho}
+g_{00}\Gamma^k_{0j}\nabla_{\rho}\mathcal{S}_{k i}{}^{j\rho}
-\Gamma^j_{j0}\nabla_{\rho}\mathcal{S}_{0i0}{}^{\rho}
\\
&&
-\frac{1}{2}g_{00}C_{jk 0}{}^{l} \mathcal{S}_{il}{}^{jk}
+\frac{1}{2}g_{00}C_{il}{}^{jk} \mathcal{S}_{jk0}{}^{l}
- C_{0j0}{}^{k}\mathcal{S}_{ik 0}{}^{j}
+ C_{0ji}{}^{k}\mathcal{S}_{0k 0}{}^{j}
\\
&=&
-g_{00} \rnabla^{j}\nabla_{\rho}\mathcal{S}_{0ij}{}^{\rho}
+g_{00}\Gamma^0_{0j}\nabla_{\rho}\mathcal{S}_{0 i}{}^{j\rho}
+g_{00}\Gamma^k_{0j}\nabla_{\rho}\mathcal{S}_{k i}{}^{j\rho}
-\Gamma^j_{j0}\nabla_{\rho}\mathcal{S}_{0i0}{}^{\rho}
\\
&&
+2 \mathcal{C}_{ik0}{}^j\mathcal{S}_{0j 0}{}^{k}
\,.
\end{eqnarray*}
Here $\rnabla$ denotes the Levi-Civita connection associated to the Riemannian family $t\mapsto g_{ij}(t,x^k)$.

If we assume that the evolution equations
\begin{equation}
\nabla_{\rho}\mathcal{S}_{0(ij)}{}^{\rho} =\mathcal{J}( {\mathcal{S}})_{0(ij)}
\label{ev_eqns}
\end{equation}
are fulfilled we derive from that formula
an equation which is satisfied by the \emph{constraint violation operator}
\begin{equation}
\Xi_i := \nabla_{\rho}\mathcal{S}_{0i0}{}^{\rho}-\mathcal{J}( \mathcal{S})_{0i0}\,.
\label{constr_op}
\end{equation}
We obtain
\begin{equation}
\Big( \delta_i{}^k(\partial_0+\frac{3}{2}\Gamma^j_{j0}-2\Gamma^0_{00})-\frac{\imagi}{2} \volform_{0i}{}^{jk}(\rnabla_{j}+2\Gamma^0_{0j})-\frac{3}{2}\Gamma^k_{0i}\Big)\Xi_k
=
-g_{00}\nabla^{\rho}\mathcal{J}( \mathcal{S})_{0i\rho}
+2 \mathcal{C}_{ik0}{}^j\mathcal{S}_{0j 0}{}^{k}
\,.
\label{prev1}
\end{equation}

Before we proceed let us compare this with the Bianchi case where $\mathcal{J}(\mathcal{S})$ vanishes.
Then the symmetric hyperbolic system
\begin{equation*}
\Big( \delta_i{}^k(\partial_0+\frac{3}{2}\Gamma^j_{j0}-2\Gamma^0_{00})-\frac{\imagi}{2} \volform_{0i}{}^{jk}(\rnabla_{j}+2\Gamma^0_{0j})-\frac{3}{2}\Gamma^k_{0i}\Big)\nabla_{\rho}\mathcal{S}_{0k0}{}^{\rho}
=
2 \mathcal{C}_{ik0}{}^j\mathcal{S}_{0j 0}{}^{k}
\,.
\end{equation*}
ensures that the constraints  are preserved, i.e.\ zero data yield the zero-solution,
if and only if
\begin{equation}
\mathcal{C}_{ij0}{}^k\mathcal{S}_{0k 0}{}^{j} =0
\quad \Longleftrightarrow \quad \mathcal{C}_{[0}{}^{\alpha\beta\gamma}\mathcal{S}_{i]\alpha\beta\gamma} =0
\,.
\end{equation}
This recovers  the Buchdahl condition \eq{Buchdahl_cond}  in self-dual language.

\subsubsection{Divergence of \texorpdfstring{$\mathcal{J}(\mathcal{S})$}{J(S)}}

In order to analyze the general case, we need to determine $\nabla^{\mu}\mathcal{J}( \mathcal{S})_{0i\mu}$ in \eq{prev1}.
This requires some computational effort.
It is convenient to collect some relations needed for this computation first.

Any self dual two form $\mathcal{V}_{\alpha\beta}$ satisfies
\begin{equation}
\mathcal{V}_{\mu\alpha}\mathcal{V}_{\nu}{}^{\alpha} = \frac{1}{4}\mathcal{V}^2 g_{\mu\nu}
\,.
\end{equation}
In any $\Lambda$-vacuum space-time a self-dual Killing form fulfills the following relations \cite{mars-senovilla},
\begin{eqnarray}
\nabla_{\mu}\mathcal{F}_{\alpha\beta} &=& -X^{\nu}\Big(\mathcal{C}_{\mu\nu\alpha\beta} + \frac{4}{3}\Lambda \mathcal{I}_{\mu\nu\alpha\beta}\Big)
\label{rln_nablaFab}
\\
&=&
 -X^{\nu}\Big(\mathfrak{S}_{\mu\nu\alpha\beta}+ Q \mathcal{F}_{\mu\nu}\mathcal{F}_{\alpha\beta}
-\frac{1}{3}( Q\mathcal{F}^2-4\Lambda )\mathcal{I}_{\mu\nu\alpha\beta}\Big)
\,,
\label{rln_nablaF}
\\
\nabla_{\mu}\mathcal{F}_{\alpha}{}^{\mu} &=& \Lambda X_{\alpha}\,,
\label{rln_divF}
\\
\nabla_{\mu}\mathcal{F}^2 &=&
-2X^{\nu}\Big(\mathcal{F}^{\alpha\beta}\mathcal{C}_{\mu\nu\alpha\beta} + \frac{4}{3}\Lambda\mathcal{F}_{\mu\nu}\Big)
\\
&=&
\frac{4}{3}(Q\mathcal{F}^2+2\Lambda) X^{\nu}\mathcal{F}_{\nu\mu}
-2X^{\nu}\mathcal{F}^{\alpha\beta}\mathfrak{S}_{\mu\nu\alpha\beta}
\label{rln_nablaF2}
\,.
\end{eqnarray}
We need to determine the gradient of $Q$.
For this recall the definition \eq{dfn_Q}-\eq{dfn_J} of $Q$. In particular it implies
\begin{equation}
 R^2=-\frac{1}{4}\mathcal{F}^2
\,,
\quad
\sigma J^ 2 -2JR +\Lambda =0
\,,
\end{equation}
%
whence
\begin{eqnarray}
\nabla_{\mu}(Q\mathcal{F}^2) &=&-12\nabla_{\mu}(JR)
\nonumber
\\
&=&  -6\Big(\frac{J}{R} +\frac{1}{\sigma} +\frac{ R}{\sigma(J\sigma-R)} \Big)\nabla_{\mu}R^2
 +6\frac{R}{\sigma}\Big(  \frac{ \Lambda}{J\sigma-R}
 +2J \Big) \nabla_{\mu}\sigma
\nonumber
\\
&=&  \frac{3}{2}\frac{J^2\sigma}{R(J\sigma- R)}\nabla_{\mu}\mathcal{F}^2
 +12 \frac{ J^2R}{J\sigma-R} X^{\alpha}\mathcal{F}_{\alpha\mu}
\nonumber
\\
&\overset{\eq{rln_nablaF2}}{=}&
Q (Q\mathcal{F}^2 -4\Lambda)X^{\alpha}\mathcal{F}_{\alpha\mu}
 -2\frac{( Q \mathcal{F}^2+2 \Lambda )(Q\mathcal{F}^2-4\Lambda)}{ Q\mathcal{F}^2+ 8 \Lambda   } \mathcal{F}^{-2} X^{\nu}\mathcal{F}^{\alpha\beta}\mathfrak{S}_{\mu\nu\alpha\beta}
\,.\phantom{xxx}
\label{rln_nablaQF2}
\end{eqnarray}

Finally, using all these relations, we compute  the divergence of $\mathcal{J}(\mathcal{S})$.
A somewhat lengthy calculation reveals that
\begin{eqnarray*}
\nabla^{\mu}\mathcal{J}( {\mathcal{S}})_{\alpha\beta\mu}
&=&
 - \frac{4}{3} \Lambda(Q\mathcal{F}^2-4\Lambda)  \frac{  5  Q\mathcal{F}^2  +4\Lambda }{Q\mathcal{F}^2 + 8\Lambda}
   \mathcal{F}^{-4} \mathcal{I}_{\alpha\beta\mu\nu}
   X^{\nu}X^{\sigma}  \mathcal{F}^{\gamma\delta}\mathcal{F}_{\rho}{}^{\mu}  {\mathcal{S}}_{\gamma\delta\sigma}{}^{\rho}
\\
&&
- 4 \Lambda  \frac{  5  Q\mathcal{F}^2  +4\Lambda }{Q\mathcal{F}^2 + 8\Lambda}
 \mathcal{F}_{\alpha\beta}\mathcal{F}^{\mu\rho}
   \mathcal{F}^{-4}  X^{\sigma}  \mathcal{F}^{\gamma\delta}  \nabla_{\mu}{\mathcal{S}}_{\rho\sigma\gamma\delta}
\\
&&
+\frac{2}{9} \mathcal{F}^{-4}(Q\mathcal{F}^2+8\Lambda ) \Big(  Q \mathcal{F}^{2}
-2\Lambda  \frac{  5  Q\mathcal{F}^2  +4\Lambda }{Q\mathcal{F}^2 + 8\Lambda}
\Big)
X^{\nu}X^{\sigma}\mathcal{F}^{\gamma\delta} \mathcal{F}_{\nu [\alpha}  {\mathcal{S}}_{ \beta]\sigma\gamma\delta}
\\
&&
- \frac{2}{3}  Q \mathcal{F}^{-2} \frac{Q \mathcal{F}^2 -4\Lambda}{Q\mathcal{F}^2 + 8\Lambda}\Big( Q\mathcal{F}^2
+8\Lambda \frac{    Q\mathcal{F}^2  -\Lambda }{Q\mathcal{F}^2 + 8\Lambda}
\Big)
 X^{\nu}  X^{\sigma} \mathcal{F}^{\gamma\delta} \mathcal{F}_{\nu[\alpha} {\mathcal{S}}_{\beta]\sigma\gamma\delta}
\\
&&
- \frac{1}{3}  \mathcal{F}^{-2}  \Big(  Q  \mathcal{F}^{2}
-2\Lambda  \frac{  5  Q\mathcal{F}^2  +4\Lambda }{Q\mathcal{F}^2 + 8\Lambda}
\Big)  \mathcal{F}^{\gamma\delta} \mathcal{F}_{[\alpha}{}^{\sigma} {\mathcal{S}}_{\beta]\sigma \gamma\delta}
\\
&&
- \frac{2}{3}  \mathcal{F}^{-2}  \Big(  Q  \mathcal{F}^{2}
-2\Lambda  \frac{  5  Q\mathcal{F}^2  +4\Lambda }{Q\mathcal{F}^2 + 8\Lambda}
\Big) \mathcal{I}_{\alpha\beta}{}^{\mu\rho}  X^{\sigma} \mathcal{F}^{\gamma\delta} \nabla_{\mu} {\mathcal{S}}_{\rho\sigma\gamma\delta}
\\
&&
+QX^{\sigma} \mathcal{F}^{\mu\rho}\nabla_{\mu} {\mathcal{S}}_{\rho\sigma\alpha\beta}
+ f_{\alpha\beta}{}^{\mu\nu\sigma\rho}\mathfrak{S}_{\mu\nu\sigma\rho}
\,,
\end{eqnarray*}
where the precise form of the generic tensor field $ f_{\alpha\beta}{}^{\mu\nu\sigma\rho}$, which depends on the metric $g$, the Killing vector field $X$ and $\mathcal{S}_{\alpha\beta\mu\nu}$ will be irrelevant for our purposes.

The right-hand side contains derivatives of $\mathcal{S}_{\mu\nu\sigma\rho}$. We observe that they can be expressed as a linear combination   of
derivatives of the form $X^{\kappa}\nabla_{\kappa}\mathcal{S}_{\alpha\beta\mu\nu}$ and
  $\nabla_{[\mu} {\mathcal{S}}_{\rho\sigma]\alpha\beta} $.
It follows from the algebraic symmetries of $\mathcal{S}_{\mu\nu\sigma\rho}$ that
\begin{equation}
\nabla_{[\mu} {\mathcal{S}}_{\rho\sigma]\alpha\beta}
=
-\frac{\imagi}{3}
\volform_{\mu\rho\sigma}{}^{\kappa}\nabla_{\lambda}{\mathcal{S}}_{\alpha\beta\kappa}{}^{\lambda}
\,.
\end{equation}
Taking this into account we obtain
\begin{eqnarray*}
\nabla^{\mu}\mathcal{J}( {\mathcal{S}})_{\alpha\beta\mu}
&=&
 - \frac{4}{3} \Lambda(Q\mathcal{F}^2-4\Lambda)  \frac{  5  Q\mathcal{F}^2  +4\Lambda }{Q\mathcal{F}^2 + 8\Lambda}
   \mathcal{F}^{-4} \mathcal{I}_{\alpha\beta\mu\nu}
   X^{\nu}X^{\sigma}  \mathcal{F}^{\gamma\delta}\mathcal{F}_{\rho}{}^{\mu}  {\mathcal{S}}_{\gamma\delta\sigma}{}^{\rho}
\\
&&
\hspace{-4em}
+4\Lambda  \frac{  5  Q\mathcal{F}^2  +4\Lambda }{Q\mathcal{F}^2 + 8\Lambda}  \mathcal{F}^{-4}
 \mathcal{F}_{\alpha\beta} X^{\sigma}\mathcal{F}_{\sigma}{}^{\kappa}
   \mathcal{F}^{\gamma\delta} \nabla_{\lambda}{\mathcal{S}}_{\gamma\delta\kappa}{}^{\lambda}
\\
&&
\hspace{-4em}
+\frac{2}{9} \mathcal{F}^{-4}(Q\mathcal{F}^2+8\Lambda ) \Big(  Q  \mathcal{F}^{2}
-2\Lambda  \frac{  5  Q\mathcal{F}^2  +4\Lambda }{Q\mathcal{F}^2 + 8\Lambda}
\Big)
X^{\nu}X^{\sigma}\mathcal{F}^{\gamma\delta} \mathcal{F}_{\nu [\alpha}  {\mathcal{S}}_{ \beta]\sigma\gamma\delta}
\\
&&
\hspace{-4em}
- \frac{2}{3}  Q\mathcal{F}^{-2} \frac{Q \mathcal{F}^2 -4\Lambda}{Q\mathcal{F}^2 + 8\Lambda}\Big( Q\mathcal{F}^2
+8\Lambda \frac{    Q\mathcal{F}^2  -\Lambda }{Q\mathcal{F}^2 + 8\Lambda}
\Big)
 X^{\nu}  X^{\sigma} \mathcal{F}^{\gamma\delta} \mathcal{F}_{\nu[\alpha} {\mathcal{S}}_{\beta]\sigma\gamma\delta}
\\
&&
\hspace{-4em}
+\frac{2}{3} \mathcal{F}^{-2}  \Big(  Q \mathcal{F}^{2}
-2\Lambda  \frac{  5  Q\mathcal{F}^2  +4\Lambda }{Q\mathcal{F}^2 + 8\Lambda}
\Big) \mathcal{I}_{\alpha\beta\sigma}{}^{\kappa}  X^{\sigma} \mathcal{F}^{\gamma\delta}
\nabla_{\lambda}{\mathcal{S}}_{\gamma\delta\kappa}{}^{\lambda}
\\
&&
\hspace{-4em}
-QX^{\sigma} \mathcal{F}_{\sigma}{}^{\kappa}\nabla_{\lambda}{\mathcal{S}}_{\alpha\beta\kappa}{}^{\lambda}
+2 \Lambda  \frac{  5  Q\mathcal{F}^2  +4\Lambda }{Q\mathcal{F}^2 + 8\Lambda}
   \mathcal{F}^{-4} \mathcal{F}_{\alpha\beta} \mathcal{F}^{\mu\rho}
    \mathcal{F}^{\gamma\delta}  \mcL_X {\mathcal{S}}_{\mu\rho\gamma\delta}
\\
&&
\hspace{-4em}
-\frac{1}{6}  \mathcal{F}^{-2}  \Big( Q  \mathcal{F}^{2}
+4\Lambda  \frac{  5  Q\mathcal{F}^2  +4\Lambda }{Q\mathcal{F}^2 + 8\Lambda}
\Big)  \mathcal{F}^{\gamma\delta} \mcL_X {\mathcal{S}}_{\alpha\beta\gamma\delta}
- \frac{1}{2}    Q  \mathcal{F}^{\gamma\delta} \mathcal{F}_{[\alpha}{}^{\sigma} {\mathcal{S}}_{\beta]\sigma \gamma\delta}  + f_{\alpha\beta}{}^{\mu\nu\sigma\rho}\mathfrak{S}_{\mu\nu\sigma\rho}
\,.
\end{eqnarray*}
Rewriting yields after another tedious computation
\begin{eqnarray*}
\nabla^{\mu}\mathcal{J}( {\mathcal{S}})_{\alpha\beta\mu}
&=&
4\Lambda  \frac{  5  Q\mathcal{F}^2  +4\Lambda }{Q\mathcal{F}^2 + 8\Lambda}  \mathcal{F}^{-4}
 \mathcal{F}_{\alpha\beta} X^{\sigma}\mathcal{F}_{\sigma}{}^{\kappa}
   \mathcal{F}^{\gamma\delta}( \nabla_{\lambda}{\mathcal{S}}_{\gamma\delta\kappa}{}^{\lambda}- \mathcal{J}(\mathcal{S})_{\gamma\delta\kappa})
\\
&&
\hspace{-6em}
+\frac{2}{3} \mathcal{F}^{-2}  \Big(  Q \mathcal{F}^{2}
-2\Lambda  \frac{  5  Q\mathcal{F}^2  +4\Lambda }{Q\mathcal{F}^2 + 8\Lambda}
\Big) \mathcal{I}_{\alpha\beta\sigma}{}^{\kappa}  X^{\sigma} \mathcal{F}^{\gamma\delta}
(\nabla_{\lambda}{\mathcal{S}}_{\gamma\delta\kappa}{}^{\lambda}- \mathcal{J}( {\mathcal{S}})_{\gamma\delta\kappa})
\\
&&
\hspace{-6em}
-QX^{\sigma} \mathcal{F}_{\sigma}{}^{\kappa}(\nabla_{\lambda}{\mathcal{S}}_{\alpha\beta\kappa}{}^{\lambda}
- \mathcal{J}( {\mathcal{S}})_{\alpha\beta\kappa})
+2 \Lambda  \frac{  5  Q\mathcal{F}^2  +4\Lambda }{Q\mathcal{F}^2 + 8\Lambda}
   \mathcal{F}^{-4} \mathcal{F}_{\alpha\beta} \mathcal{F}^{\mu\rho}
    \mathcal{F}^{\gamma\delta}  \mcL_X {\mathcal{S}}_{\mu\rho\gamma\delta}
\\
&&
\hspace{-6em}
- \frac{1}{6}   \Big( Q  \mathcal{F}^{2}
+4\Lambda  \frac{  5  Q\mathcal{F}^2  +4\Lambda }{Q\mathcal{F}^2 + 8\Lambda}
\Big)  \mathcal{F}^{-2}  \mathcal{F}^{\gamma\delta} \mcL_X {\mathcal{S}}_{\alpha\beta\gamma\delta}
\\
&&
\hspace{-6em}
+ \frac{1}{2}    Q \mathcal{U}_{[\alpha}{}^{\sigma\gamma\delta} {\mathcal{S}}_{\beta]\sigma \gamma\delta}  + f_{\alpha\beta}{}^{\mu\nu\sigma\rho}\mathfrak{S}_{\mu\nu\sigma\rho}
\,.
\end{eqnarray*}
For this step we used the following relation, which follows from the self-duality of the fields involved,
\begin{eqnarray*}
\mathcal{I}_{\alpha\beta\mu\nu}  X^{\nu}  X^{\sigma}\mathcal{F}_{\rho} {}^{\mu}
   \mathcal{F}^{\gamma\delta}  {\mathcal{S}}_{\gamma\delta\sigma}{}^{\rho}
=
 X^{\nu}X^{\sigma}   \mathcal{F}^{\gamma\delta} \mathcal{F}_{\nu[\alpha}
   {\mathcal{S}}_{\beta]\sigma\gamma\delta}
\,.
\end{eqnarray*}

Altogether we have shown that \eq{prev1} can be written as
\begin{eqnarray*}
&&
\hspace{-2em}
\Big( \delta_i{}^k(\partial_0+\frac{3}{2}\Gamma^j_{j0}-2\Gamma^0_{00})-\frac{\imagi}{2} \volform_{0i}{}^{jk}(\rnabla_{j}+2\Gamma^0_{0j})-\frac{3}{2}\Gamma^k_{0i}\Big)\Xi_k
\nonumber
\\
&=&
-g_{00}\Big[
4\Lambda  \frac{  5  Q\mathcal{F}^2  +4\Lambda }{Q\mathcal{F}^2 + 8\Lambda}  \mathcal{F}^{-4}
 \mathcal{F}_{0i} X^{\sigma}\mathcal{F}_{\sigma}{}^{\kappa}
   \mathcal{F}^{\gamma\delta}( \nabla_{\lambda}{\mathcal{S}}_{\gamma\delta\kappa}{}^{\lambda}- \mathcal{J}(\mathcal{S})_{\gamma\delta\kappa})
\\
&&
+\frac{2}{3} \mathcal{F}^{-2}  \Big(  Q  \mathcal{F}^{2}
-2\Lambda  \frac{  5  Q\mathcal{F}^2  +4\Lambda }{Q\mathcal{F}^2 + 8\Lambda}
\Big) \mathcal{I}_{0i\sigma}{}^{\kappa}  X^{\sigma} \mathcal{F}^{\gamma\delta}
(\nabla_{\lambda}{\mathcal{S}}_{\gamma\delta\kappa}{}^{\lambda}- \mathcal{J}( {\mathcal{S}})_{\gamma\delta\kappa})
\\
&&
-QX^{\sigma} \mathcal{F}_{\sigma}{}^{\kappa}(\nabla_{\lambda}{\mathcal{S}}_{0i\kappa}{}^{\lambda}
- \mathcal{J}( {\mathcal{S}})_{0i\kappa})
+2 \Lambda  \frac{  5  Q\mathcal{F}^2  +4\Lambda }{Q\mathcal{F}^2 + 8\Lambda}
   \mathcal{F}^{-4} \mathcal{F}_{0i} \mathcal{F}^{\mu\rho}
    \mathcal{F}^{\gamma\delta}  \mcL_X {\mathcal{S}}_{\mu\rho\gamma\delta}
\\
&&
- \frac{1}{6}   \Big( Q  \mathcal{F}^{2}
+4\Lambda  \frac{  5  Q\mathcal{F}^2  +4\Lambda }{Q\mathcal{F}^2 + 8\Lambda}
\Big)  \mathcal{F}^{-2}  \mathcal{F}^{\gamma\delta} \mcL_X {\mathcal{S}}_{0i\gamma\delta} \Big]
 + f_{0i}{}^{\mu\nu\sigma\rho}\mathfrak{S}_{\mu\nu\sigma\rho}
\,.
\end{eqnarray*}
Let us plug in the evolution equations $\nabla_{\rho}\mathcal{S}_{0(ij)}{}^{\rho} =\mathcal{J}( {\mathcal{S}})_{0(ij)}$,
\begin{eqnarray*}
&&
\hspace{-2em}
\Big[\Big( \delta_i{}^p(\partial_0+\frac{3}{2}\Gamma^j_{j0}-2\Gamma^0_{00})-\frac{\imagi}{2} \volform_{0i}{}^{jp}(\rnabla_{j}+2\Gamma^0_{0j})-\frac{3}{2}\Gamma^p_{0i}\Big)
+Q\Big( X^{j} \mathcal{F}_{0j}\delta_i{}^p
+\frac{\imagi}{2}\volform_{0ij}{}^{p}X^{\sigma} \mathcal{F}_{\sigma}{}^{j}\Big)
\\
&&
-
8\Lambda g^{00} \mathcal{F}^{-4} \frac{  5  Q\mathcal{F}^2  +4\Lambda }{Q\mathcal{F}^2 + 8\Lambda} \mathcal{F}_{0i} \Big(2X^{j} \mathcal{F}_{0j}
   \mathcal{F}_0{}^{p}
+\imagi  \volform_{0jk}{}^{p} X^{\sigma}\mathcal{F}_{\sigma}{}^{k}
   \mathcal{F}_0{}^{j}\Big)
\\
&&
-\frac{4}{3} g^{00}\mathcal{F}^{-2} \Big(  Q \mathcal{F}^{2}
-2\Lambda  \frac{  5  Q\mathcal{F}^2  +4\Lambda }{Q\mathcal{F}^2 + 8\Lambda}
\Big) \Big(2  \mathcal{I}_{0i0j}  X^{j} \mathcal{F}_0{}^{p}
+\imagi\volform_{0kl}{}^{p} \mathcal{I}_{0i\sigma}{}^{l}  X^{\sigma} \mathcal{F}_0{}^{k}
\Big)\Big]
\times \Xi_p
\\
&=&
-g_{00}\Big[32 \Lambda  \frac{  5  Q\mathcal{F}^2  +4\Lambda }{Q\mathcal{F}^2 + 8\Lambda}
   \mathcal{F}^{-2} \mathcal{F}_{0i}  \mathcal{F}_0{}^{k}
+ \frac{2}{3}   \Big( Q \mathcal{F}^{2}
+4\Lambda  \frac{  5  Q\mathcal{F}^2  +4\Lambda }{Q\mathcal{F}^2 + 8\Lambda}
\Big) \delta_i{}^k
\Big] \mathcal{F}^{-2} \mathcal{F}_0{}^{j} \mcL_X {\mathcal{S}}_{0j0k}
\\
&&
  + f_{0i}{}^{\mu\nu\sigma\rho}\mathfrak{S}_{\mu\nu\sigma\rho}
\,.
\end{eqnarray*}
Employing one more time the self-duality of the fields involved this can be written as
\begin{eqnarray}
&&
\hspace{-2em}
\Big[\Big( \delta_i{}^p(\partial_0+\frac{3}{2}\Gamma^j_{j0}-2\Gamma^0_{00})-\frac{\imagi}{2} \volform_{0i}{}^{jp}(\rnabla_{j}+2\Gamma^0_{0j})-\frac{3}{2}\Gamma^p_{0i}\Big)
\nonumber
\\
&&
+Q\Big( \frac{2}{3}X^{j} \mathcal{F}_{0j}\delta_i{}^p
-\frac{5}{6}X_0 \mathcal{F}_{i}{}^p
-\frac{1}{6}X^{p} \mathcal{F}_{0i}
- \frac{1}{6}X_{i} \mathcal{F}_{0}{}^{p}
\Big)
\nonumber
\\
&&
-
\frac{2}{3}\Lambda  \mathcal{F}^{-4} \frac{  5  Q\mathcal{F}^2  +4\Lambda }{Q\mathcal{F}^2 + 8\Lambda} \Big(36g^{00}X^{j} \mathcal{F}_{0j}   \mathcal{F}_0{}^{p} \mathcal{F}_{0i}
-2\mathcal{F}^2 X^p \mathcal{F}_{0i}
-2  \mathcal{F}^2  X_i\mathcal{F}_0{}^{p}
\nonumber
\\
&&
- \mathcal{F}^2X_{0} \mathcal{F}_{i}{}^p
- \mathcal{F}^2\delta_{i}{}^p  X^{j} \mathcal{F}_{0j}
\Big)\Big]\times \Xi_p
\nonumber
\\
&=&
-g_{00}\Big[32 \Lambda  \frac{  5  Q\mathcal{F}^2  +4\Lambda }{Q\mathcal{F}^2 + 8\Lambda}
   \mathcal{F}^{-2} \mathcal{F}_{0i}  \mathcal{F}_0{}^{k}
+ \frac{2}{3}   \Big( Q  \mathcal{F}^{2}
+4\Lambda  \frac{  5  Q\mathcal{F}^2  +4\Lambda }{Q\mathcal{F}^2 + 8\Lambda}
\Big) \delta_i{}^k
\Big]
\nonumber
\\
&&
\qquad \times \mathcal{F}^{-2} \mathcal{F}_0{}^{j} \mcL_X {\mathcal{S}}_{0j0k}
  + f_{0i}{}^{\mu\nu\sigma\rho}\mathfrak{S}_{\mu\nu\sigma\rho}
\,.
\label{constr_prop}
\end{eqnarray}

\subsubsection{Generalized Buchdahl condition and its realization}
\label{sec:realizationBuchdahlphys}

We deduce from \eq{constr_prop} that the analog of the Buchdahl condition for the MST equation \eq{MST_evolution_eqn}, which 
is necessary for
the
preservation of the constraints under evolution, adopts the form
\begin{equation}
 f_i{}^j  \mathcal{F}_0{}^{k} \mcL_X {\mathcal{S}}_{0j0k}
  + f_{0i}{}^{\mu\nu\sigma\rho}\mathfrak{S}_{\mu\nu\sigma\rho} =0
\,.
\label{Buchdahl_gen}
\end{equation}
Since \eq{constr_prop} is symmetric hyperbolic -- its principal part has the same structure as the principal part of the system discussed in the specific case in \Sectionref{sec_constraints} -- the validity of the Buchdahl condition is also sufficient for the preservation of the constraints.
Similar to how one proceeds to construct solutions to the Bianchi equation, let us  assume that the MST of the background metric vanishes,
\begin{equation}
\mathfrak{S}_{\mu\nu\sigma\rho} =0
\,.
\label{assumption1}
\end{equation}
%
(It also follows from the considerations in Section~\ref{sec_lin_MST}  and the relevance of spacetimes with vanishing MST that this is a reasonable ansatz to realize \eq{Buchdahl_gen}.)

In contrast to the Bianchi equation, though,  this is not sufficient to satisfy the \emph{generalized  Buchdahl condition}  \eq{Buchdahl_gen},
whence we assume, in addition to  \eq{assumption1}, that
\begin{equation}
 \mcL_X {\mathcal{S}}_{\alpha\beta\mu\nu} =0
\,.
\label{assumption2b}
\end{equation}
Recall that the Lie derivative of the MST always vanishes, $\mcL_X\mathfrak{S}_{\alpha\beta\mu\nu}=0$.
Having the expectation in mind (cf.\ Remark~\ref{expectation}) that  a solution to the MST equation on a given background with vanishing MST  can be  interpreted
as the linearized MST of some perturbed metric, \eq{assumption2b}  seems to be a very natural condition.


%

In contrast to \eq{assumption1}, though,   condition \eq{assumption2b} involves the solution and not just the background spacetime.
Fortunately, this  spacetime condition  can be realized by an appropriate choice of the initial data as will be discussed next.

Let us consider again \eq{MST_evolution_eqn}.
Note that it follows from \eq{rln_nablaFab} and \eq{rln_nablaQF2}
that
\begin{equation}
\mcL_X \mathcal{F}_{\alpha\beta }= 0
\,,\quad
\mcL_X (Q\mathcal{F}^2) = 0
\,,
\end{equation}
whence the Lie derivative w.r.t.\ the Killing vector  $X$  of the coefficients in \eq{MST_evolution_eqn} vanishes.
That implies that once we have solved the evolution equations, the relation
\begin{equation}
\nabla_{\rho}\mcL_X \mathcal{S}_{0(ij)}{}^{\rho} =\mcL_X\nabla_{\rho} \mathcal{S}_{0(ij)}{}^{\rho} =\mathcal{J}( \mcL_X {\mathcal{S}})_{0(ij)}
\label{evolution_Lie}
\end{equation}
holds automatically, i.e.\ the Lie derivative satisfies an identical system of equations.
The condition \eq{assumption2b} can therefore be realized by an appropriate choice of the initial data, namely
\begin{equation}
 \mcL_X {\mathcal{S}}_{\alpha\beta\mu\nu}|_{\Sigma} =0
\,.
\label{assumption2b_initial}
\end{equation}

%

\begin{remark}
{\rm
In fact, in order to fulfill the Buchdahl condition it  suffices if, in addition to \eq{assumption1},
 certain components of the Lie derivative vanish, namely
\begin{equation}
 \mathcal{F}_0{}^{k} \mcL_X {\mathcal{S}}_{0j0k} =0
\quad \Longleftrightarrow \quad \mcL_X (  \mathcal{F}^{\mu\nu} {\mathcal{S}}_{\alpha\beta\mu\nu}) =0
\label{assumption2}
\end{equation}
holds.
However, one would need to make sure that \eq{assumption2} follows from the evolution equations for appropriately chosen
initial data sets, and it does not seem to be possible to derive a homogeneous system of equations for $\mcL_X(\mathcal{F}^{\mu\nu}\mathcal{S}_{\alpha\beta\mu\nu})$.
}
\end{remark}


Let us consider a Cauchy problem for \eq{MST_evolution_eqn} in a $\Lambda$-vacuum spacetime $( M , g, X)$ with vanishing MST and which satisfies \eq{ineqs}
(at least in some neighborhood of the Cauchy surface).
Then the evolution equations \eq{MST_evolution}
 form a regular symmetric hyperbolic system for which well-posedness results are available.

The Cauchy problem is therefore reduced to the issue to construct initial data sets  which satisfy  both the constraint equations and
our requirement that the Lie derivative of $\mathcal{S}$ vanishes,
\begin{eqnarray}
\mcL_X \mathcal{S}_{0i0j}|_{\Sigma} &=& 0
\,,
\label{Lie_deriv_on Cauchy_surface}
\\
\nabla_{j} \mathcal{S}_{0i0}{}^{j} -\mathcal{J}(  {\mathcal{S}})_{0i0}|_{\Sigma}&=& 0
\,.
\end{eqnarray}
In this paper  we are particularly interested in the construction solutions  from a spacelike $\scri$, so we will not attempt  to solve this system here.
Instead we will consider its analog on a spacelike $\scri$ in more detail, where it is in fact  simpler since e.g.\ \eq{Lie_deriv_on Cauchy_surface} is always an inner equation on $\scri$ (otherwise one would have to eliminate the transverse derivative via the evolution equations).

\section{Bianchi-like equation in a conformally rescaled spacetime}
\label{sec_conf_MST_eqn}

\subsection{Conformally rescaled spacetime}

In view of Penrose's notion of a smooth conformal structure  at infinity, let us conformally rescale the  spacetime $( M , g)$,
\begin{equation}
g\mapsto \widetilde g = \Theta^2 g\,, \quad  M  \overset{\phi}{\mapsto}\widetilde M \,, \quad \Theta|_{\phi}( M )>0
\,.
\label{conf_resc}
\end{equation}
In this ``unphysical'' spacetime Einstein's vacuum field equations are most conveniently replaced by Friedrich's conformal field equations \cite{F3}.
Since we will rarely need these equations, we will not discuss them here and refer to the literature. Nevertheless, two of these equations will be needed later:
Set  $\widetilde s:= \frac{1}{4}\Box_{\widetilde g} \Theta + \frac{1}{24}\widetilde R \Theta$ and denote by $\widetilde L_{\mu\nu}$ the Schouten
tensor of $\widetilde g$. Then the following equations hold in any spacetime $(\widetilde M , \widetilde g, \Theta)$ which arises from a $\Lambda$-vacuum spacetime $( M , g)$,
\begin{eqnarray}
\widetilde\nabla_{\mu}\widetilde \nabla_{\nu}\Theta &=& -\Theta\widetilde L_{\mu\nu} +\widetilde s\widetilde g_{\mu\nu}
\,,
\label{CFE1}
\\
 \widetilde\nabla_{\mu}\Theta\widetilde\nabla^{\mu}\Theta &=&2\Theta \widetilde s-  \frac{\Lambda}{3}
\,.
\label{CFE2}
\end{eqnarray}

A Killing vector field $X$ in $( M , g)$  is mapped to  a conformal Killing vector field $\widetilde  X$
 in $(\widetilde  M , \widetilde g, \Theta)$ (which we identify with $X$)  which, in addition, satisfies \cite{ttpKIDs}
\begin{equation}
\widetilde X^{\mu}\widetilde\nabla_{\mu}\Theta = \frac{1}{4}\Theta\widetilde\nabla_{\mu}\widetilde X^{\mu}
\,.
\label{Killing_conf_eqn}
\end{equation}
Here and henceforth objects associated with the conformally rescaled metric $\widetilde g$ will be decorated with a $\widetilde{\enspace}$.
The indices of those objects will be raised and lowered with $\widetilde g$.

Due to the simple behavior of \eq{MST_evolution_eqn} under conformal rescalings of the metric, one easily shows \cite{mpss} that in $(\widetilde  M , \widetilde g,\Theta, \widetilde X)$ the conformally rescaled MST
\begin{equation}
\widetilde{\mathcal{T}}_{\mu\nu\sigma}{}^{\rho} := \Theta^{-1}\mathcal{S}_{\mu\nu\sigma}{}^{\rho}
\end{equation}
satisfies the equation
\begin{eqnarray}
\widetilde\nabla_{\rho} \widetilde{\mathcal{T}}_{\alpha\beta\mu}{}^{\rho}
\,=\, \mathcal{J}( \widetilde{\mathcal{T}})_{\alpha\beta\mu} &\equiv&
-4 \Lambda  \frac{  5  Q\mathcal{F}^2  +4\Lambda }{Q\mathcal{F}^2 + 8\Lambda}
 \mathcal{U}_{\alpha\beta\mu\nu}
   \mathcal{F}^{-4} X^{\sigma}g^{\gamma\kappa}g^{\delta\varkappa} \mathcal{F}_{\kappa\varkappa}  \widetilde{\mathcal{T}}_{\gamma\delta\sigma}{}^{\rho}
\nonumber
\\
&& +  QX^{\sigma}\Big( \frac{2}{3}
 \mathcal{I}_{\alpha\beta\mu\rho} g^{\gamma\kappa} g^{\delta\varkappa} \mathcal{F}_{\kappa\varkappa}  \widetilde{\mathcal{T}}_{\gamma\delta\sigma}{}^{\rho}
- \mathcal{F}_{\mu\rho} \widetilde{\mathcal{T}}_{\alpha\beta\sigma}{}^{\rho}
\Big)
\label{sym_hyp0}
\;,
\phantom{xxx}
\end{eqnarray}
where the fields on the right-hand side need to be expressed in terms of the ``unphysical'' fields,
\begin{eqnarray}
\mathcal{F}_{\mu\nu} &=& \Theta^{-3}\widetilde{\mathcal{H}}_{\mu\nu}  + \Theta^{-2}\widetilde{ \mathcal{F}}_{\mu\nu}
\;,
\label{asympt_exp_F}
\\
\mathcal{F}^2 &=& \Theta^{-2} \widetilde{\mathcal{H}}^2 + 2\Theta^{-1} {\widetilde{ \mathcal{F}}}_{\alpha\beta}\widetilde{\mathcal{H}}^{\alpha\beta}  + \widetilde{\mathcal{F}}^2
\label{asympt_exp_F2}
\;,
\\
\mathcal{I}_{\alpha\beta\mu\nu} &=&\Theta^{-4} \widetilde {\mathcal{I}}_{\alpha\beta\mu\nu}
\;,
\\
\mathcal{U}_{\alpha\beta\mu\nu} &=&   -\Theta^{-6}\Big( \widetilde{\mathcal{H}}_{\alpha\beta}\widetilde{\mathcal{H}}_{\mu\nu}
-\frac{1}{3}\widetilde{\mathcal{H}}^2 \widetilde{\mathcal{I}}_{\alpha\beta\mu\nu} \Big)
\nonumber
\\
&&
-\Theta^{-5}\Big(
\widetilde{ \mathcal{F}}_{\alpha\beta}\widetilde{\mathcal{H}}_{\mu\nu}
+
\widetilde{\mathcal{H}}_{\alpha\beta} \widetilde{ \mathcal{F}}_{\mu\nu}
-\frac{2}{3} {\widetilde{ \mathcal{F}}}_{\kappa\varkappa}\widetilde{\mathcal{H}}^{\kappa\varkappa}\widetilde{\mathcal{I}}_{\alpha\beta\mu\nu}  \Big)
\nonumber
\\
&&
- \Theta^{-4}\Big( \widetilde{\mathcal{F}}_{\alpha\beta}\widetilde{\mathcal{F}}_{\mu\nu}
-\frac{1}{3}\widetilde{\mathcal{F}}^2 \widetilde{\mathcal{I}}_{\alpha\beta\mu\nu} \Big)
\;.
\phantom{xx}
\label{expansion_U}
\end{eqnarray}

As in Section~\ref{sec_constr_ev}
we introduce \emph{Gaussian normal coordinates} (near some Cauchy surface say) where
\begin{equation}
\widetilde g_{00} =-1\,, \quad \widetilde g_{0i}=0
\,,
\label{gauge1}
\end{equation}
such that the system splits into a symmetric hyperbolic system of evolution equations $\widetilde\nabla_{\rho} \widetilde{\mathcal{T}}_{0(ij)}{}^{\rho}
= \mathcal{J}( \widetilde{\mathcal{T}})_{0(ij)} $
and a system of constraint equations $\widetilde\nabla_{\rho} \widetilde{\mathcal{T}}_{0i0}{}^{\rho}=\mathcal{J}( \widetilde{\mathcal{T}})_{0i0}$.

\subsection{Behavior of the MST equation near a spacelike \texorpdfstring{$\scri$}{J}}
\label{sec_asymp_behavior}

\subsubsection{Expansions near \texorpdfstring{$\scri$}{J}}
\label{sec_exp}

Let us consider spacetimes which admit a smooth conformal completion at infinity \`a la Penrose \cite{p1,p2}.
By this it is meant that $( M , g)$ admits a conformal rescaling \eq{conf_resc} such that $(\widetilde  M , \widetilde g, \Theta)$ admits a representation
of null infinity
\begin{equation}
\scri =\{ \Theta=0,\enspace \mathrm{d}\Theta\ne 0\} \cap \partial\phi( M )
\end{equation}
through which $\widetilde g $ and $\Theta$ can be smoothly extended.
$\scri$  is a smooth hypersurface consisting of two subsets $\scri^-$ and $\scri^+$, distinguished by the absence  of endpoints of future  and past
causal curves in $( M , g)$, respectively.

The causal character of $\scri$ is determined by the sign of the cosmological constant (cf.\ \eq{CFE2}).
A positive cosmological constant
\begin{equation}
\Lambda >0
\label{pos_lambda}
\end{equation}
yields a spacelike $\scri$.
From now on  we will assume that \eq{pos_lambda} holds. We further take  $\scri^-$ to be a connected component of past null infinity
and restrict, if necessary, $(\widetilde  M , \widetilde g, \Theta)$ to the domain of dependence of $\scri^-$.
Our aim is to construct solutions to \eq{sym_hyp0} by prescribing appropriate data on the Cauchy surface $\scri^-$.

In \cite{mpss}  the leading order behavior of the coefficients  in \eq{sym_hyp0} at $\scri^-$ has been computed, and it has been shown that it
is actually a Fuchsian system. It is one of the main purposes of this work to analyze this system.
It turns out that for this one  needs to know the next-to-leading order terms, which
have not been determined in \cite{mpss}.

In Gaussian coordinates we have (as a consequence of \eq{CFE2})
\begin{equation}
\partial_0 \Theta |_{\scri^-} \,=\, \sqrt{\frac{\Lambda}{3}}
\;.
\label{nablaTheta}
\end{equation}
The gauge freedom which arises from the artificially introduced conformal factor $\Theta$ can be employed \cite{F_lambda, ttp2}
to achieve that
\begin{equation}
\widetilde s |_{\scri^-}\,=\,0
\;.
\label{gauge2}
\end{equation}
Then we have by \eq{CFE1}
\begin{equation}
  \label{eq:gauge44}
 \partial_0\widetilde g_{ij}|_{\scri^-}=0\;, \quad \partial_0\partial_0\Theta|_{\scri^-}=0
\;.
\end{equation}
In particular,
\begin{equation}
 \Theta\,=\, \sqrt{\frac{\Lambda}{3}}\, t + O(t^3)
\label{expansion_Theta}
\;.
\end{equation}
In addition, there remains the gauge freedom to conformally rescale the initial 3-manifold which we shall employ later.

We denote the induced 3-metric on $\scri^-$ by $\mathfrak{h}$, the volume form of $\mathfrak{h}$ by $\epsilon_{ijk}$, its covariant derivative by $\mcD$.
It follows from \eq{Killing_conf_eqn} that the conformal Killing vector $\widetilde X$ has no transverse component on $\scri$.
It induces a conformal Killing vector on $(\scri^-, \mathfrak{h})$ which we denote by $Y$.
Moreover, let $Y$ and $N$ denote divergence and curl of $Y$,
\begin{equation}
\label{eq:deffN}
f:=\mcD_i Y^i
\,, \quad  N_i := \epsilon_{ijk}\mcD^j Y^k
\,.
\end{equation}

In the gauge \eq{gauge1} and \eq{gauge2} we find  the following expansions,
\begin{eqnarray}
\widetilde X^i &=& Y^i + O(\Theta^2)
\;,
\label{expansion_X}
\\
\widetilde X^0 &=& \frac{1}{3}\sqrt{\frac{3}{\Lambda}}f\Theta +  O(\Theta^2)
\;,
\label{expansion_X0}
\\
\widetilde{\mathcal{H}}_{0i} &=& - \sqrt{\frac{\Lambda}{3}} Y_i + O(\Theta^2)
\;,
\\
\widetilde{\mathcal{H}}_{ij} &=& \imagi\sqrt{\frac{\Lambda}{3}} \epsilon_{ijk}Y^k  + O(\Theta^2)
\;,
\\
\widetilde{\mathcal{H}}^2 &=& - 4 \frac{\Lambda}{3} |Y|^2   + O(\Theta^2)
\;,
\\
 {\widetilde{ \mathcal{F}}}_{\alpha\beta}\widetilde{\mathcal{H}}^{\alpha\beta}  &=&
2\imagi\sqrt{\frac{\Lambda}{3}} Y_kN^k + O(\Theta)
\;,
\\
\widetilde{\mathcal{F}}_{0i} &=& \frac{\imagi}{2} N_i + O(\Theta)
\;,
\\
\widetilde{\mathcal{F}}_{ij} &=&\frac{1}{2} \epsilon_{ijk} N^k + O(\Theta)
\;.
\label{expansion_F}
\end{eqnarray}
and  \cite{mpss}
\begin{equation}
Q= O(\Theta^4)
\,.
\label{asympt_Q}
\end{equation}
Moreover, we have the  useful relations,
\begin{eqnarray}
\widetilde{\mathcal{H}}^{\gamma\delta}
\widetilde{\mathcal{T}}_{\gamma\delta\sigma}{}^{\rho} &=& 4 \widetilde X^{\gamma}\widetilde\nabla^{\delta}\Theta
\widetilde{\mathcal{T}}_{\gamma\delta\sigma}{}^{\rho}
 \,=\,  4  \sqrt{\frac{\Lambda}{3}} Y^{k}
\widetilde{\mathcal{T}}_{0  k \sigma}{}^{\rho}   + (O(\Theta^2) \widetilde{\mathcal{T}})_{\sigma}{}^{\rho}
\;,
\phantom{xxx}
\label{HT_relation}
\\
\widetilde{\mathcal{F}}^{\gamma\delta}
\widetilde{\mathcal{T}}_{\gamma\delta\sigma}{}^{\rho}
&=&2 \widetilde\nabla^{\gamma}\widetilde X ^{\delta}
\widetilde{\mathcal{T}}_{\gamma\delta\sigma}{}^{\rho}
\,=\, - 2i N^k \widetilde{\mathcal{T}}_{0k\sigma}{}^{\rho}
 + (O(\Theta) \widetilde{\mathcal{T}})_{\sigma}{}^{\rho}
\;.
\end{eqnarray}

\subsubsection{Evolution equations and remaining gauge freedom}

We analyze the behavior of the evolutionary part of the system \eq{sym_hyp0} at $\scri^-$.  In Gaussian coordinates  it  is given by the $\alpha\beta\mu=0(ij)$-components.
We denote by $\widetilde\rnabla$ the Levi-Civita connection associated to the  family $t\mapsto \widetilde g_{ij}(t,x^k)$ of \emph{conformally rescaled}
Riemannian metrics,
 its volume form is denoted by $\widetilde \reta_{ijk}$.

For the left-hand side of \eq{sym_hyp0} we find
\begin{equation}
\widetilde\nabla_{\rho} \widetilde{\mathcal{T}}_{0(ij)}{}^{\rho}
\,=\,
\partial_0\widetilde{\mathcal{T}}_{0i0j}
- i\widetilde\reta_{(i}{}^{kl} \widetilde\rnabla_{|k} \widetilde{\mathcal{T}}_{0|j)0l}
 + (O(\Theta) \widetilde{\mathcal{T}})_{0(ij)}
\;,
\end{equation}
while the right-hand side satisfies
\begin{eqnarray*}
\mathcal{J}( \widetilde{\mathcal{T}})_{0(ij)}
&=&
\sqrt{\frac{\Lambda}{3}} |Y|^{-2}\Theta^{-1}
\Big(
 \frac{3}{2}  Y_{(i} Y^k\widetilde{\mathcal{T}}_{|0|j)0k}
-  3|Y|^{-2} Y_{i}Y_jY^k  Y^l \widetilde{\mathcal{T}}_{0k0l}
+\frac{1}{2}\mathfrak{h}_{ij} Y^{k}Y^l  \widetilde{\mathcal{T}}_{0 k0l}
\Big)
\\
&& +2\imagi|Y|^{-4}Y_pN^p\Big(
 \frac{9}{8}  Y_{(i} Y^k\widetilde{\mathcal{T}}_{|0|j)0k}
-  3|Y|^{-2} Y_{i}Y_jY^k  Y^l \widetilde{\mathcal{T}}_{0k0l}
+\frac{1}{4}\mathfrak{h}_{ij} Y^{k}Y^l  \widetilde{\mathcal{T}}_{0 k0l}
\Big)
\\
&&
- \frac{3}{4}\imagi |Y|^{-2}\Big(
  (Y_{(i} N^k
+N_{(i} Y^k ) \widetilde{\mathcal{T}}_{|0|j)0k}
- 3|Y|^{-2}( Y_{i}Y_j  N^k
+ N_{(i}Y_{j)}  Y^{k} ) Y^l\widetilde{\mathcal{T}}_{0 k0l}
\\
&&
+\frac{1}{3}\mathfrak{h}_{ij} Y^{k}N^l  \widetilde{\mathcal{T}}_{0 k0l}
-  \frac{2}{3}f |Y|^{-2} Y_{(i}\widetilde\reta_{j)p}{}^{k}Y^p Y^l \widetilde{\mathcal{T}}_{0k0l}
\Big)
 + (O(\Theta) \widetilde{\mathcal{T}})_{0(ij)}
\;.
\end{eqnarray*}

Combined we end up with a symmetric hyperbolic system of the form (set $\mathcal{E}_{ij} :=\widetilde{\mathcal{T}}_{0 i0j}$ and observe that $\mathcal{E}_{ij}$ contains all independent components of the MST),
\begin{eqnarray}
&&\hspace{-3em}
\partial_0\mathcal{E}_{ij}
- \imagi\widetilde\reta_{(i}{}^{kl} \mcD_{|k|}\mathcal{E}_{j)l}
\nonumber
\\
&=&\sqrt{\frac{\Lambda}{3}} |Y|^{-2}\Theta^{-1}
\Big(
 \frac{3}{2}  Y_{(i} Y^k\mathcal{E}_{j)k}
-  3|Y|^{-2} Y_{i}Y_jY^k  Y^l \mathcal{E}_{kl}
+\frac{1}{2}\mathfrak{h}_{ij} Y^{k}Y^l  \mathcal{E}_{kl}
\Big)
\nonumber
\\
&& +2\imagi |Y|^{-4}Y_pN^p\Big(
 \frac{9}{8}  Y_{(i} Y^k\mathcal{E}_{j)k}
-  3|Y|^{-2} Y_{i}Y_jY^k  Y^l \mathcal{E}_{kl}
+\frac{1}{4}\mathfrak{h}_{ij} Y^{k}Y^l  \mathcal{E}_{ kl}
\Big)
\nonumber
\\
&&
- \frac{3}{4}\imagi |Y|^{-2}\Big(
  \big(Y_{(i} N^k
+N_{(i} Y^k \big)\mathcal{E}_{j)k}
- 3|Y|^{-2}\big( Y_{i}Y_j  N^k
+ N_{(i}Y_{j)}  Y^{k} \big) Y^l \mathcal{E}_{ kl}
\nonumber
\\
&&
+\frac{1}{3}\mathfrak{h}_{ij} Y^{k}N^l  \mathcal{E}_{ kl}
-  \frac{2}{3}f |Y|^{-2} Y_{(i}\epsilon_{j)p}{}^{k}Y^p Y^l \mathcal{E}_{kl}
\Big)
 + (O(\Theta) \mathcal{E})_{ij}
\;.
\label{shs}
\end{eqnarray}

To simplify the analysis of \eq{shs} we exploit the above mentioned remaining gauge freedom, which is
to conformally rescale the initial 3-manifold $(\scri^-, \mathfrak{h})$.

We would like to achieve that the conformal Killing vector  $Y$ and its curl $N$ are parallel, i.e.\ that the cross product vanishes,
\begin{equation}
0 \,\overset{!}{=}\,(Y \times N)_i \,=\,\epsilon_{ijk}Y^jN^k\,=\, 2Y^j\partial_{[i}Y_{j]}
\;.
\label{vanishing_cross}
\end{equation}
For this purpose let us consider  a conformal rescaling of the induced Riemannian metric $ \mathfrak{h}$,
\begin{equation}
\mathfrak{h} \,\mapsto \,\widehat {\mathfrak{h}} \,:=\, \omega^2 \mathfrak{h}
\;.
\end{equation}
Then the following relations hold on  the  rescaled Riemannian manifold $(\scri, \widehat { \mathfrak{h}})$ (we denote the associated Levi-Civita covariant derivative by $\widehat \mcD$),
\begin{eqnarray}
|\widehat Y|^2 &=& \omega^2 |Y|^2 \;,
\label{conf_trans1}
\\
\widehat f &=& f + \frac{3}{2}\omega^{-2} Y^j\partial_j\omega^2
\;,
\label{conf_trans2}
\\
 2\widehat Y^j\widehat \mcD_{[i}\widehat Y_{j]}  &=& \partial_{i}|\widehat Y|^2  -  \frac{2}{3} \widehat f \widehat Y_i
\nonumber
\\
&=&   \partial_{i}(\omega^2 | Y|^2)  -  \frac{2}{3} \omega^2 \Big( f + \frac{3}{2}\omega^{-2} Y^j\partial_j\omega^2\Big)  Y_i
\nonumber
\\
&=& \partial_{i}(\omega^2 | Y|^2)  -  \frac{2}{3} \omega^2 f Y_i    -  Y_i   Y^j\partial_j\omega^2
\;.
\label{conf_trans3}
\end{eqnarray}
Let us now make the gauge choice
\begin{equation}
\omega^2 \,=\, |Y|^{-2}
\;.
\end{equation}
Then, taking further into account that $Y$ is a conformal Killing vector,  \eq{conf_trans1}-\eq{conf_trans3}  simply become
\begin{eqnarray}
|\widehat Y|^2 &=& 1\;,
\label{conf_trans1B}
\\
\widehat f &=& f - \frac{3}{2}|Y|^{-2} Y^j\partial_j|Y|^2
\,=\,   f - 3|Y|^{-2} Y^jY^k\mcD _{(j}Y_{k)}\,=\, 0
\;,
\phantom{xxx}
\label{conf_trans2B}
\\
 2\widehat Y^j\widehat \mcD_{[i}\widehat Y_{j]}
&=& -\frac{2}{3} |Y|^{-2} Y_i   \Big(   f   - \frac{3}{2}  |Y|^{-2} Y^j\partial_j|Y|^{2}  \Big)  \,=\,0
\;,
\label{conf_trans3B}
\end{eqnarray}
so in particular \eq{vanishing_cross} holds.

 We therefore can and will impose the gauge condition
\begin{equation}
 |Y|^2\,=\, 1\;, \quad f\,=\, 0\;, \quad N = \lambda Y
\label{add_gauge}
\end{equation}
for some real function $\lambda = \lambda(x^i)$. In particular, $Y$ is a proper Killing vector in this gauge.
It  simplifies  the evaluation of \eq{shs} significantly which now reads (using also \eq{expansion_Theta}),
\begin{equation}
\label{eq:leadingevoleq}
\partial_t\mathcal{E}_{ij}
- \imagi\widetilde\reta_{(i}{}^{kl} \mcD_{|k|}\mathcal{E}_{j)l}
=\Big(\frac{1}{t} + \frac{\imagi}{2}\lambda\Big)
\Big(
 \frac{3}{2}  Y_{(i}\delta_{j)}{}^k
-  3Y_{i}Y_jY^k  +\frac{1}{2}\mathfrak{h}_{ij} Y^{k}
\Big)Y^l  \mathcal{E}_{kl}
 + (O(t) \mathcal{E})_{ij}
\;.
\end{equation}

\subsubsection{Constraint equations}

Next, we compute the asymptotic behavior of the constraint equations, i.e.\ of the  $\alpha\beta\mu=0i0$-components
 of \eq{sym_hyp0} at $\scri^-$.  In a gauge where  \eq{gauge1} and \eq{gauge2} hold we obtain
 for the right-hand side
\begin{eqnarray}
\mathcal{J}( \widetilde{\mathcal{T}})_{0i0}
&=&
-\frac{1}{2} \imagi \sqrt{\frac{\Lambda}{3}}  \Theta^{-1}
 |Y|^{-2}  \epsilon_{ij}{}^{l} Y^j  Y^{k} \widetilde{\mathcal{T}}_{0  k 0l}
+ \frac{1}{2} f  |Y|^{-4}  (Y_iY_j)_{\mathrm{tf}}  Y^{k} \widetilde{\mathcal{T}}_{0  k 0}{}^{j}
\nonumber
\\
&&
-\frac{1}{4} |Y|^{-2}\epsilon_{ij}{}^l
Y^j  N^k \widetilde{\mathcal{T}}_{0k0l}  \
+\frac{3}{4}  |Y|^{-4} \epsilon_{jk}{}^{l}
Y_iN^jY^k Y^{m}
\widetilde{\mathcal{T}}_{0  l 0m}
\nonumber
\\
&&
+\frac{1}{2}  |Y|^{-4}  \epsilon_{ij}{}^{k}
 Y_mN^m Y^{j}Y^l
\widetilde{\mathcal{T}}_{0  k 0l}
 + (O(\Theta) \widetilde{\mathcal{T}})_{0i0}
\;,
\end{eqnarray}
where we have used the various expansions derived in Section~\ref{sec_exp}.
For the left-hand side of \eq{sym_hyp0} we find
\begin{equation}
\widetilde\nabla_{\rho} \widetilde{\mathcal{T}}_{0i0}{}^{\rho}
\,=\,
\mcD_{j} \widetilde{\mathcal{T}}_{0i0}{}^{j}
 + (O(\Theta) \widetilde{\mathcal{T}})_{0i0}
\;,
\end{equation}
and combined, again with $\mathcal{E}_{ij} \equiv\widetilde{\mathcal{T}}_{0 i0j}$,
\begin{eqnarray}
\mcD_{j} \mathcal{E}_i{}^j
&=&
-\frac{1}{2} \imagi \sqrt{\frac{\Lambda}{3}}  \Theta^{-1}
 |Y|^{-2}  \epsilon_{ij}{}^{l} Y^j  Y^{k} \mathcal{E}_{  k l}
+ \frac{1}{2} f  |Y|^{-4}  (Y_iY_j)_{\mathrm{tf}}  Y^{k}\mathcal{E}_{  k }{}^{j}
\nonumber
\\
&&
-\frac{1}{4} |Y|^{-2}\epsilon_{ij}{}^l
Y^j  N^k \mathcal{E}_{kl}
+\frac{3}{4}  |Y|^{-4}\epsilon_{jk}{}^{l}
Y_iN^jY^k Y^{m} \mathcal{E}_{ l m}
\nonumber
\\
&&
+\frac{1}{2}  |Y|^{-4}  \epsilon_{ij}{}^{k}
 Y_mN^m Y^{j}Y^l
\mathcal{E}_{ k l}
 + (O(\Theta)\mathcal{E})_{i}
\;.
\end{eqnarray}
Imposing, in addition, the gauge condition \eq{add_gauge}, this becomes
\begin{equation}
\mcD_{j} \mathcal{E}_i{}^j
+ \frac{\imagi}{2}
\Big( \frac{1}{t}
+\frac{\imagi}{2}\lambda   \Big) \epsilon_{ij}{}^{l}  Y^{j}Y^k
\mathcal{E}_{ k l}
 + (O(t)\mathcal{E})_{i} \,=\, 0
\;.
\label{asympt_constraint}
\end{equation}

We sum up the result of Section~\ref{sec_asymp_behavior} by the following
\begin{lemma}
Let $(\widetilde M , \widetilde g, \Theta, \widetilde X)$ be a solution to the conformal field equations with cosmological constant $\Lambda>0$
which admits a conformal Killing vector field $\widetilde X$ which satisfies \eq{Killing_conf_eqn}, and which admits a smooth $\scri$ where  $|\widetilde X|^2|_{\scri}> 0$.
Assume further that the gauge conditions \eq{gauge1}, \eq{gauge2} and \eq{add_gauge} hold. Then the evolution and constraint part of \eq{sym_hyp0}
show a Fuchsian  behavior near  $\scri$ as given by  \eq{eq:leadingevoleq} and \eq{asympt_constraint}, respectively.
\end{lemma}

\begin{remark}
{\rm
The inequalities \eq{ineqs} hold  near $\scri$ as long as $\widetilde X$ has no zeros on $\scri$. This follows from the expansions
computed in \cite{mpss}.
}
\end{remark}

\subsection{Admissible data sets on a spacelike \texorpdfstring{$\scri$}{J}}

\subsubsection{Preservation of the constraints from \texorpdfstring{$\scri$}{J}}
\label{sec:preservconstrscri}

For the time being let us consider again the physical spacetime $( M , g, X)$.
Let us assume that $\mathcal{S}_{\mu\nu\sigma\rho}$
fulfills the evolution equations  \eq{MST_evolution}
and that \eq{assumption1}-\eq{assumption2b} hold.
Then \eq{constr_prop} becomes
\begin{eqnarray*}
&&
\hspace{-2em}
\Big[\Big( \delta_i{}^p(\partial_0+\frac{3}{2}\Gamma^j_{j0}-2\Gamma^0_{00})-\frac{\imagi}{2} \volform_{0i}{}^{jp}(\widetilde \rnabla_{j}+2\Gamma^0_{0j})-\frac{3}{2}\Gamma^p_{0i}\Big)
\\
&&
+Q\Big( \frac{2}{3}X^{j} \mathcal{F}_{0j}\delta_i{}^p
-\frac{5}{6}X_0 \mathcal{F}_{i}{}^p
-\frac{1}{6}X^{p} \mathcal{F}_{0i}
- \frac{1}{6}X_{i} \mathcal{F}_{0}{}^{p}
\Big)
\\
&&
-
\frac{2}{3}\Lambda  \mathcal{F}^{-4} \frac{  5  Q\mathcal{F}^2  +4\Lambda }{Q\mathcal{F}^2 + 8\Lambda} \Big(36g^{00}X^{j} \mathcal{F}_{0j}   \mathcal{F}_0{}^{p} \mathcal{F}_{0i}
-2\mathcal{F}^2 X^p \mathcal{F}_{0i}
-2  \mathcal{F}^2  X_i\mathcal{F}_0{}^{p}
\\
&&
- \mathcal{F}^2X_{0} \mathcal{F}_{i}{}^p
- \mathcal{F}^2\delta_{i}{}^p  X^{j} \mathcal{F}_{0j}
\Big)\Big]
\Xi_p
=
0
\,.
\end{eqnarray*}
To transform this equation into the unphysical spacetime,
note that the algebraic properties of $\mathcal{S}_{\mu\nu\sigma\rho}$ imply that
\begin{equation}
\nabla_{\lambda}{\mathcal{S}}_{0p0}{}^{\lambda}
- \mathcal{J}( {\mathcal{S}})_{0p0} = \Theta \Big( \widetilde\nabla_{\lambda}{\widetilde{\mathcal{T}}}_{0p0}{}^{\lambda}
- \mathcal{J}( \widetilde{{\mathcal{T}}})_{0p0}\Big)
=:\Theta\widetilde\Xi_p
\,.
\end{equation}
We further employ the  relations \eq{asympt_exp_F}-\eq{expansion_U}  and \eq{expansion_X}- \eq{asympt_Q}
and take  the behavior of the Levi-Civita connection under conformal transformations into account to  obtain, in Gaussian normal coordinates \eq{gauge1},
%
%
a homogeneous equations for $\widetilde\Xi_i$,
%
\begin{equation}
\Big[\Big( \delta_i{}^k\partial_0-\frac{\imagi}{2} \widetilde\reta_{i}{}^{jk}\widetilde\rnabla_j \Big)
+\frac{1}{4}\ (\delta_{i}{}^k + 13|Y|^{-2}Y_iY^k) t^{-1}+O(1) \Big]
 \widetilde\Xi_k
=
0
\,.
\label{eqn_const_vio}
\end{equation}
%

\subsubsection{Realization of the Buchdahl condition at \texorpdfstring{$\scri$}{J}}
\label{sec:realizationBuchdahl}

We need to make sure that the generalized Buchdahl condition \eq{Buchdahl_gen} holds, so let us translate it into
the unphysical spacetime $(\widetilde  M , \widetilde g, \Theta, \widetilde X)$.

Under conformal rescaling we have (set $\widetilde F := \frac{1}{4}\widetilde \nabla_{\kappa} \widetilde X^{\kappa}$)
\begin{equation}
\mcL_X \mathcal{S}_{\alpha\beta\mu}{}^{\nu}
=
\Theta \mcL_{\widetilde X} \widetilde{\mathcal{T}}_{\alpha\beta\mu}{}^{\nu}+
\widetilde X^{\kappa}\widetilde \nabla_{\kappa} \Theta\widetilde{\mathcal{T}}_{\alpha\beta\mu}{}^{\nu}
=
\Theta( \mcL_{\widetilde X}+ \widetilde F)\widetilde{\mathcal{T}}_{\alpha\beta\mu}{}^{\nu}
\,,
\end{equation}
as follows from  \eq{Killing_conf_eqn}.
In the conformally rescaled spacetime  the Buchdahl condition \eq{Buchdahl_gen} therefore adopts the form
\begin{equation}
\widetilde f^{(1)}{}_i{}^{\mu\nu\sigma\rho}(\mcL_{\widetilde X} +
\widetilde F)\widetilde{\mathcal{T}}_{\mu\nu\sigma\rho}
  + \widetilde  f^{(2)}_{i}{}^{\mu\nu\sigma\rho}\widetilde{\mathfrak{T}}_{\mu\nu\sigma\rho} =0
\,,
\label{gen_Buchdahl_conf}
\end{equation}
where $\widetilde{\mathfrak{T}}_{\mu\nu\sigma\rho}$ denotes the rescaled MST of the background spacetime $(\widetilde  M , \widetilde g, \Theta, \widetilde X)$.

We assume that the background spacetime $(\widetilde  M , \widetilde g, \Theta, \widetilde X)$ has a vanishing (rescaled) MST, and
in order to make sure that \eq{gen_Buchdahl_conf} is fulfilled
we will further make sure that
\begin{equation}
\label{eq:buchdahlspecial}
 (\mcL_{\widetilde X}+
\widetilde F)\widetilde{\mathcal{T}}_{\mu\nu\sigma\rho}=0
\,.
\end{equation}

The analog of \eq{evolution_Lie} in the conformally rescaled spacetime reads
\begin{eqnarray}
&&\hspace{-3em}\widetilde \nabla_{\rho}((\mcL_{\widetilde X}  +
\widetilde F)\widetilde{\mathcal{T}}_{0(ij)}{}^{\rho})
\nonumber
\\
&=&
\widetilde \nabla_{\rho}(\Theta^{-1}\mcL_X \mathcal{S}_{0(ij)}{}^{\rho})
\nonumber
\\
&=&
\Theta^{-1} \widetilde \nabla_{\rho}\mcL_X \mathcal{S}_{0(ij)}{}^{\rho}
-\Theta^{-2}\widetilde \nabla_{\rho}\Theta \mcL_X \mathcal{S}_{0(ij)}{}^{\rho}
\nonumber
\\
&=&
\mcL_{\widetilde X} \mathcal{J}(\widetilde{ \mathcal{T}})_{0(ij)}
+ \Theta^{-1}\mcL_{\widetilde X} \Theta \mathcal{J}(\widetilde{ \mathcal{T}})_{0(ij)}
\nonumber
\\
&=&
\mcL_{ X}(\Theta^{-1} \mathcal{J}(\mathcal{S})_{0(ij)})
+ \Theta^{-1}\mcL_{\widetilde X} \Theta \mathcal{J}(\widetilde{ \mathcal{T}})_{0(ij)}
\nonumber
\\
&=&
\Theta^{-1} \mathcal{J}(\mcL_{ X}\mathcal{S})_{0(ij)}
\nonumber
\\
&=&
 \mathcal{J}( (\mcL_{\widetilde X}+
\widetilde F)\widetilde{\mathcal{T}})_{0(ij)}
\,,
\label{Buchdahl_vio}
\end{eqnarray}
so $ (\mcL_{\widetilde X}+
\widetilde F)\widetilde{\mathcal{T}}_{\mu\nu\sigma\rho}$ satisfies the same equation as $\widetilde{\mathcal{T}}_{\mu\nu\sigma\rho}$.

Let us determine the expansion of $( \mcL_{\widetilde X}+ \widetilde F)\widetilde{\mathcal{T}}_{\alpha\beta\mu\nu}$ near $\scri$.
Assuming that the evolution equations hold, we find
\begin{eqnarray*}
( \mcL_{\widetilde X}+ \widetilde F)\widetilde{\mathcal{T}}_{0i0j} &=&
\widetilde X^{0} \widetilde\nabla_{0} \widetilde{\mathcal{T}}_{0i0j}
+ \widetilde X^{k} \widetilde\nabla_{k} \widetilde{\mathcal{T}}_{0i0j}
+ \widetilde F \widetilde{\mathcal{T}}_{0i0j}
\\
&&
+ 2\widetilde{\mathcal{T}}_{0k 0(i} \widetilde\nabla_{j)} \widetilde X^{k}
+2 \widetilde{\mathcal{T}}_{0i0 j} \widetilde\nabla_{0} \widetilde X^{0}
- 2\widetilde{\mathcal{T}}_{0(ij)k } \widetilde\nabla_{0} \widetilde X^{k}
\\
&=&
\widetilde X^{0} \mathcal{J}( \widetilde{\mathcal{T}})_{0(ij)}
- \widetilde X^{0} \widetilde\nabla_{k} \widetilde{\mathcal{T}}_{0(ij)}{}^{k}
+ \widetilde X^{k} \widetilde\nabla_{k} \widetilde{\mathcal{T}}_{0i0j}
+ \widetilde F \widetilde{\mathcal{T}}_{0i0j}
\\
&&
+ 2\widetilde{\mathcal{T}}_{0k 0(i} \widetilde\nabla_{j)} \widetilde X^{k}
+2 \widetilde{\mathcal{T}}_{0i0 j} \widetilde\nabla_{0} \widetilde X^{0}
- 2\widetilde{\mathcal{T}}_{0(ij)k } \widetilde\nabla_{0} \widetilde X^{k}
\,.
\end{eqnarray*}
In Gaussian normal coordinates we have the expansions \eq{expansion_X}, \eq{expansion_X0} and
\begin{equation}
\widetilde F =\frac{1}{3} f +   O(\Theta)
\,,
\label{eq:FFff}
\end{equation}
whence
\begin{eqnarray*}
( \mcL_{\widetilde X}+ \widetilde F)\mathcal{E}_{ij}
&=&
f|Y|^{-2} \Big(
 \frac{1}{2}  Y_{(i} Y^k\mathcal{E}_{j)k}
-  |Y|^{-2} Y_{i}Y_jY^k  Y^l \mathcal{E}_{kl}
+\frac{1}{6}\mathfrak{h}_{ij} Y^{k}Y^l  \mathcal{E}_{kl}
\Big)
\\
&&
+(\mcL_Y+f)\mathcal{E}_{ij}
+(O(\Theta )\mathcal{E})_{ij}
+(O(\Theta)\mcD \mathcal{E})_{ij}
\,.
\end{eqnarray*}
In a gauge where \eq{add_gauge} holds  this becomes
\begin{equation}
 \mcL_{\widetilde X}\mathcal{E}_{ij}
=
\mcL_Y\mathcal{E}_{ij}
+(O(\Theta )\mathcal{E})_{ij}
+(O(\Theta)\mcD \mathcal{E})_{ij}
\,.
\label{Lie_deriv_conf}
\end{equation}
In Section~\ref{sec_constraints} it will be analyzed how initial data need to be chosen in order to satisfy the constraint equations and the Buchdahl condition,
and for this the equations \eq{eqn_const_vio} and \eq{Buchdahl_vio}
will be relevant.


\section{Fuchsian analysis near \texorpdfstring{${\scri^-}$}{Scri-}}
\label{sec:fuchsian}

\subsection{Preliminaries and the main result}
\label{sec:prelim}
\newcommand{\hnabla}{\mcD}
\newcommand{\gmetr}{\widetilde{g}}
\newcommand{\gvolf}{\widetilde{\eta}}
\newcommand{\gnabla}{\widetilde{\nabla}}
\newcommand{\gstnabla}{\gnabla}
\newcommand{\gChrist}{\widetilde{{\Gamma}}}
\newcommand{\XKill}{\widetilde {X}}
\newcommand{\Tfield}{T}
\newcommand{\Tcheckfield}{\Tfield}
\newcommand{\Cfield}{\underline C}
\newcommand{\Ffield}{\underline F}
\newcommand{\SigmaF}{\Sigma}
\newcommand{\Dfield}{\widetilde \Xi}
\newcommand{\Mtilde}{{\widetilde M}}
\newcommand{\unitnormal}{{\widetilde N}}
\newcommand{\frameh}[2]{{e}_{\mathbf{[#1]}}^{#2}}
\newcommand{\frameD}[2]{{e}_{\mathbf{[#1]}#2}}

For the purpose of  this whole section, we shall introduce certain small variations of previous conventions and notations.
Pick $\delta>0$ and an orientable $3$-dimensional differentiable manifold $\SigmaF$. We refer to $\Mtilde=(-\delta,\delta)\times \SigmaF$ as the {conformal (or unphysical) spacetime}
and to $M=(0,\delta)\times \SigmaF$ as the {physical spacetime}; we always identify the physical spacetime with this subset of the conformal spacetime explicitly. As before, both manifolds are equipped with conformally related Lorentzian metrics;
 in this section we only deal with the conformal (unphysical) metric $\gmetr_{\mu\nu}$ on $M$. 
Let $t$ be the parameter on the $(-\delta,\delta)$-factor of $\Mtilde$. We refer to the $t=const$-hypersurface for any $t\in (-\delta,\delta)$ as $\SigmaF_t$, i.e.,
\[\SigmaF_t=\{t\}\times\SigmaF\]
which is clearly a subset of $\Mtilde$ diffeomorphic to $\SigmaF$. We assume that all these hypersurfaces are spacelike Cauchy surfaces of $\Mtilde$. Notice that $\SigmaF_0$ agrees with ${\scri^-}$. We assume the existence of a conformal Killing vector field $\XKill^\mu$ of $\gmetr_{\mu\nu}$ on $\Mtilde$.

Let $\unitnormal^\mu$ be the future-pointing unit normal of the surfaces $\SigmaF_t$ in $\Mtilde$ with respect to $\gmetr_{\mu\nu}$. It will be convenient for the following discussion to adopt a slightly different index convention than before: Tensor indices $\mu,\nu,\ldots$ are still considered as abstract spacetime indices, while  $i,j,\ldots$ shall now denote abstract indices which have been projected orthogonally into the hypersurfaces $\SigmaF_t$ with respect to $\unitnormal^\mu$. Correspondingly,  the index $0$ denotes projections onto $\unitnormal^\mu$. Interpreting all indices in this coordinate invariant manner has several advantages for the following discussion. It is only a slight shift of the view point and it is in full consistency with the conventions in previous sections when tensors are expressed in terms of Gaussian coordinate frames. 
Most of the tensor fields we are dealing with in this section are completely intrinsic to $\SigmaF_t$, i.e., \emph{fully spatial}, and will henceforth carry indices $i,j,\ldots$ exclusively. A particular important example is the tensor $\mathcal E_{ij}$ derived from the MST.

The metric induced on $\SigmaF_t$ from the conformal metric $\gmetr_{\mu\nu}$ is therefore denoted by $\gmetr_{ij}$ (we shall sometimes write $\gmetr_{ij}(t)$ when the particular value of $t$ is relevant or when we want to emphasize the fact that this metric is time dependent). 
On any $\SigmaF_t$, this Riemannian metric $\gmetr_{ij}$ determines a volume form $\gvolf_{ijk}$ and a Levi-Civita covariant derivative
$\gnabla_k$.
The consistent use of our abstract index conventions above makes  it is unnecessary to introduce a special symbol for the Levi-Civita connection of the Riemannian metric $\gmetr_{ij}$.

The metric induced on ${\scri^-}=\SigmaF_0$ will play a particular role in the following and is denoted by $\hmetr_{ij}=\gmetr_{ij}(0)$ as before. Via Lie transport along $\unitnormal^\mu$, this metric $\hmetr_{ij}$ can be dragged to any surface $\SigmaF_t$; the resulting field on $\Mtilde$ shall be referred to as $\hmetr_{ij}$ as well for simplicity. We shall do the same for all quantities derived from $\hmetr_{ij}$, in particular, for the volume form $\hvolf_{ijk}$ and the covariant derivative $\hnabla_k$ 
associated with $\hmetr_{ij}$. Since such  fields are therefore by definition \emph{invariant} under Lie transport along $\unitnormal^\mu$ we say that they are \emph{time-independent}. \emph{Any} field on $\Mtilde$ (or $M$) that is invariant in this way shall be referred to as \emph{time-independent}; otherwise we call it \emph{time-dependent}. 

According to the discussion in \Sectionref{sec_conf_MST_eqn} we shall now make certain assumptions about the behavior of various quantities at $t=0$; recall that most of these conditions constitute no loss of generality because they can always be achieved by an appropriate choice of gauge.
First we assume that 
\begin{equation}
  \label{eq:metrics}
  \gmetr_{ij}(t)=\hmetr_{ij}+O(t^2),\quad \widetilde K_{ij}(t)=O(t^2),
\end{equation}
where, in this coordinate invariant sense, the $O$-symbol is defined with respect to any time-independent Riemannian reference metric uniformly on $\SigmaF$ in all of what follows.
Second we assume that the conformal Killing vector field $\XKill^\mu$ can be written as
\begin{equation}
  \label{eq:XrelY}
  \XKill^0(t)=O(t^2),\quad \XKill^i(t)=Y^i+O(t^2)
\end{equation}
according to \Eqref{expansion_X} where $Y^i$ is a conformal Killing vector field of $\hmetr_{ij}$ without zeros on $\SigmaF_0$.
We shall  interpret $Y^i$ as a time-independent field on $\Mtilde$.
In agreement with \Eqsref{add_gauge} and \eqref{eq:deffN} we also assume
\begin{equation}
  \label{eq:gaugeY}
\hmetr_{ij}Y^i Y^j=1,\quad \hvolf\indices{_{i}^{k}_{l}} \hnabla_{k} Y^{l}=\lambda Y_{i},\quad \hnabla_{l} Y^{l}=0,
\end{equation}
where
the quantity $\lambda$ is some, in principle known, smooth time-independent function.

Following the earlier discussion,  the MST
is  represented by a
time-dependent purely spatial complex symmetric $\gmetr_{ij}$-trace-free $(0,2)$-tensor field
$\mathcal E_{ij}$
as follows. 
According to \Eqref{eq:leadingevoleq}, the MST evolution equations 
are
\begin{equation}
  \label{eq:MarsSimonRed}
  t \gnabla_0\mathcal E_{ij}
  -\imagi t \gvolf\indices{_{m}_{n}_{(i}}\gmetr^{mk}\gmetr^{nl}
  \gnabla_{|k|}\mathcal E_{j)l}
  =\Tfield\indices{^k^l_i_j}\mathcal E_{kl}.
\end{equation}
They are formally singular at $t=0$ and hence only make sense for $t>0$, i.e.,  on the subset $M$ of $\Mtilde$.
Near $t=0$, the smooth time-dependent field $\Tfield\indices{^k^l_i_j}$ is given as
\begin{equation}
  \label{eq:expansionS}
  \Tfield\indices{^k^l_i_j}(t)=\Tfield\indices{^k^l_i_j_{(0)}}+t\frac{\imagi\lambda}2 \Tfield\indices{^k^l_i_j_{(0)}}
+t^2 \Tfield\indices{^k^l_i_j_{(1)}}(t)
\end{equation}
with
\begin{equation}
  \label{eq:2}
  \Tfield\indices{^k^l_i_j_{(0)}}
  =\frac 32 Y_{(i}\,\delta\indices{_{j)}^{\!\!\!(k}}\,Y^{l)}-3Y_iY_jY^kY^l
  +\frac 12 \hmetr_{ij} Y^kY^l +\frac 12 \hmetr^{kl} Y_iY_j-\frac 16 \hmetr_{ij}\hmetr^{kl},
\end{equation}
which is hence time-independent, and with some known smooth time-dependent field $\Tfield\indices{^k^l_i_j_{(1)}}$. According to \Eqref{asympt_constraint}, the constraint equations take the form
\begin{equation}
  \label{eq:MarsSimonRedConstr}
 0=\hnabla_{l}\mathcal E\indices{_i^l}+\frac{\imagi}2\left(\frac 1t+\frac \imagi2\right) \hvolf\indices{_{i}_{j}^{l}} Y^j Y^k\mathcal E_{kl}-t\Cfield\indices{^k^l_i}(t)\mathcal E_{kl}=:\Dfield_i,
\end{equation}
on $M$,
where  $\Cfield\indices{^k^l_i}$
 is some smooth time-dependent tensor field on $M$ which is also known.
All index operations in \Eqsref{eq:metrics} -- \eqref{eq:MarsSimonRedConstr} are performed with the metric $\hmetr_{ij}$.

Before we continue, a few remarks are in place.
Recall that \Eqsref{eq:MarsSimonRed} -- \eqref{eq:MarsSimonRedConstr} had been derived in previous sections assuming Gauss coordinates and further particular gauge choices. Once these equations have been derived, however, we can forget about this and consider them as fully invariant tensorial equations. 
Indeed, for large parts of the analysis in the following section we only need to impose \Eqsref{eq:metrics} -- \eqref{eq:gaugeY}. Only for specific steps of our discussion, we explicitly need to introduce Gauss coordinates. One obtains a Gauss coordinate system within the general setup above by imposing the following restrictions on the foliation $\Sigma_t$ with respect to the time function $t$. We pick spatial coordinates $x^i$ on each leaf $\Sigma_t$ and make the additional assumption that $\unitnormal^\mu=\partial_t^\mu$ and $\unitnormal_\mu=dt_\mu$ for the corresponding spacetime coordinates $(t,x^i)$.

The reader will notice that \eqref{eq:2} looks significantly different from the corresponding terms in \Eqref{eq:leadingevoleq} (inside the second pair of brackets there). As one can easily check, however, both expressions are equivalent \emph{if} $\mathcal E_{ij}$ is symmetric and $\gmetr_{ij}$-trace-free. Since in some of the intermediate steps of our arguments below we will allow $\mathcal E_{ij}$ to be non-symmetric and non-trace-free, the expression in \Eqref{eq:2} is more suitable than the one in \Eqref{eq:leadingevoleq}. The field in  \Eqref{eq:2} is by construction explicitly symmetric and $\hmetr_{ij}$-trace-free with respect to both pairs of indices. 
The field in \Eqref{eq:expansionS} is {identically} symmetric and $\gmetr_{ij}$-trace-free with respect to both pairs of indices.

The overall goal is to solve an initial value problem for \Eqref{eq:MarsSimonRed} with data prescribed at $t=0$ such that \Eqref{eq:MarsSimonRedConstr} is satisfied identically on $M$. As noticed before, \Eqref{eq:MarsSimonRed} is formally singular at $t=0$ and hence  the initial value problem in the standard sense does not make sense. Instead we consider a \keyword{singular initial value problem}. The \emph{Fuchsian method}  shall allow us to prove the main result of this section, \Theoremref{thm:maintheorem}, below. Before we can state this theorem, however, we must introduce some further notation and terminology.

We say that an open subset $\Omega$ of $M$ with compact closure in $\Mtilde$ is a \keyword{lens-shaped region} (cf.\ Section~3.1 in \cite{Friedrich:2000wp}) if its boundary is the union of two smooth spacelike hypersurfaces in $\Mtilde$ with respect to the conformal metric $\gmetr_{\mu\nu}$ and if the boundaries of these two hypersurfaces coincide and are smooth. When one of these two spacelike hypersurfaces is called $S$ in $\Mtilde$, then we say that $\Omega$ is a \emph{lens-shaped region with respect to} $S$. For any $t\in (-\delta,\delta)$, we define $\Omega_t=\Omega\cap\SigmaF_t$; we allow $\Omega_t$ to be empty. We shall also write $\Omega_{(t_0,t)}$ to denote the intersection of $\Omega$ with the spacetime slab $(t_0,t)\times\SigmaF$ with $0<t_0<t<\delta$. 

In anticipation of the results in \Sectionref{sec:spectraldecomp}, we define the following time-independent fields (index operations are performed with respect to $\hmetr_{ij}$)
\begin{align*}
  E_{kl(1)}&=\frac 1{\sqrt 6}\left(3Y_k Y_l-\hmetr_{kl}\right),\\
  E_{kl(2)}&=\frac 1{\sqrt 2} \left(\frameD{1}{k} \frameD{2}{l}+\frameD{1}{l} \frameD{2}{k}\right),\quad E_{kl(3)}=\frac 1{\sqrt 2}\left(\frameD{1}{k} \frameD{1}{l}-\frameD{2}{k} \frameD{2}{l}\right)\\
  E_{kl(4)}&=\frac 1{\sqrt 2} \left(Y_{k} \frameD{1}{l}+Y_{l} \frameD{1}{k}\right),\quad E_{kl(5)}=\frac 1{\sqrt 2} \left(Y_{k} \frameD{2}{l}+Y_{l} \frameD{2}{k}\right),
\end{align*}
where $(\frameh{1}i, \frameh{2}i,Y^i)$ is any time-independent, $Y^i$-invariant $\hmetr_{ij}$-orthonormal frame.

\begin{theorem}[Singular initial value problem of the MST equations]
  \label{thm:maintheorem}
  Pick $\delta>0$ and an orientable $3$-dimensional differentiable manifold $\SigmaF$. Equip
 $\Mtilde=(-\delta,\delta)\times \SigmaF$ with a smooth Lorentzian metric $\gmetr_{\mu\nu}$ with a conformal Killing vector field $\XKill^\mu$ without zeros on ${\scri^-}=\SigmaF_0$ and Gaussian coordinates $(t,x^i)$ as before. 
Suppose that the MST of $(M,\gmetr_{\mu\nu},\XKill^\mu)$ vanishes and that \Eqsref{eq:metrics} -- \eqref{eq:gaugeY} hold with respect to our coordinate system. 
Pick any non-empty open subset $S_0$ of $\SigmaF$ with compact closure and non-empty smooth boundary, and, any non-empty lens-shaped region $\Omega$ with respect to the subset $\{t=0\}\times S_0$ of $\partial M\subset\Mtilde$ where $M=(0,\delta)\times \SigmaF\subset\Mtilde$.

Then, for any smooth complex time-independent functions $c_1$, $c_2$ and $c_3$ on $\Omega$ with the property
\[\mcL_Y c_1=\mcL_Y c_2=\mcL_Y c_3=0,\]
where $Y^i$ is defined by $\XKill^\mu$ through \Eqref{eq:XrelY},
there is a smooth solution $\mathcal E_{ij}$ of \Eqsref{eq:MarsSimonRed} -- \eqref{eq:MarsSimonRedConstr}, i.e., of the full Mars-Simon equation \eq{sym_hyp0}, of the form
\begin{equation}
  \label{eq:mainsolformula}
  \begin{split}
    \mathcal E_{ij}(t,x)=& \,c_1(x) E_{ij(1)}(x) t^{-1}
    +\imagi\lambda c_1(x) E_{ij(1)}(x)+c_{2}(x)  E_{ij(2)}(x)
    +c_{3}(x)  E_{ij(3)}(x)\\
    & -\frac{2 \imagi}{\sqrt 3}  \hnabla_{\frameh{2}{}} c_1(x) E_{ij(4)}(x)
    +\frac{2 \imagi}{\sqrt 3}  \hnabla_{\frameh{1}{}} c_1(x)  E_{ij(5)}(x)
    +E_{ij}(t,x),
  \end{split}
\end{equation}
for every $(t,x)\in \Omega$
provided $\delta>0$ is sufficiently small.
Here, $E_{ij}$ is some smooth complex symmetric $\gmetr_{ij}$-trace-free field which can be extended smoothly through $t=0$ and $\lim_{t\searrow 0} E_{ij}(t,x)=0$ for all points $x$. Any two smooth solutions $\mathcal E_{ij}$ and $\tilde{\mathcal E}_{ij}$ of this form given by the same data $c_1$, $c_2$ and $c_3$ are identical on $\Omega$.
\end{theorem}

The remainder of this section is devoted to the proof of this theorem.

\subsection{Spectral analysis of the principal part matrix}
\label{sec:spectraldecomp}

An essential first step in the proof of \Theoremref{thm:maintheorem} is a detailed analysis of the field $\Tfield\indices{^k^l_i_j_{(0)}}$ defined in \Eqref{eq:2}. This will lead naturally to the quantities $E_{ij(1)},\ldots, E_{ij(5)}$ above and to the structure of the leading-order term in \Eqref{eq:mainsolformula}. 

The whole discussion in this subsection only involves time-independent fields. It can therefore be carried out on the abstract Cauchy surface $\SigmaF$ without any reference to time $t$. For this whole subsection, complex conjugates of complex fields are denoted with a bar. All index operations are performed with the metric $\hmetr_{ij}$. We pick an arbitrary point $x\in\SigmaF$ and refrain from writing $x$ in the following formulas for simplicity. We consider the $9$-dimensional complex vector space of $(0,2)$-tensors of $T_x\SigmaF$; we do not  impose any symmetry or trace-free conditions at this stage yet.
On this vector space we have an inner product
\begin{equation}
  \label{eq:scalarproduct}
  \left({  E}_{ij},  \tilde E_{ij}\right)\mapsto\bar{E}^{ij} \tilde E_{ij}.
\end{equation}
The quantity $\Tfield\indices{_i_j^k^l_{(0)}}$ defined in \Eqref{eq:2} can be considered as an endomorphism of this vector space which we find to be self-adjoint
\[\Tfield\indices{_i_j^k^l_{(0)}}=\Tfield\indices{^k^l_i_j_{(0)}}.\]
It is thus diagonalizable, the eigenvalues are real and the respective eigenspaces are mutually orthogonal.
Using any $\hmetr_{ij}$-orthonormal basis $(\frameh{1}i, \frameh{2}i,Y^i)$ of $T_x\SigmaF$ (recall \Eqref{eq:gaugeY}) we can show:
\begin{description}
\item [Eigenvalue $-1$]: This eigenspace is spanned by the tensor
  \begin{equation}
    \label{eq:eigenspacem1}
    E_{kl(1)}=\frac 1{\sqrt 6}\left(3Y_k Y_l-\hmetr_{kl}\right),
  \end{equation}
  and is therefore a $1$-dimensional subspace.
\item [Eigenvalue $0$]: This eigenspace is spanned by all antisymmetric $(0,2)$-tensors ($3$-dimensional), all pure $\hmetr_{ij}$-trace $(0,2)$-tensors ($1$-dimensional), and, all symmetric $\hmetr_{ij}$-trace-free $(0,2)$-tensors $e_{kl}$  with the property $e_{kl} Y^l=0$ ($2$-dimensional). The latter subspace is spanned by
  \begin{equation}
    \label{eq:eigenspacem2}
    E_{kl(2)}=\frac 1{\sqrt 2} \left(\frameD{1}{k} \frameD{2}{l}+\frameD{1}{l} \frameD{2}{k}\right),\quad E_{kl(3)}=\frac 1{\sqrt 2}\left(\frameD{1}{k} \frameD{1}{l}-\frameD{2}{k} \frameD{2}{l}\right).
  \end{equation}
  In total this eigenspace is therefore $6$-dimensional.
\item [Eigenvalue $3/4$]: This eigenspace is spanned by
\begin{equation}
  \label{eq:thirdeigenspace}
   E_{kl(4)}=\frac 1{\sqrt 2} \left(Y_{k} \frameD{1}{l}+Y_{l} \frameD{1}{k}\right),\quad E_{kl(5)}=\frac 1{\sqrt 2} \left(Y_{k} \frameD{2}{l}+Y_{l} \frameD{2}{k}\right),
\end{equation}
and is therefore $2$-dimensional.
\end{description}

Now, when we restrict the map $\Tfield\indices{_i_j^k^l_{(0)}}$  to the $5$-dimensional  subspace of symmetric, $\hmetr_{ij}$-trace-free  $(0,2)$-tensors, it is still a self-adjoint endomorphism and the analogue eigenspace decomposition can be carried out. The only difference is that the eigenspace of the eigenvalue $0$ is now reduced to the two dimensional subspace spanned by \Eqref{eq:eigenspacem2}.
The self-adjoint property implies that all eigenspaces are mutually orthogonal with respect to the scalar product \Eqref{eq:scalarproduct}. Thanks to the normalizations chosen above we find
\begin{equation}
  \label{eq:orthonormality}
  E_{ij(p)}\bar E^{ij}_{(q)}=\delta_{pq},\quad\text{for all $p,q=1,\ldots,5$}.
\end{equation}
The collection of fields $E_{ij(1)}$ to $E_{ij(5)}$ therefore constitutes an orthonormal basis of the complex vector space of symmetric $\hmetr_{ij}$-trace-free  $(0,2)$-tensors at $x\in T_x\SigmaF$.

Since \Theoremref{thm:maintheorem} only refers to open proper  subsets $S_0$ of $\SigmaF$ with compact closure we can assume without loss of generality that $\SigmaF$ is a compact orientable $3$-dimensional manifold and therefore parallelizable. We may therefore {assume} that $(\frameh{1}i, \frameh{2}i, Y^i)$ is a \emph{global} orthonormal frame on $\SigmaF$ and then construct $E_{ij(1)}$ to $E_{ij(5)}$ at \emph{each} $x\in\SigmaF$ as above. In this way we may interpret these as smooth fields on $\SigmaF$ which satisfy all the properties above at each $x\in\SigmaF$. We may even consider $E_{ij(1)}$ to $E_{ij(5)}$ as smooth time-independent fields on $\Mtilde$ which satisfy all the properties above at each $(t,x)\in \Mtilde$.
Without loss of generality we can assume additionally that the orthonormal frame $(\frameh{1}i, \frameh{2}i, Y^i)$ is invariant under Lie transport along $Y^i$ globally on $\SigmaF$. This implies that
\begin{equation}
  \label{eq:Liedereigenfields}
  \mathcal L_Y E_{ij(p)}=0
\end{equation}
everywhere on $\Mtilde$.

The remainder of this subsection is now devoted to some further technical properties of the fields $E_{ij(p)}$ which turn out to be useful later.
Given \eqref{eq:eigenspacem1}, the following quantity related to the curl of $E_{jl(1)}$ can be written as
\begin{align*}
\hvolf\indices{_{(i}^k^l}\hnabla_{|k|} E_{j)l(1)}
&=\sqrt{\frac 32}\hvolf\indices{_{(i}^k^l}(\hnabla_{|k|} Y_{j)})Y_{l}
+\sqrt{\frac 32}\hvolf\indices{_{(i}^k^l}Y_{j)} \hnabla_{k} Y_{l}\\
&=\sqrt{\frac 32}\hvolf\indices{_{(i}^k^l}(\hnabla_{|k|} Y_{j)})Y_{l}
+\sqrt{\frac 32}\lambda Y_i Y_j,
\end{align*}
where we have used the second condition of \Eqref{eq:gaugeY}.
The third condition there
implies that $\hnabla_{k} Y_{j}$ is $\hmetr_{ij}$-trace-free. Since $Y^i$ is a conformal Killing vector field of $\hmetr_{ij}$, it therefore follows that
\begin{equation}
  \label{eq:conformalKVF}
  \hnabla_{k} Y_{j}=\hnabla_{[k} Y_{j]}
  =\frac 12\hvolf_{nkj}\hvolf^{nml}\hnabla_{m} Y_{l}
  =\frac \lambda 2\hvolf_{nkj} Y^n.
\end{equation}
We plug this into the expression above:
\begin{align*}
  \hvolf\indices{_{(i}^k^l}\hnabla_{|k|} E_{j)l(1)}
&=\frac 12\sqrt{\frac 32}\hvolf\indices{_{i}^k^l}(\hnabla_{k} Y_{j})Y_{l}
+\frac 12\sqrt{\frac 32}\hvolf\indices{_{j}^k^l}(\hnabla_{k} Y_{i})Y_{l}
+\sqrt{\frac 32}\lambda Y_i Y_j\\
&=\frac 12\sqrt{\frac 32}\hvolf\indices{_{i}^k^l}\left(\frac \lambda 2\hvolf_{nkj} Y^n\right)Y_{l}
+\frac 12\sqrt{\frac 32}\hvolf\indices{_{j}^k^l}\left(\frac \lambda 2\hvolf_{nki} Y^n\right)Y_{l}
+\sqrt{\frac 32}\lambda Y_i Y_j.
\end{align*}
The first term is
\begin{align*}
  \frac \lambda4\sqrt{\frac 32} \hmetr_{mi}\hvolf\indices{^k^{m}^l}\hvolf_{knj} Y^n Y_{l}
  =\frac \lambda4\sqrt{\frac 32} \hmetr_{mi}\left(\delta\indices{^{m}_n}\delta\indices{^l_j}-\delta\indices{^{m}_j}\delta\indices{^l_n}\right) Y^n Y_{l}
  =\frac \lambda4\sqrt{\frac 32} \left(Y_i Y_{j}-\hmetr_{ij}\right).
\end{align*}
Since this is symmetric in $i$ and $j$, it follows
\begin{equation}
  \label{eq:blubbb5}
  \hvolf\indices{_{(i}^k^l}\hnabla_{|k|} E_{j)l(1)}
=\frac \lambda2\sqrt{\frac 32} \left(Y_i Y_{j}-\hmetr_{ij}\right) +\sqrt{\frac 32}\lambda Y_i Y_j
=\frac 32\lambda E_{ij(1)},
\end{equation}
using \Eqref{eq:eigenspacem1}.

Another useful identity can be derived for the divergence of $E_{jl(1)}$. Using again \Eqref{eq:eigenspacem1}, we find
\[\hnabla^k E_{ik(1)}=\sqrt{\frac 32} \hmetr^{jk} \hnabla_j(Y_i Y_k)
=\sqrt{\frac 32} Y^j \hnabla_j Y_i,\]
which follows from the third relation in \Eqref{eq:gaugeY}. \Eqref{eq:conformalKVF} yields
\begin{equation}
  \label{eq:divergenceE1}
  \hnabla^k E_{ik(1)}=\sqrt{\frac 32} Y^j \frac \lambda 2\hvolf_{nji} Y^n=0.
\end{equation}

Finally, we discuss the following
\begin{equation*}
  \hvolf\indices{_{(i}^k^l} E_{j)l(1)}
  =\frac 1{\sqrt 6}\hvolf\indices{_{(i}^k^l}(3 Y_{j)} Y_l-\hmetr_{{j)}l})
  =\frac 1{\sqrt 6}\left(3\hvolf\indices{_{(i}^k^l} Y_{j)} Y_l-\hvolf\indices{_{(i}^k_{j)}}\right)
  =\sqrt{\frac 32} \hvolf\indices{_{(i}^k^l} Y_{j)} Y_l,
\end{equation*}
using \Eqref{eq:eigenspacem1}. If now we assume in addition to the above that the orthonormal frame $(\frameh{1}i, \frameh{2}i, Y^i)$ is oriented according to
$\hvolf\indices{_k_{i}_l} \frameh{1}k \frameh{2}iY^{l} =1$, then
\Eqsref{eq:eigenspacem1} -- \eqref{eq:thirdeigenspace} imply
\begin{equation}
  \label{eq:blubbbbb4}
  \hvolf\indices{_{i}^k^l} E_{jl(1)} \bar E^{ij}_{(p)}
  =
  \begin{cases}
    0, & p=1,2,3,\\
    \frac{\sqrt{3}}2\, \frameh{2}k,& p=4,\\
    -\frac{\sqrt{3}}2\, \frameh{1}k,& p=5,
  \end{cases}
\end{equation}
and,
\begin{equation}
  \label{eq:blubbbbb44}
  \hvolf\indices{^{i}_k_l} Y^{j}Y^l E_{ij(p)}
=
\begin{cases}
  0, & p=1,2,3,\\
  \frac{\sqrt{2}}2\, \frameD{2}{k},& p=4,\\
  -\frac{\sqrt{2}}2\, \frameD{1}{k},& p=5.
\end{cases}
\end{equation}

\subsection{The singular initial
  value problem of the evolution equations}
\label{sec:SIVPevol}

Next we are going to discuss the singular initial value problem of the evolution equations \Eqref{eq:MarsSimonRed} with ``data'' prescribed at $t=0$. This is an application of the Fuchsian method developed in \cite{Ames:2013uh,Ames:2012vz} for general quasilinear symmetric hyperbolic systems.
The main first step of the analysis of any {singular initial value problem} is to split the unknown into two parts. The first part is some \emph{explicitly known} function -- in most cases a generalized power series expansion about $t=0$ some of whose coefficients constitute  free data -- which is supposed to describe the \emph{leading-order} behavior of the unknown at the singular time $t=0$. The second part, i.e., the difference of the original unknown and this leading-order term, is called the \emph{remainder}. It is considered as the  \emph{unknown} of the singular initial value problem. The aim is then to show that given the leading-order term, there exists a remainder which is uniquely determined by the equation and which decays with some sufficiently \emph{high order} at $t=0$ relative to the leading-order term in some suitable sense.
We remark that this decomposition also applies to the \emph{standard regular} initial value problem. There the leading-order term corresponds to a (truncated) Taylor series of the solution about the initial time, and the free Cauchy data can be identified with some of its coefficients.

In the case of  \Eqref{eq:MarsSimonRed}, 
the main tasks are therefore, (i), to
derive appropriate leading-order terms (see \Sectionref{sec:deriveLOT}), and, (ii),  to analyze the equation for the remainder. This subsection now is devoted to (ii).
In a first step, we do not yet explicitly specify a leading-order term (and hence the free data). Instead we work with the following equation
\begin{equation}
  \label{eq:MarsSimonRedMod}
  t\gnabla_0 E_{ij}
  -\imagi t \gvolf\indices{^{m}^{(l} _{(i}}\delta\indices{_{j)}^{k)}}\gnabla_{m} E_{kl}=\Tfield\indices{^k^l_i_j} E_{kl}+\Ffield_{ij}
\end{equation}
where the unknown is a complex time-dependent purely spatial $(0,2)$-tensor field $E_{ij}$. The time-dependent source term field $\Ffield_{ij}$ is considered as given in most of the following discussion. It will eventually describe the contribution from the leading-order term of the singular initial value problem to the original equation \eqref{eq:MarsSimonRed}, and the unknown $E_{ij}$ will be identified with the remainder of the singular initial value problem; see the end of \Sectionref{sec:deriveLOT}. 
In this subsection it is convenient to perform index operations with the metric $\gmetr_{ij}$. Notice that the second term in \Eqref{eq:MarsSimonRedMod} differs from the corresponding term in \Eqref{eq:MarsSimonRed}; however, both terms are the same in the eventual case of interest when $E_{ij}$ is symmetric and $\gmetr_{ij}$-trace-free.
For the time being we allow $\Ffield_{ij}$ and $E_{ij}$ to be \emph{any} smooth time-dependent purely spatial $(0,2)$-tensor fields.  The purpose of the following now is to derive precise conditions on the source term field which guarantee the existence of a uniquely determined solution $E_{ij}$ with a sufficient decay at $t=0$.

Under all the conditions before, \Eqref{eq:MarsSimonRedMod} is a linear symmetric hyperbolic system with smooth coefficients (the regularity of the source term field $\Ffield_{ij}$ has not been fixed yet). We shall therefore follow the analysis of  this class of equations in \cite{Ames:2013uh,Ames:2012vz}. Notice however that in contrast to these references, the spatial domain $\SigmaF$ here may not be a torus. 
In fact, the techniques presented below shall have a particular emphasis on localization and ``domain of dependence'' arguments; the torus will still play an intermediate role as we will see.

We derive now the fundamental energy estimates for solutions of \Eqref{eq:MarsSimonRedMod}. In contrast to the estimate in \cite{Ames:2013uh,Ames:2012vz}, we shall here take the tensorial character of the unknown $ E_{ij}$ explicitly into account and, in addition, localize the estimate.
To this end, we suppose  that the field $\Ffield_{ij}$ in \Eqref{eq:MarsSimonRedMod} is defined on  some open subset $\Omega$ of $M$ (not necessarily a lens-shaped region) and is sufficiently smooth there (we specify the minimal regularity later). Suppose also that $E_{ij}$ is a sufficiently smooth solution of  \Eqref{eq:MarsSimonRedMod} defined on $\Omega$.
We find
\begin{equation*}
  t\gstnabla_0\bar{  E}^{ij}
  +\imagi t \gvolf\indices{^{m}_{(l} ^{(i}}\delta\indices{^{j)}_{k)}}\gstnabla_{m}\bar{ E}^{kl}
=\Tcheckfield\indices{_k_l^i^j}\bar { E}^{kl}+\bar \Ffield^{ij},
\end{equation*}
where, as before, complex conjugates are denoted with a bar.
Hence, for any
$\mu\in\R$, we have
\begin{align*}
   t\gstnabla_0&\left(t^{-2\mu} \bar{  E}^{ij} E_{ij}\right)\\
  =&-2\mu t^{-2\mu}\bar{  E}^{ij} E_{ij}\\
  &+t^{-2\mu} E_{ij}\left(-\imagi t \gvolf\indices{^{m}_{(l} ^{(i}}\delta\indices{^{j)}_{k)}}\gstnabla_{m}\bar{ E}^{kl}
+\Tcheckfield\indices{_k_l^i^j}\bar { E}^{kl}+\bar \Ffield^{ij}\right)\\
&+t^{-2\mu}\bar{  E}^{ij}\left(
+\imagi t \gvolf\indices{^{m}^{(l} _{(i}}\delta\indices{_{j)}^{k)}}\gstnabla_{m} E_{kl}
+\Tcheckfield\indices{^k^l_i_j} E_{kl}+\Ffield_{ij}\right)\\
=&-2\mu t^{-2\mu}\bar{  E}^{ij} E_{ij}\\
&+t^{-2\mu}\left( E^{ij}\Tcheckfield\indices{^k^l_i_j}\bar { E}_{kl}
  +\bar{  E}^{ij}\Tcheckfield\indices{^k^l_i_j} E_{kl}\right)\\
  &+t^{-2\mu}\left( E_{ij}\bar \Ffield^{ij}+\bar{  E}^{ij} \Ffield_{ij}\right)
-\imagi t^{-2\mu+1} E_{ij} \gvolf\indices{^{m}_{(l} ^{(i}}\delta\indices{^{j)}_{k)}}\gstnabla_{m}\bar{ E}^{kl}
+\imagi t^{-2\mu+1}\bar{  E}^{ij}\gvolf\indices{^{m}^{(l} _{(i}}\delta\indices{_{j)}^{k)}}\gstnabla_{m} E_{kl}.  
\end{align*}
The last two terms can be written as follows
\begin{align*}
  &-\imagi t^{-2\mu+1} E_{ij} \gvolf\indices{^{m}_{(l} ^{(i}}\delta\indices{^{j)}_{k)}}\gstnabla_{m}\bar{ E}^{kl}
+\imagi t^{-2\mu+1}\bar{  E}^{ij}\gvolf\indices{^{m}^{(l} _{(i}}\delta\indices{_{j)}^{k)}}\gstnabla_{m} E_{kl}\\
=&
  -\imagi t^{-2\mu+1} E_{ij} \gvolf\indices{^{m}_{(l} ^{(i}}\delta\indices{^{j)}_{k)}}\gnabla_{m}\bar{ E}^{kl}
+\imagi t^{-2\mu+1}\bar{  E}^{ij}\gvolf\indices{^{m}^{(l} _{(i}}\delta\indices{_{j)}^{k)}}\gnabla_{m} E_{kl}\\
=&-\imagi t^{-2\mu+1} E_{ij} \gvolf\indices{^{m}_{(l} ^{(i}}\delta\indices{^{j)}_{k)}}\gnabla_{m}\bar{ E}^{kl}
+\gnabla_{m}\left(\imagi t^{-2\mu+1}\bar{  E}^{ij}\gvolf\indices{^{m}^{(l} _{(i}}\delta\indices{_{j)}^{k)}} E_{kl}\right)\\
&-\imagi t^{-2\mu+1}\left(\gnabla_{m}\bar{  E}^{ij}\right)\gvolf\indices{^{m}^{(l} _{(i}}\delta\indices{_{j)}^{k)}} E_{kl}\\
=&\gnabla_{m}\left(\imagi t^{-2\mu+1}\bar{  E}^{ij}\gvolf\indices{^{m}^{(l} _{(i}}\delta\indices{_{j)}^{k)}} E_{kl}\right)
=\gstnabla_{m}\left(\imagi t^{-2\mu+1}\bar{  E}^{ij}\gvolf\indices{^{m}^{(l} _{(i}}\delta\indices{_{j)}^{k)}} E_{kl}\right).
\end{align*}
We have therefore found the following identity
\begin{equation}
  \label{eq:7}
  \begin{split}
    t\gstnabla_0\left(t^{-2\mu} \bar{  E}^{ij} E_{ij}\right)
  =&
 -t^{-2\mu}\bar{  E}^{ij}\underbrace{\left(2\mu\delta\indices{_i^k}\delta\indices{_j^l}
-\Tcheckfield\indices{_i_j^k^l}-\Tcheckfield\indices{^k^l_i_j}\right)}_{=: M\indices{^k^l_i_j}} E_{kl}\\
  &+t^{-2\mu}\left(\bar \Ffield^{ij} E_{ij}+\bar{  E}^{ij} \Ffield_{ij}\right)
  +\gstnabla_{m}\left(\imagi t^{-2\mu+1}\bar{  E}^{ij}\gvolf\indices{^{m}^{(l} _{(i}}\delta\indices{_{j)}^{k)}} E_{kl}\right).
  \end{split}
\end{equation}

If $\delta>0$ is now sufficiently small, the linear map $M\indices{^k^l_i_j}$ on the complex vector space of $(0,2)$-tensors at any given $(t,x)\in \Omega$ is positive definite with respect to the scalar product
\begin{equation}
  \left({  E}_{ij},  \tilde E_{ij}\right)\mapsto\bar{  E}_{}^{ij} \tilde E_{ij},
\label{inner_prod}
\end{equation}
provided
 the linear map
\begin{equation}
  \label{eq:5}
  M\indices{^k^l_i_j_{(0)}}
:=2\mu\delta\indices{_i^k}\delta\indices{_j^l}-\Tfield\indices{_i_j^k^l_{(0)}}-\Tfield\indices{^k^l_i_j_{(0)}}
\end{equation}
is positive definite at $t=0$. 
Recall that index operations are performed with the metric $\gmetr_{ij}$ in this subsection and hence the  inner product in \eq{inner_prod} is distinct from the one in \Eqref{eq:scalarproduct}.

Using that $\gmetr_{ij}$ equals the metric $\hmetr_{ij}$ at $t=0$ according to \Eqref{eq:metrics},
we conclude that the endomorphism $M\indices{^k^l_i_j_{(0)}}$ is self-adjoint, and it is therefore positive definite at each $(0,x)\in\SigmaF_0$ provided that $\mu$ is strictly larger than the largest eigenvalue of the endomorphism  $\Tfield\indices{_i_j^k^l_{(0)}}$. According to the results in \Sectionref{sec:spectraldecomp}, \Eqref{eq:7} therefore implies the inequality
\begin{equation}
  \label{eq:12}
    t\gstnabla_0\left(t^{-2\mu} \bar{  E}^{ij} E_{ij}\right)
  \le t^{-2\mu}\left(\bar \Ffield^{ij} E_{ij}+\bar{  E}^{ij} \Ffield_{ij}\right)
  +\gstnabla_{m}\left(\imagi t^{-2\mu+1}\bar{  E}^{ij}\gvolf\indices{^{m}^{(l} _{(i}}\delta\indices{_{j)}^{k)}} E_{kl}\right),
\end{equation}
for any $(t,x)\in \Omega$ provided that $\mu$ is any constant larger than $3/4$ and $\delta$ is any sufficiently small positive constant.

Let us now, for the time being, pick
$\Omega=M$ and $\SigmaF=T^3$. Given this, let us, first, replace the spatial covariant derivative $\gnabla_m$ (defined with respect to $\gmetr_{ij}$) by $\hnabla_m$ (defined with respect to $\hmetr_{ij}$) in \Eqref{eq:7}; this yields an additional (tensorial) contribution of $O(t)$ to $M\indices{^k^l_i_j}$ in \Eqref{eq:7}. For any sufficiently small $\delta>0$, this new map $M\indices{^k^l_i_j}$ is therefore positive definite under the same conditions as above. Second, noting that $\gstnabla_0$ acts on a scalar function in \Eqref{eq:7}, it can therefore be interpreted as the directional derivative along $\unitnormal^\mu$. When we now introduce arbitrary coordinates $(t,x^i)$ on $\Mtilde$, for example (but not necessarily) Gauss coordinates, for the same time function $t$ as above and hence obtain that $\unitnormal^\mu=\alpha \partial_t^\mu+\beta^\mu$ for a strictly positive function $\alpha$ and a purely spatial vector field $\beta^i$, the spatial derivative associated with the $\beta^i$-term yields an additional contribution to the last and to the first term on the right side of \Eqref{eq:7}. The contribution to the first term is $O(t)$ and does therefore not change the positivity criterion of the new map $M\indices{^k^l_i_j}$.
Third, we integrate over the spatial domain $\SigmaF_t=T^3$ with respect to the volume element of $\hmetr_{ij}$ at any $t\in (0,\delta)$. The last term in \Eqref{eq:7} (including the additional contribution obtained in the second step above) then disappears as a consequence of Stokes' theorem.
If $(e_a^i)$ ($a,b,\ldots=1,2,3$) is any smooth $\gmetr_{ij}$-orthonormal frame  intrinsic to each surface $\SigmaF_t$ (for example, consider the frame in \Sectionref{sec:spectraldecomp}) and we denote by
 $E(t,x)$ the $9$-dimensional complex vector of frame components of $E_{ij}(t,x)$ and by $\Ffield(t,x)$ the corresponding vector for $\Ffield_{ij}(t,x)$, then
\begin{equation}
  \label{eq:step1}
  t \frac d{dt} \|t^{-\mu} E(t,\cdot)\|_{L^2(T^3,\hmetr_{ij})}^2\le 2C\,\mathrm{Re}\,\left<t^{-\mu}  E(t,\cdot), t^{-\mu} \Ffield(t,\cdot)\right>_{L^2(T^3,\hmetr_{ij})},
\end{equation}
for some uniform constant $C>0$ (determined by the function $\alpha$ above).
In this notation, the metric $\hmetr_{ij}$ determines the volume element with respect to which the norm and scalar product are defined.
An application of the Cauchy-Schwarz inequality and integration with respect to $t$ over any interval $(t_0,t)$ with $0<t_0<t<\delta$ yield
\begin{equation}
  \label{eq:step2}
  \|t^{-\mu} E(t,\cdot)\|_{L^2(T^3,\hmetr_{ij})}\le C\left(\|t^{-\mu} E(t_0,\cdot)\|_{L^2(T^3,\hmetr_{ij})}+ \int_{t_0}^t \|s^{-\mu} \Ffield(s,\cdot)\|_{L^2(T^3,\hmetr_{ij})} s^{-1} ds\right);
\end{equation}
see Section~7.2 in \cite{Ringstrom:2009cj} for some of the basic technical steps going from \eqref{eq:step1} to \eqref{eq:step2}.
By adapting the constant $C$ we can conclude that
\[\|t^{-\mu} E(t,\cdot)\|_{L^2(T^3)}\le C\left(\|t^{-\mu} E(t_0,\cdot)\|_{L^2(T^3)}+ \int_{t_0}^t \|s^{-\mu} \Ffield(s,\cdot)\|_{L^2(T^3)} s^{-1} ds\right),\]
for any $t_0>0$ and $t\in [t_0,\delta)$. These norms are now defined with respect to the Euclidean flat metric on $T^3$.
Thanks to this estimate, we are now in the position to apply the methods in \cite{Ames:2013uh,Ames:2012vz} (see Props.~2.10 and 2.12 in \cite{Ames:2012vz}, and, Prop.~3.5 in \cite{Ames:2013uh}). to establish the following fundamental existence result.

\begin{proposition}[$T^3$-existence and uniqueness: Evolution equations]
  \label{prop:fundamentalexistence}
  Pick any sufficiently small $\delta>0$, any integer $q>3/2+1$ and set $\SigmaF=T^3$. Equip
 $\Mtilde=(-\delta,\delta)\times \SigmaF$ with a smooth Lorentzian metric $\gmetr_{\mu\nu}$ with a conformal Killing vector field $\XKill^\mu$ without zeros on ${\scri^-}=\SigmaF_0$. Suppose that \Eqsref{eq:metrics} -- \eqref{eq:gaugeY} hold.
Consider any complex field $\Ffield_{ij}$ on $M=(0,\delta)\times\SigmaF$ whose orthonormal frame components $\Ffield_{ab}$ are in $X_{\delta,\nu,q}(T^3)$ for some constant $\nu>3/4$. Then \Eqref{eq:MarsSimonRedMod} has a classical solution $E_{ij}$ defined on $M$ whose orthonormal frame components $E_{ab}$ are in $X_{\delta,\mu,q}(T^3)$ for any constant $\mu<\nu$. Moreover, all functions
$t\unitnormal(E_{ab})$ are in $X_{\delta,\mu,q-1}$ for the same $\mu$.
This solution $E_{ij}$ is uniquely determined within the class of all fields ${\tilde E}_{ij}$ defined on $M$ whose orthonormal frame components ${\tilde E}_{ab}$ are continuously differentiable and which satisfy
\[\lim_{t\searrow 0} \|t^{-\eta} {\tilde E}(t)\|_{L^2(T^3)}=0\]
for some constant $\eta>3/4$.
\end{proposition}

The Banach spaces $X_{\delta,\mu,q}(T^3)$ are defined \cite{Ames:2013uh,Ames:2012vz} as the completion of the set of smooth functions $f$ on $(0,\delta)\times T^3$ with respect to the norm
\[\|f\|_{\delta,\mu,q}:=\sup_{t\in (0,\delta)} \|t^{-\mu} f(t,\cdot)\|_{H^q(T^3)},\]
where $H^q(T^3)$ is the standard $L^2$-Sobolev space on $T^3$ (with respect to the flat metric) of differentiability order $q$. Notice that in general the exponent $\mu$ is allowed to be any smooth function on $T^3$; for the purpose of our studies here it is sufficient to work with constant exponents $\mu$ exclusively.
The interested reader can find the details in the references above and in \cite{Ames:2016uy,Beyer:2015fhs}.

\Propref{prop:fundamentalexistence} is our fundamental existence and uniqueness result. However, several of its aspects are not fully satisfying. The first issue we address now is the restriction $\SigmaF=T^3$. We thereby obtain a localized existence and uniqueness result.

To this end, we shall first determine the characteristics of the hyperbolic system \Eqref{eq:MarsSimonRedMod}. The characteristic surfaces are expected to bound the domain of dependence of solutions of the singular initial value problem. By definition, the normal to any characteristic surface at $(t,x)$ in $M$ is a spacetime covector $(\xi_0,\xi_m)$ for which the determinant of the principal symbol of \Eqref{eq:MarsSimonRedMod}
\[\xi_0 \delta\indices{_i^k}\delta\indices{_j^l}-i\xi_m\gvolf\indices{^{m}^{(l} _{(i}}\delta\indices{_{j)}^{k)}}\]
vanishes; this is interpreted as an endomorphism on the complex vector space of $(0,2)$-tensors at $(t,x)$. For any choice of spatial covector $\xi_m$, this is the case if and only if $\xi_0$ is an eigenvalue of the linear map
\begin{equation}
  \label{eq:characteristicmap}
  i\xi_m\gvolf\indices{^{m}^{(l} _{(i}}\delta\indices{_{j)}^{k)}}.
\end{equation}
By straightforward calculations we show that this map is diagonalizable on our $9$-dimensional complex vector space. The eigenspace of the eigenvalue $0$ is $5$-dimensional, and the eigenspace of each of the  four eigenvalues $\pm |\xi|/2$ and $\pm |\xi|$ is $1$-dimensional where $|\xi|=(\gmetr^{ij}\xi_i\xi_j)^{1/2}$. Solutions to \Eqref{eq:MarsSimonRedMod} therefore propagate at either speed $0$, half of the speed of light or the full speed of light.

Consider now any non-empty open subset $S_0$ of $\SigmaF_0$ with compact closure and smooth non-empty boundary and pick a {lens-shaped region} $\Omega$  with respect to $S_0$.
Suppose first now that $\Ffield_{ij}$ vanishes identically on $\Omega$, and that $ E_{ij}$ is a solution of \Eqref{eq:MarsSimonRedMod} defined on $\Omega$ whose orthonormal frame components (chosen as in \Propref{prop:fundamentalexistence}) are $C^1$-functions on $\Omega$ and, for any $t$ and $t_0$ with  $0<t_0<t<\delta$, extend as $C^1$-functions to the boundary of $\Omega_{(t_0,t)}$.
Integrating \eqref{eq:12} over $\Omega_{(t_0,t)}$ with respect to the volume element of the metric $\gmetr_{\mu\nu}$ assuming $\mu>3/4$ and applying Stokes' theorem, we get
\begin{equation}
  \label{eq:adlfkjasd}
  \int_{\Omega_t} t^{-2\mu} \bar{  E}^{ij} E_{ij} \text{Vol}_{\gmetr_{ij}(t)}
\le \int_{\Omega_{t_0}} t_0^{-2\mu} \bar{  E}^{ij} E_{ij} \text{Vol}_{\gmetr_{ij}(t_0)}.
\end{equation}
Recall here that the boundary of the lens-shaped region is spacelike everywhere and hence all remaining boundary integrals implied by Stokes' theorem  have a ``good'' sign thanks to the properties of the characteristics established above. 
 If now we assume that
\[\lim_{t\searrow 0} \int_{\Omega_{t}} t^{-2\mu} \bar{  E}^{ij}(t,x) E_{ij}(t,x) \text{Vol}_{\gmetr_{ij}(t)}=0,\]
and we take the limit $t_0\searrow 0$ in \Eqref{eq:adlfkjasd}, we conclude that $E_{ij}(t,x)=0$ for all $(t,x)\in\Omega$.
Because $\gmetr_{ij}$ is a smooth Riemannian metric near $t=0$, this is the case if and only if
\[\lim_{t\searrow 0} \int_{\Omega_{t}} t^{-2\mu} \bar{  E}^{ij}(t,x) E_{ij}(t,x) dx=0,\]
where $dx$ denotes the flat volume element. 
This completes the proof of the following lemma.

\begin{lemma}
  \label{lem:localuniqueness}
  Pick  any sufficiently small $\delta>0$ and orientable $3$-dimensional differentiable manifold $\SigmaF$. Equip
 $\Mtilde=(-\delta,\delta)\times \SigmaF$ with a smooth Lorentzian metric $\gmetr_{\mu\nu}$ with a conformal Killing vector field $\XKill^\mu$ without zeros on ${\scri^-}=\SigmaF_0$. 
Suppose that
\Eqsref{eq:metrics} -- \eqref{eq:gaugeY} hold. 
Pick any non-empty open subset $S_0$ of $\SigmaF$ with compact closure and non-empty smooth boundary, and, any lens-shaped region $\Omega$ with respect to the subset $\{t=0\}\times S_0$ of $\partial M\subset\Mtilde$ where $M=(0,\delta)\times \SigmaF\subset\Mtilde$.
  Suppose that $\Ffield_{ij}$ vanishes identically on $\Omega$, and that $E_{ij}$ is a solution of \Eqref{eq:MarsSimonRedMod} defined on $\Omega$ whose orthonormal frame components are $C^1$-functions on $\Omega$, and, for any $t$ and $t_0$ with  $0<t_0<t<\delta$, extend as $C^1$-functions to the boundary of $\Omega_{(t_0,t)}$.
If in addition
\begin{equation}
  \label{eq:decayintegral}
  \lim_{t\searrow 0} \int_{\Omega_{t}} t^{-2\mu} \bar{  E}^{ij}(t,x) E_{ij}(t,x) dx=0
\end{equation}
for some $\mu>3/4$,
then $E_{ij}(t,x)=0$ for all $(t,x)\in\Omega$.
\end{lemma}

Due to the linearity of the equations, this lemma yields conditions under which the values of the solution $E_{ij}$ on $\Omega$ are guaranteed to be independent of the values of the source term $\Ffield_{ij}$ outside of $\Omega$. 

We remark that this local uniqueness statement here is more general than Lemma~4.14 in \cite{mpss} because only the class of solutions of the MST evolution equations which extend as $C^1$-functions through $t=0$ is considered there. We will see below that the evolution equations admit solutions which violate this property. Only towards the end of this whole section we will see how to restrict the class of solutions appropriately in order to establish smoothness through $t=0$; see \Sectionref{sec:smoothextendibility}.

In consistency with the localized character of \Theoremref{thm:maintheorem} (and therefore \Lemref{lem:localuniqueness}) we shall henceforth ignore the dynamics outside of $\Omega$ now. In particular, since $\Omega$ is a proper subset of $M$,  it is no loss of generality to pick $\SigmaF=T^3$ as in \Propref{prop:fundamentalexistence} now in all of what follows. Suppose that $\Ffield_{ij}$ is any smooth field on $\Omega$ which can be extended smoothly to $M=(0,\delta)\times T^3$ such that the vector of its frame components satisfies
\[\lim_{t\searrow 0} \|t^{-\nu} {\Ffield}(t,\cdot)\|_{H^q(T^3)}=0\]
for some $\nu>3/4$ and for all positive integers $q$. This is equivalent to the condition that this extended field is in $X_{\delta,\nu,q}(T^3)$ for some $\nu>3/4$ for all positive integers $q$. \Propref{prop:fundamentalexistence} therefore implies the existence of a classical (in fact smooth) solution $E_{ij}$ on $M$.  Since the vector $E$ of its orthonormal frame components is in $X_{\delta,\mu,q}$ for \emph{any} $\mu<\nu$, it is therefore possible to satisfy \Eqref{eq:decayintegral} for \emph{some} $\mu\in (3/4,\nu)$. \Lemref{lem:localuniqueness} thus implies that the restriction of $E_{ij}$ to $\Omega$ is unaffected by the extension of $\Ffield_{ij}$ to $M=(0,\delta)\times T^3$.

\begin{proposition}[Local existence and uniqueness: Evolution equations]
  \label{prop:localisedexistence}
  Pick any sufficiently small $\delta>0$ and orientable $3$-dimensional differentiable manifold $\SigmaF$. Equip
 $\Mtilde=(-\delta,\delta)\times \SigmaF$ with a smooth Lorentzian metric $\gmetr_{\mu\nu}$ with a conformal Killing vector field $\XKill^\mu$ without zeros on ${\scri^-}=\SigmaF_0$. 
Suppose that
\Eqsref{eq:metrics} -- \eqref{eq:gaugeY} hold. 
Pick any non-empty open subset $S_0$ of $\SigmaF$ with compact closure and non-empty smooth boundary, and, any lens-shaped region $\Omega$ with respect to the subset $\{t=0\}\times S_0$ of $\partial M\subset\Mtilde$ where $M=(0,\delta)\times \SigmaF\subset\Mtilde$.
Consider any smooth complex field $\Ffield_{ij}$ on $\Omega$ which can be extended smoothly to $(0,\delta)\times T^3$ (considering $S_0$ as a subset of $T^3$) such that the
 orthonormal frame components of the extended field are in $X_{\delta,\nu,q}(T^3)$ for some constant $\nu>3/4$ and every positive integer $q$. Then \Eqref{eq:MarsSimonRedMod} has a smooth solution $E_{ij}$ defined on $\Omega$ that satisfies
\begin{equation}
  \label{eq:conddecay}
  \lim_{t\searrow 0}\int_{\Omega_{t}} t^{-2\mu}\bar{  E}^{ij}(t,x) E_{ij}(t,x) dx=0 
\end{equation}
for any $\mu<\nu$. The same estimate holds for any spatial derivative of any order of the frame components of $E_{ij}$.
If there is any other solution $\tilde E_{ij}$ on $\Omega$ whose orthonormal frame components are $C^1$-functions on $\Omega$, and, for any $t$ and $t_0$ with  $0<t_0<t<\delta$, extend as $C^1$-functions to the boundary of $\Omega_{(t_0,t)}$, and, has the property
\[\lim_{t\searrow 0}\int_{\Omega_{t}} t^{-2\mu}\bar{\tilde E}^{ij}(t,x) \tilde E_{ij}(t,x) dx=0\]
for some $\mu>3/4$,
then
\[\tilde E_{ij}(t,x)=E_{ij}(t,x)\]
for every $(t,x)\in\Omega$.
\end{proposition}

The following is a direct consequence of the symmetry and trace-free-ness of all terms in \Eqsref{eq:MarsSimonRed} and \eqref{eq:expansionS}.
\begin{corollary}
  \label{cor:tracelesssymmetric}
  In addition to the hypothesis of \Propref{prop:localisedexistence}, suppose that the field $\Ffield_{ij}$ is $\gmetr_{ij}$-traceless and symmetric on $\Omega$. Then the solution $ E_{ij}$ asserted by \Propref{prop:localisedexistence} is $\gmetr_{ij}$-traceless and symmetric.
\end{corollary}
More  specifically, this corollary is proved by first decomposing the unknown into its symmetric trace-free part, its antisymmetric part and its pure trace part. Due to the symmetry and trace-free-ness of all terms in the equations, the first part can be handled as above, while the equations for the second and third parts become trivial. All solutions compatible with condition \eqref{eq:conddecay}  therefore have vanishing antisymmetric and pure trace parts.

\subsection{The leading-order term}
\label{sec:deriveLOT}

According to \Sectionref{sec:spectraldecomp} there are real time-independent symmetric, $\hmetr_{ij}$-trace-free purely spatial tensor fields $E_{ij(1)}, \ldots, E_{ij(5)}$ which form an orthonormal basis of the complex vector space of symmetric and $\hmetr_{ij}$-trace-free purely spatial tensors at any $(t,x)$ in $\Mtilde$. 
Any complex symmetric  $\gmetr_{ij}$-trace-free $(0,2)$-tensor field (in particular the unknown of \eqref{eq:MarsSimonRed}) can therefore be expanded in $M$ as follows
\begin{equation}
  \label{eq:decomposeeigen}
  \mathcal E_{ij}(t,x)= \sum_{p=1}^5 f_p(t,x) \left(E_{ij(p)}(x)+O(t^2)\right),
\end{equation}
for complex functions $f_1,\ldots, f_5$ (whose regularity shall be specified later). The $O(t^2)$-correction is necessary here because the fields $E_{ij(p)}$ are $\hmetr_{ij}$-, and therefore not $\gmetr_{ij}$-trace-free.

Now we assume Gaussian coordinates $(t,x^i)$ where $t$ is the time function introduced earlier. Hence $\unitnormal^\mu=\partial_t^\mu$.
We plug \eqref{eq:decomposeeigen} into \eqref{eq:MarsSimonRed} with \eqref{eq:expansionS}, noticing that due to \Eqsref{gauge1} and \eqref{eq:metrics} the additional terms picked up when $\gnabla_0$ is expressed in terms of $\partial_t^\mu$ contribute smoothly to the last term of \eqref{eq:expansionS} only. Then the properties of the fields $E_{ij(p)}$ established before imply the system (index operations are performed with the metric $\hmetr_{ij}$ for this whole subsection)
\begin{align}
  \label{eq:f1evol}
   t\partial_t f_1
  + f_1  =&
     \imagi t\sum_{p=1}^5 \gvolf\indices{_{m}_{n} _{i}}\gmetr^{mk}\gmetr^{nl}\gnabla_{k} (f_p E_{jl(p)}) E^{ij}_{(1)}
     -t\frac{\imagi\lambda}2 f_1+t^2 \sum_{p=1}^5 \Tfield^p_1 f_p,\\
  \label{eq:f2evol}
  t\partial_t f_2
   =&
     \imagi t\sum_{p=1}^5 \gvolf\indices{_{m}_{n} _{i}}\gmetr^{mk}\gmetr^{nl}\gnabla_{k} (f_p E_{jl(p)}) E^{ij}_{(2)}
  +t^2 \sum_{p=1}^5 \Tfield^p_2 f_p,\\
  \label{eq:f3evol}
t\partial_t f_3
   =&
     \imagi t\sum_{p=1}^5 \gvolf\indices{_{m}_{n} _{i}}\gmetr^{mk}\gmetr^{nl}\gnabla_{k} (f_p E_{jl(p)}) E^{ij}_{(3)}
  +t^2 \sum_{p=1}^5 \Tfield^p_3 f_p,\\
  \label{eq:f4evol}
  t\partial_t f_4 -\frac 34 f_4
   =&
     \imagi t\sum_{p=1}^5 \gvolf\indices{_{m}_{n} _{i}}\gmetr^{mk}\gmetr^{nl}\gnabla_{k} (f_p E_{jl(p)}) E^{ij}_{(4)}
      +\frac 34 t\frac{\imagi\lambda}2 f_4
  +t^2 \sum_{p=1}^5 \Tfield^p_4 f_p,\\
  \label{eq:f5evol}
  t\partial_t f_5 -\frac 34 f_5
   =&
     \imagi t\sum_{p=1}^5 \gvolf\indices{_{m}_{n} _{i}}\gmetr^{mk}\gmetr^{nl}\gnabla_{k} (f_p E_{jl(p)}) E^{ij}_{(5)}
      +\frac 34 t\frac{\imagi\lambda}2 f_5
      +t^2 \sum_{p=1}^5 \Tfield^p_5 f_p,
\end{align}
for smooth, in principle known functions $\Tfield^p_q(t,x)$ defined for all $p,q=1,\ldots 5$.
The aim is now to derive expansions of the solutions of this system which shall  determine  the leading-order term of our singular initial value problem eventually. 

It is ok to use slightly loose and imprecise language in a first step now because we will eventually justify the resulting expressions fully rigorously.
With this in mind we suppose now that $f_1=O(t^{-1})$, and, $f_2, f_3,f_4,f_5=O(1)$ at $t=0$, possibly, with additional $\log t$ factors which we control formally by incorporating an arbitrarily small positive constant $\eta>0$ into the following arguments. Under this assumption,
\Eqref{eq:f1evol} can be simplified as follows
\[t\partial_t f_1
  + f_1  =
     \imagi t\hvolf\indices{_{i}^k^l}\hnabla_{k} (f_1 E_{jl(1)}) E^{ij}_{(1)}
     -t\frac{\imagi\lambda}2 f_1+O(t^{1-\eta}),
\]
where we recall that $\hvolf_{ijk}$ and $\hnabla_k$ are the volume form and covariant derivative associated with $\hmetr_{ij}$. 
The family of functions
\[f_1(t,x)=c_1(x) t^{-1}+ \imagi \hvolf\indices{_{i}^k^l}\hnabla_{k} (c_1(x) E_{jl(1)}(x)) E^{ij}_{(1)}(x)
     -\frac{\imagi\lambda}2 c_1(x)\]
given by an arbitrary (time independent) complex function $c_1$ satisfies this equation, i.e., it represents the leading terms of a formal expansion of \Eqref{eq:f1evol}.
Given this, \eqref{eq:f2evol} and \eqref{eq:f3evol} become (we refrain from writing the arguments $x$ and $t$ now)
\[t\partial_t f_{2,3}
   =
     \imagi \hvolf\indices{_{i}^k^l}\hnabla_{k} (c_1 E_{jl(1)}) E^{ij}_{(2,3)}
     +O(t^{1-\eta}),
\]
for which we find
\[f_{2,3}=\imagi \hvolf\indices{_{i}^k^l}\hnabla_{k} (c_1 E_{jl(1)}) E^{ij}_{(2,3)}\log t
+c_{2,3},\]
for arbitrary complex functions $c_2(x)$ and $c_3(x)$.
Finally, \eqref{eq:f4evol} and \eqref{eq:f5evol} are written as
\[ t\partial_t f_{4,5} -\frac 34 f_{4,5}
   =
     \imagi \hvolf\indices{_{i}^k^l}\hnabla_{k} (c_1 E_{jl(1)}) E^{ij}_{(4,5)}
      +O(t^{1-\eta}),
\]
which leads to
\[f_{4,5}=-\frac{4\imagi}3 \hvolf\indices{_{i}^k^l}\hnabla_{k} (c_1 E_{jl(1)}) E^{ij}_{(4,5)}
+c_{4,5} t^{3/4},\]
for arbitrary complex functions $c_4(x)$ and $c_5(x)$.

These formal leading-order expressions for the functions $f_1,\ldots,f_5$ can be simplified significantly using
\begin{equation}
\label{eq:bhsd2}
\imagi \hvolf\indices{_{i}^k^l}\hnabla_{k} (c_1 E_{jl(1)}) E^{ij}_{(p)}
=\imagi \hvolf\indices{_{i}^k^l} E_{jl(1)} E^{ij}_{(p)} \hnabla_{k} c_1+
\frac {3\imagi}2\lambda c_1 \delta_{p,1},
\end{equation}
which holds for every $p=1,\ldots,5$ as a consequence of \Eqsref{eq:blubbb5} and \eqref{eq:orthonormality},
and using \Eqref{eq:blubbbbb4}. Observe in particular that the logarithmic terms drop out thanks to \Eqref{eq:blubbbbb4}. Given all this and \Eqref{eq:decomposeeigen}, we set
\begin{align}
  \mathcal E_{kl,*}(t,x):=&
  \left(c_1(x) t^{-1}+\imagi\lambda c_1(x)
  \right) E_{kl(1)}(x)
+c_{2}(x)  E_{kl(2)}(x)
+c_{3}(x)  E_{kl(3)}(x)\notag\\
&+\left(c_{4}(x) t^{3/4}-\frac{2\imagi}{\sqrt 3}  \hnabla_{\frameh{2}{}} c_1(x)
\right) E_{kl(4)}(x) \notag\\
&+\left(c_{5}(x) t^{3/4}+\frac{2\imagi}{\sqrt 3}  \hnabla_{\frameh{1}{}} c_1(x)
\right) E_{kl(5)}(x) \notag\\
\begin{split}
=&\,\,c_1(x) t^{-1}E_{kl(1)}(x)\\
&+\imagi\lambda c_1(x) E_{kl(1)}(x)+c_{2}(x)  E_{kl(2)}(x)
+c_{3}(x)  E_{kl(3)}(x)\\
& -\frac{2\imagi}{\sqrt 3}  \hnabla_{\frameh{2}{}} c_1(x) E_{kl(4)}(x)
+\frac{2\imagi}{\sqrt 3}  \hnabla_{\frameh{1}{}} c_1(x)  E_{kl(5)}(x)\\
&+t^{3/4}\left(c_{4}(x) E_{kl(4)}(x)+c_{5}(x) t^{3/4} E_{kl(5)}(x)\right),
\label{eq:formalLOT}
\end{split}
\end{align}
which is determined by arbitrary time-independent complex functions $c_1,\ldots, c_5$. Anticipating the following results we shall refer to this field as the \keyword{leading-order term} of our singular initial value problem, and the functions $c_1,\ldots, c_5$ as the (singular) \keyword{data}.

Let us now assume as in \Propref{prop:localisedexistence} that $\SigmaF$ is any $3$-dimensional orientable manifold, $\delta>0$ is sufficiently small, $S_0$ is a non-empty open subset of $\SigmaF_0$ with compact closure and smooth non-empty boundary, and $\Omega$ is a lens-shaped region with respect to $S_0$. Let us further suppose that $c_1,\ldots, c_5$ are smooth complex functions on $S_0$. Since $\Omega$ is a proper subset of $M$, we can assume without loss of generality that $\SigmaF=T^3$ and that $c_1,\ldots, c_5$ have been extended as smooth functions to $T^3$ (where as before we consider $S_0$ as a subset of $T^3$). The field $\mathcal E_{ij,*}$ defined in \Eqref{eq:formalLOT} can then be interpreted as a smooth field for all $(t,x)\in (0,\delta)\times T^3$. When we now write $\mathcal E_{ij}=\mathcal E_{ij,*}+E_{ij}$
and plug this into \Eqref{eq:MarsSimonRed}, we obtain \Eqref{eq:MarsSimonRedMod} with
\begin{equation}
  \label{eq:FfieldLOT}
  \Ffield_{ij}:= -t\partial_t\mathcal E_{ij,*}
  +\imagi t \gvolf\indices{^{m}^{(l} _{(i}}\delta\indices{_{j)}^{k)}}\gnabla_{m}\mathcal E_{kl,*}+\Tfield\indices{^k^l_i_j}\mathcal E_{kl,*},
\end{equation}
which is therefore also a smooth field on $(0,\delta]\times T^3$. We check easily that the frame components satisfy
\[\Ffield_{ab}\in X_{\delta,1,\infty}(T^3).\]
The following proposition is then a consequence of \Propref{prop:localisedexistence} and Corollary~\ref{cor:tracelesssymmetric}.

\begin{proposition}
\label{prop:LOTsols}
Consider the  hypothesis of \Propref{prop:localisedexistence}, introduce Gauss coordinations and let $\Ffield_{ij}$ be given by \Eqsref{eq:formalLOT} and \eqref{eq:FfieldLOT} on $\Omega$ for arbitrary smooth time-independent complex functions $c_1,\ldots, c_5$. The solution $E_{ij}$ asserted by \Propref{prop:localisedexistence} for $\nu=1$ 
gives rise to a smooth solution $\mathcal E_{ij}=\mathcal E_{ij,*}+E_{ij}$ of the Mars-Simon evolution equations \eqref{eq:MarsSimonRed} on $\Omega$. In particular, the field $\mathcal E_{ij}$ is symmetric and $\gmetr_{ij}$-trace-free.
\end{proposition}

\subsection{The constraints}
\label{sec_constraints}

We have now constructed smooth solutions $\mathcal E_{ij}$ of the singular initial value problem of the evolution equations \eqref{eq:MarsSimonRed} on certain subsets $\Omega$ of $M$. Through the relation $\mathcal{E}_{ij} =\widetilde{\mathcal{T}}_{0 i0j}$, this determines the field $\widetilde{\mathcal{T}}_{\mu\nu\sigma\rho}$ on $\Omega$ which is supposed to be a MST. The next question we need to address is whether, given any such solution, the constraint violation quantities $\Dfield_i$ (see \Eqref{eq:MarsSimonRedConstr}) vanish identically on $\Omega$. Only if this is the case, this field $\widetilde{\mathcal{T}}_{\mu\nu\sigma\rho}$ is a solution of the \emph{full} Mars-Simon equations \eqref{sym_hyp0} on $\Omega$ and therefore a MST.

First we need to make sure
that the Buchdahl condition, see \Sectionsref{sec:realizationBuchdahlphys} and \ref{sec:realizationBuchdahl}, holds. To this end, we shall now assume that the background spacetime has a vanishing MST. The Buchdahl condition then reduces to \Eqref{eq:buchdahlspecial}. This means that we shall from now on only accept those solutions asserted by \Propref{prop:LOTsols} for which the associated tensor field $r_{\mu\nu\sigma\rho}=(\mcL_{\XKill}+
\widetilde F)\widetilde{\mathcal{T}}_{\mu\nu\sigma\rho}$ vanishes identically on $\Omega$. According to the discussion in \Sectionref{sec:realizationBuchdahl}, this tensor field satisfies the same evolution equations as $\widetilde{\mathcal{T}}_{\mu\nu\sigma\rho}$ and has the same algebraic properties as $\widetilde{\mathcal{T}}_{\mu\nu\sigma\rho}$. When \Lemref{lem:localuniqueness} is applied to the tensor field $\tilde r_{ij}=r_{0i0j}$ instead of $E_{ij}$, it follows that $\tilde r_{ij}$, and thereby $r_{\mu\nu\sigma\rho}$, vanish identically on $\Omega$ if the vector $\tilde r$ 
of its orthonormal frame components satisfies
\begin{equation}
  \label{eq:localuniqueness2}
  \lim_{t\searrow 0}\|t^{-\mu} \tilde r(t,\cdot)\|_{L^2(\Omega_t)}=0
\end{equation}
for some $\mu>3/4$. We conclude from \Eqsref{Lie_deriv_conf}, \eqref{eq:FFff} and \eqref{add_gauge} that this is the case if the orthonormal frame components of $\mcL_Y\mathcal{E}_{ij}$ vanish in this same sense in the limit $t\searrow 0$. Because of \Eqref{eq:Liedereigenfields}, this is the case for the class of solutions given by \Propref{prop:LOTsols}, if the Lie derivatives of the data $c_1,\ldots, c_5$ with respect to $Y^i$ vanish.

\begin{lemma}
  \label{lem:buchdahlevol}
  Consider the hypothesis of \Propref{prop:LOTsols}. Suppose in addition that the MST of the background spacetime vanishes identically on $\Omega$ and that
\[\mcL_Y c_1=\ldots=\mcL_Y c_5=0.\]
Then the solution $\mathcal E_{ij}$ asserted by \Propref{prop:LOTsols} of the evolution equations \eqref{eq:MarsSimonRed} satisfies the Buchdahl condition \Eqref{gen_Buchdahl_conf} on $\Omega$.
\end{lemma}
Recall \Sectionref{sec:generalizedbuchner}  where we established the constraint propagation system and discussed the role of the Buchdahl condition.
Close to $t=0$, this system takes the form
\begin{equation}
  \label{eq:aldjalsjdaskdjwe}
  t\gnabla_0 \Dfield_i-\frac \imagi2 t \gvolf\indices{_{i}^{j}^{k}}\gnabla_{j} \Dfield_k
+\frac 14\left(\delta\indices{_i^k}+13 Y_i Y^k\right)\Dfield_k+O(t) \Dfield_i=0,
\end{equation}
where index operations are performed with the metric $\gmetr_{ij}$.

Let us now make the same assumptions as in \Lemref{lem:buchdahlevol} where $\mathcal E_{ij}=\mathcal E_{ij,*}+E_{ij}$ is the smooth solution asserted by \Propref{prop:LOTsols}. The fields $\Dfield_i$ are determined by \Eqref{eq:MarsSimonRedConstr} from $\mathcal E_{ij}$, which are therefore also smooth fields on $\Omega$.
With similar arguments as in \Sectionref{sec:SIVPevol}, we establish from \Eqref{eq:aldjalsjdaskdjwe} that
\begin{align*}
  t\gstnabla_0(t^{-2\mu} \Dfield_i\bar \Dfield^i)
  =&-\gstnabla_{j}\left(\frac \imagi2 t^{-2\mu+1}\hvolf\indices{^{i}^{j}_{k}} \Dfield_i \bar \Dfield^k\right)\\
&-\frac 12 t^{-2\mu}\bar \Dfield^i\left((1+4\mu)\delta\indices{_i^k}+13 Y_i Y^k\right)\Dfield_k+O(t^{-2\mu}\Dfield_i \bar\Dfield^i).
\end{align*}
By choosing a time-dependent orthonormal frame with respect to $\gmetr_{ij}$ which is smooth through $t=0$ and which has the property that one of the frame vector fields agrees with $Y^i$ at $t=0$, we see that the map $(1+4\mu)\delta\indices{_i^k}+13 Y_i Y^k$ on the complex vector space of $(0,1)$-tensors at any $(t,x)$ in $\Omega$ is positive definite if $\mu>-1/4$.
Similar to our discussion in \Sectionref{sec:SIVPevol}, we then perform an integration over 
 any spacetime slab $\Omega_{(t_0,t)}\subset M$ with $0<t_0\le t<\delta$ with respect to the spacetime metric $\gmetr_{\mu\nu}$. As a consequence of a characteristic analysis of \Eqref{eq:aldjalsjdaskdjwe}, which establishes that all characteristic eigenvalues are $0$ or $\pm |\xi|/2$, all additional boundary terms which arise from the application of Stokes' theorem over the lens-shaped region have a
``good'' sign if $\delta>0$ is sufficiently small.  We conclude that
the constraint violation quantities $\Dfield_i$ vanish identically on $\Omega$ provided
$\lim_{t\searrow 0}\|t^{-\mu} \Dfield(t,\cdot)\|_{L^2(\Omega_t)}=0$ for some $\mu>-1/4$ where $\Dfield$ is the vector of orthonormal frame components.

When we now calculate the fields  $\Dfield_i$ using \Eqsref{eq:MarsSimonRedConstr} for any solution $\mathcal E_{ij}=\mathcal E_{ij,*}+E_{ij}$ asserted by \Propref{prop:LOTsols} with the additional hypothesis of \Lemref{lem:buchdahlevol},
we find that $\lim_{t\searrow 0}\|t^{-\mu} \Dfield(t,\cdot)\|_{L^2(\Omega_t)}=0$ for some $\mu>-1$ provided
\[\mcL_Y c_1=0.\]
This is therefore always the case under the hypothesis of \Lemref{lem:buchdahlevol}. Using  \Eqref{eq:blubbbbb44} we then show that $\lim_{t\searrow 0}\|t^{-\mu} \Dfield(t,\cdot)\|_{L^2(\Omega_t)}=0$ for some $\mu>-1/4$ provided
\[c_4=c_5=0.\]
We have therefore established the following result.

\begin{proposition}
  \label{prop:fullmarssimon}
  Consider the hypothesis of \Propref{prop:LOTsols}. Suppose in addition that the MST of the background spacetime vanishes identically on $\Omega$ and that
\[\mcL_Y c_1=\mcL_Y c_2=\mcL_Y c_3=0,\quad c_4=c_5=0.\]
Then the solution $\mathcal E_{ij}$ asserted by \Propref{prop:LOTsols} of the evolution equations \eqref{eq:MarsSimonRed} on $\Omega$ is a solution of the \emph{full} Mars-Simon equations on $\Omega$.
\end{proposition}

\subsection{Smooth extendibility through \texorpdfstring{${\scri^-}$}{Scri-}}
\label{sec:smoothextendibility}
Any of the solutions of the Mars-Simon equations constructed in \Propref{prop:fullmarssimon} is smooth on the lens-shaped region $\Omega$. If the datum $c_1$ does not vanish, however, it is singular in the limit $t\searrow 0$.
In fact, it has been observed in \cite{mpss} that such a divergent behavior of the rescaled Mars-Simon tensor at ${\scri^-}$ needs to be expected
and is in fact generic (unless certain components of the radiations field and the Cotton-York tensor of the induced metric on ${\scri^-}$ vanish).
 We now finally wish to show that if we subtract the leading singular term $c_1 t^{-1}E_{kl(1)}$ from the solution $\mathcal E_{ij}$, see \Eqref{eq:formalLOT}, the resulting field  extends smoothly through $t=0$. Observe that the $t^{3/4}$-terms in \Eqref{eq:formalLOT} are not present as a consequence of the hypothesis of \Propref{prop:fullmarssimon}.

In fact, we attempt to show now that the remainder field $E_{ij}$ can be written as a truncated Taylor series of arbitrary order $L>0$ for all $(t,x)\in \Omega$, i.e., that
\begin{equation}
  \label{eq:Tayloransatz}
  E_{ij}(t,x)=\sum_{\ell=1}^L E_{ij,\ell}(x) t^\ell+\tilde E_{ij}(t,x)
\end{equation}
for some new remainder $\tilde E_{ij}(t,x)$ with the property that $\lim_{t\searrow 0}\|t^{-\mu} \tilde E(t,\cdot)\|_{H^q(\Omega_t)}=0$ for any $\mu<L+1$ and any positive integer $q$, formed from the orthonormal frame components of $\tilde E_{ij}$.

To this end let us consider any solution $\mathcal E_{ij}$ asserted by \Propref{prop:LOTsols} under the additional conditions of \Propref{prop:fullmarssimon}. Then set $E_{ij}(t,x)=\mathcal E_{ij}-\mathcal E_{ij,*}$. Recall that $E_{ij}(t,x)$ satisfies \Eqref{eq:MarsSimonRedMod} with $\Ffield_{ij}$ given by \Eqref{eq:FfieldLOT}. Plugging \Eqref{eq:Tayloransatz} into \eqref{eq:MarsSimonRedMod}, we find that $\tilde E_{ij}$ satisfies the same equation \eqref{eq:MarsSimonRedMod}, just with a different source term
\begin{equation}
  \label{eq:defhatFij}
  \hat \Ffield_{ij}(t,x):=\Ffield_{ij}-t\partial_t \sum_{\ell=1}^L E_{ij,\ell}(x) t^\ell
  +i t \gvolf\indices{^{m}^{(l} _{(i}}\delta\indices{_{j)}^{k)}}\gnabla_{m}\sum_{\ell=1}^L E_{kl,\ell}(x) t^\ell+\Tfield\indices{^k^l_i_j}\sum_{\ell=1}^L E_{kl,\ell}(x) t^\ell.
\end{equation}
We know that $\Ffield_{ij}$ is a smooth field through $t=0$ and that it can hence be written  as
\[\Ffield_{ij}=\sum_{\ell=1}^L \Ffield_{ij,\ell}(x) t^\ell+t^{L+1}\,\tilde\Ffield_{ij}(t,x)\]
for some smooth fields $\Ffield_{ij,\ell}(x)$ and $\tilde\Ffield_{ij}(t,x)$ through $t=0$.
According to \Eqref{eq:expansionS} we can also write
\[\Tfield\indices{^k^l_i_j}(t,x)=\Tfield\indices{^k^l_i_j_{(0)}}(x)+t\,\tilde \Tfield\indices{^k^l_i_j}(t,x)\]
for another smooth field  $\tilde\Tfield\indices{^k^l_i_j}(t,x)$ through $t=0$.
All this yields that
\begin{align*}
  \hat \Ffield_{ij}&=\sum_{\ell=1}^L \left[\Ffield_{ij,\ell} t^\ell
-\ell E_{ij,\ell} t^\ell
  +i t^{\ell+1} \gvolf\indices{^{m}^{(l} _{(i}}\delta\indices{_{j)}^{k)}}\gnabla_{m} E_{kl,\ell} +\Tfield\indices{^k^l_i_j_{(0)}} E_{kl,\ell} t^\ell +t^{\ell+1}\,\tilde \Tfield\indices{^k^l_i_j} E_{kl,\ell} \right]\\
&\quad+t^{L+1}\,\tilde\Ffield_{ij}\\
&=\sum_{\ell=1}^L \left[\Ffield_{ij,\ell}
-\ell E_{ij,\ell}
  +i \gvolf\indices{^{m}^{(l} _{(i}}\delta\indices{_{j)}^{k)}}\gnabla_{m} E_{kl,{\ell-1}} +\Tfield\indices{^k^l_i_j_{(0)}} E_{kl,\ell} +\,\tilde \Tfield\indices{^k^l_i_j} E_{kl,\ell-1} \right]t^{\ell}\\
&\quad+t^{L+1}\left(\tilde\Ffield_{ij}+i \gvolf\indices{^{m}^{(l} _{(i}}\delta\indices{_{j)}^{k)}}\gnabla_{m} E_{kl,L}+\tilde \Tfield\indices{^k^l_i_j} E_{kl,L} \right).
\end{align*}
Hence, $\lim_{t\searrow 0}\|t^{-\nu} \hat F(t,\cdot)\|_{H^q(\Omega_t)}=0$ for $\nu=L+1$ and all $q$, formed from the frame components of $\hat \Ffield_{ij}$  -- and hence \Propref{prop:localisedexistence} applied to the equation for $\tilde E_{ij}$ discussed above establishes that the uniquely determined solution $\tilde E_{ij}$ on $\Omega$ has the required properties -- provided the following finite hierarchy of linear algebraic equations has a solution
\begin{align*}
  \left(\Tfield\indices{^k^l_i_j_{(0)}}-\delta\indices{^k_i}\delta\indices{^l_j}\right) E_{kl,1} &=-\Ffield_{ij,1},\\
  \left(\Tfield\indices{^k^l_i_j_{(0)}}-\ell \delta\indices{^k_i}\delta\indices{^l_j}\right) E_{kl,\ell} &=-\Ffield_{ij,\ell}-i \gvolf\indices{^{m}^{(l} _{(i}}\delta\indices{_{j)}^{k)}}\gnabla_{m} E_{kl,{\ell-1}}-\,\tilde \Tfield\indices{^k^l_i_j} E_{kl,\ell-1}
\end{align*}
for all $\ell=2,\ldots,L$. Indeed it does. The first equation has a unique solution $E_{kl,1}$ since $1$ is not an eigenvalue of $\Tfield\indices{^k^l_i_j_{(0)}}$, see \Sectionref{sec:spectraldecomp}. Given this solution $E_{kl,1}$, the second equation for $\ell=2$ has a unique solution $E_{kl,2}$ since $2$ is not an eigenvalue of $\Tfield\indices{^k^l_i_j_{(0)}}$. If $\ell$ is any integer $2\, \ldots, L$, given any solution $E_{kl,\ell-1}$, the second equation for $\ell$ has a unique solution $E_{kl,\ell}$ since $\ell$ is not an eigenvalue of $\Tfield\indices{^k^l_i_j_{(0)}}$.

We have therefore now fully established \Theoremref{thm:maintheorem}.


\section*{Acknowledgments}
TTP acknowledges financial support  by the Austrian Science Fund (FWF):  P~28495-N27.
TTP wishes to thank the Department of Mathematics and Statistics, University of Otago, for hospitality and
support during work on this paper.


\begin{thebibliography}{[10]}
\bibitem{aik1} S. Alexakis, A.D. Ionescu, S. Klainerman: \emph{Uniqueness of smooth stationary black holes in vacuum: Small perturbations of the Kerr spaces},
Comm. Math. Phys. \textbf{299} (2010) 89--127.
\bibitem{Ames:2012vz}
E.~Ames, F.~Beyer, J.~Isenberg, and P.~G. LeFloch.
\newblock {Quasilinear Hyperbolic Fuchsian Systems and AVTD Behavior in
  $T^2$-Symmetric Vacuum Spacetimes}.
\newblock {\em Ann. Henri Poincar{\'e}}, 14(6):1445--1523, 2013.
\bibitem{Ames:2013uh}
E.~Ames, F.~Beyer, J.~Isenberg, and P.~G. LeFloch.
\newblock {Quasilinear Symmetric Hyperbolic Fuchsian Systems in Several Space
  Dimensions}.
\newblock In M.~Agranovsky, M.~Ben-Artzi, G.~J. Galloway, L.~Karp,
  V.~Maz{\textquoteright}ya, S.~Reich, D.~Shoikhet, G.~Weinstein, and
  L.~Zalcman, editors, {\em Complex Analysis and Dynamical Systems V}. American
  Mathematical Society, Providence, Rhode Island, 2013.
\bibitem{Ames:2016uy}
E.~Ames, F.~Beyer, J.~Isenberg, and P.~G. LeFloch.
\newblock {A class of solutions to the Einstein equations with AVTD behavior in generalized wave gauges}.
\newblock {\em J. Geom. Phys.}, 121:42--71, 2017.
\bibitem{Andersson:2001fa}
L.~Andersson and A.~D. Rendall.
\newblock {Quiescent cosmological singularities}.
\newblock {\em Commun. Math. Phys.}, 218(3):479--511, 2001.
\bibitem{bc} R. Beig, P.T. Chru\'sciel: \emph{Shielding linearised gravity},  Phys. Rev. D \textbf{95} (2017) 064063.
\bibitem{Beyer:2010foa}
F.~Beyer and P.~G. LeFloch.
\newblock {Second-order hyperbolic Fuchsian systems and applications}.
\newblock {\em Class. Quantum Grav.}, 27(24):245012, 2010.
\bibitem{Beyer:2011ce}
F.~Beyer and P.~G. LeFloch.
\newblock {Second-order hyperbolic Fuchsian systems: Asymptotic behavior of
  geodesics in Gowdy spacetimes}.
\newblock {\em Phys. Rev. D}, 84(8):084036, 2011.
\bibitem{beyer11}
 F. Beyer and J. Hennig, 
 \emph{Smooth Gowdy-symmetric generalized Taub-NUT solutions},
 Class.\ Quantum Grav.\ {\bf 29}, 245017 (2012)
\bibitem{Beyer:2015fhs}
F.~Beyer and P.~G. LeFloch.
\newblock {Self-gravitating fluid flows with Gowdy symmetry near cosmological singularities}.
\newblock {\em Commun. Part. Diff. Eq.}, 42(8):1199--1248, 2017.
\bibitem{buchdahl} H.A. Buchdahl: \emph{On  the  compatibility  of  relativistic  wave  equations  for  particles  of  higher  spin  in  the
presence of a gravitational field},
Nuovo Cimento {\bf 10} (1958) 96--103.
\bibitem{ChoquetBruhat:2006fc}
Y.~Choquet-Bruhat and J.~Isenberg.
\newblock {Half polarized $U(1)$-symmetric vacuum spacetimes with AVTD
  behavior}.
\newblock {\em J. Geom. Phys.}, 56(8):1199--1214, 2006.
\bibitem{c_BHs} P.T. Chru\'sciel: \emph{The geometry of black holes},
lecture notes (2015), \url{http://homepage.univie.ac.at/piotr.chrusciel/teaching/BlackHoles/BlackHolesViennaJanuary2015.pdf}.
\bibitem{Claudel:1998tt}
C.~M. Claudel and K.~P. Newman.
\newblock {The Cauchy problem for quasi-linear hyperbolic evolution problems
  with a singularity in the time}.
\newblock {\em Proc. R. Soc. London A}, 454(1972):1073--1107, 1998.
\bibitem{Damour:2002exa}
T.~Damour, M.~Henneaux, A.~D. Rendall, and M.~Weaver.
\newblock {Kasner-like behaviour for subcritical Einstein-matter systems}.
\newblock {\em Ann. Henri Poincar{\'e}}, 3(6):1049--1111, 2002.
\bibitem{F_lambda}  H. Friedrich: \textit{Existence and structure of past asymptotically simple solutions of Einstein's field equations with positive cosmological constant}, J. Geom. Phys. \textbf{3} (1986) 101--117.
 \bibitem{F3} H. Friedrich: \textit{Conformal Einstein evolution}, in: \textit{The conformal structure of space-time -- Geometry, analysis, numerics}, J. Frauendiener, H. Friedrich (eds.),  Berlin, Heidelberg: Springer, 2002,  1--50.
\bibitem{Friedrich:2000wp}
H.~Friedrich and A.~D. Rendall.
\newblock {The Cauchy Problem for the Einstein Equations}.
\newblock In {\em Einstein{\textquoteright}s Field Equations and Their Physical
  Implications}, pages 127--223. Springer Berlin Heidelberg, Berlin,
  Heidelberg, 2000.
\bibitem{lrr-2008-1}
M.~Henneaux, D.~Persson, and P.~Spindel.
\newblock {Spacelike singularities and hidden symmetries of gravity}.
\newblock {\em Living Rev. Relativity}, 11(1), 2008.
\bibitem{hintz_vasy} P. Hintz, A. Vasy: \emph{The global non-linear stability of the Kerr-de Sitter family of black holes},
(2016),   arXiv:1606.04014 [math.DG].
\bibitem{IK} A.D. Ionescu, S. Klainerman: \emph{On the uniqueness of smooth, stationary black holes in vacuum},
Invent. Math. \textbf{175} (2009) 35--102.
\bibitem{Isenberg:1999ba}
J.~Isenberg and S.~Kichenassamy.
\newblock {Asymptotic behavior in polarized $T^2$-symmetric vacuum
  space{\textendash}times}.
\newblock {\em J. Math. Phys.}, 40(1):340, 1999.
\bibitem{Isenberg:2002ku}
J.~Isenberg and V.~Moncrief.
\newblock {Asymptotic behaviour in polarized and half-polarized $U(1)$
  symmetric vacuum spacetimes}.
\newblock {\em Class. Quantum Grav.}, 19(21):5361, 2002.
\bibitem{israel}
W. Israel: \emph{Differential forms in general relativity}, Commun. of the Dublin Institute for Advanced Studies, Series A, \textbf{19} (1970) 1--100.
\bibitem{mars} M. Mars: \textit{A spacetime characterization of the Kerr metric}, Class. Quantum Grav. \textbf{16} (1999) 2507--2523.
\bibitem{mars1}  M. Mars: \emph{Uniqueness properties of the {K}err metric}, Class. Quantum   Grav. \textbf{17} (2000) 3353--3373.
\bibitem{mpss} M. Mars, T.-T. Paetz, J.M.M. Senovilla, W. Simon: \emph{Characterization of (asymptotically) Kerr-de Sitter-like spacetimes at null infinity},  Class. Quantum Grav. \textbf{33} (2016) 155001.
\bibitem{mars-senovilla} M. Mars, J.M.M. Senovilla: \textit{A spacetime characterization of the Kerr-NUT-(A)de Sitter and related metrics},
Ann. Henri Poincar\'e {\bf 16} (2015) 1509--1550.
\bibitem{mars-senovilla-null} M. Mars, J.M.M. Senovilla: \textit{Spacetime characterizations of $\Lambda$-vacuum metrics with a null Killing 2-form},
Class. Quantum Grav. \textbf{33} (2016) 195004.
\bibitem{ttpKIDs} T.-T. Paetz: \emph{KIDs prefer special cones}, Class. Quantum Grav. \textbf{31} (2014) 085007.
\bibitem{ttp2} T.-T. Paetz: \emph{Killing Initial Data on space-like conformal boundaries}, (2014),
J. Geom. Phys. \textbf{106} (2016) 51--69.
\bibitem{kerr_char} T.-T. Paetz: \emph{ Algorithmic characterization results for the Kerr-NUT-(A)dS space-time. I. A space-time approach },   J. Math. Phys. \textbf{58} (2017) 042501.
\bibitem{p1} R. Penrose: \textit{Asymptotic properties of fields and space-time}, Phys. Rev. Lett. \textbf{10} (1963) 66--68.
\bibitem{p2} R. Penrose: \textit{Zero rest-mass fields including gravitation: Asymptotic behavior},  Proc. R. Soc. Lond. A \textbf{284} (1965) 159--203.
\bibitem{Rendall:2000ki}
A.~D. Rendall.
\newblock {Fuchsian analysis of singularities in Gowdy spacetimes beyond
  analyticity}.
\newblock {\em Class. Quantum Grav.}, 17(16):3305--3316, 2000.
\bibitem{Ringstrom:2009cj}
H.~Ringstr{\"o}m.
\newblock {\em {The Cauchy problem in General Relativity}}.
\newblock ESI Lectures in Mathematics and Physics. European Mathematical
  Society, Z{\"u}rich, Switzerland, 2009.
\bibitem{simon} W. Simon: \emph{Characterizations of the Kerr metric},  Gen. Rel. Grav. \textbf{16} (1984) 465--476.
 \bibitem{wald} R.M. Wald: \textit{General relativity},  Chicago and London: The University of Chicago Press, 1984.
\bibitem{wald2} R.M. Wald: \textit{Spin-two fields and general covariance},  Phys. Rev. \textbf{D 33}  (1986) 3613--3625.
\end{thebibliography}
\end{document}